\documentclass[prb,aps,tightenlines,twocolumn,floatfix,superscriptaddress,showpacs,preprintnumbers,citeautoscript,10pt]{revtex4-1}
\usepackage{amsmath}
\usepackage{amssymb}
\usepackage{mathrsfs}
\usepackage{bm}
\usepackage{xfrac}
\usepackage{graphicx}
\usepackage{url}
\usepackage{tikz}
\usepackage{multirow}
\usepackage{braket}
\usepackage{courier}
\usepackage[version=4]{mhchem}
\usepackage[T1]{fontenc}
\usepackage{scalerel}
\usepackage{soul}
\usepackage{algorithm}
\usepackage{algpseudocode}
\usepackage{tabularx}
\usepackage{xspace}
\usepackage{microtype}
\usepackage{upgreek}

\algnewcommand{\Or}{\textbf{ or }}
\algnewcommand{\And}{\textbf{ and }}
\algnewcommand{\Break}{\textbf{break}}
\algrenewcommand\algorithmicforall{\textbf{foreach}}
\def\dd{\mathrm{d}}
\newcommand{\RN}[1]{\textup{\uppercase\expandafter{\romannumeral#1}}}
\newcommand{\BRN}[1]{\textbf{\textup{\uppercase\expandafter{\romannumeral#1}}}}
\definecolor{dartmouthgreen}{rgb}{0.05, 0.5, 0.06}
\newcommand{\tphin}[1]{\widetilde{\phi}_{#1}}
\newcommand{\tphi}[1]{\widetilde{\phi}_{#1}(\bm r)}
\newcommand{\tphib}[1]{\widetilde{\phi}_{#1}(\overline{\bm r})}
\newcommand{\trho}[1]{\widetilde{\rho}_{#1}(\bm r)}
\newcommand{\trhon}[1]{\widetilde{\rho}_{#1}}
\newcommand{\trhonb}[1]{\widetilde{\bm \rho}_{#1}}
\newcommand{\trhob}[1]{\widetilde{\rho}_{#1}(\overline{\bm r})}

\newcommand{\terhonb}[1]{\widetilde{\bm\rho}^{b}_{#1}}
\newcommand{\tv}[1]{\widetilde{v}_{#1}(\bm r)}
\newcommand{\tvn}[1]{\widetilde{v}_{#1}}
\newcommand{\tvnb}[1]{\widetilde{\bm v}_{#1}}
\newcommand{\tvb}[1]{\widetilde{v}_{#1}(\overline{\bm r})}
\newcommand{\tC}[1]{\widetilde{\bm C}_{#1}}
\newcommand{\gC}[1]{\bm C_{#1}}
\newcommand{\xmega}[1]{\Omega_{#1}}
\newcommand{\xto}[2]{#1 \rightarrow #2}
\newcommand{\rbar}{\overline{\bm r}}

\newcommand{\rpr}{R_{\rm pair}}
\newcommand{\rpe}{R_{\rm PE}}
\newcommand{\rme}{R_{\rm ME}}
\newcommand{\npe}{N_{\rm PE}}
\newcommand{\nme}{N_{\rm ME}}
\newcommand{\rpes}{R_{\rm PE}^{\rm s}}
\newcommand{\rmes}{R_{\rm ME}^{\rm s}}
\newcommand{\rpep}{R_{\rm PE}^{\rm ns}}
\newcommand{\rmep}{R_{\rm ME}^{\rm ns}}
\newcommand{\exx}{E_{\rm xx}}
\newcommand{\exxr}{E_{\rm xx}^{\rm ref}}

\newcommand{\dxx}[1]{\widetilde{D}_{\rm xx}^{#1}(\bm r)}
\newcommand{\dxxb}[1]{\widetilde{D}_{\rm xx}^{#1}(\overline{\bm r})}
\newcommand{\dxxr}[1]{\widetilde{D}_{\rm xx}^{#1, \rm ref}(\bm r)}
\newcommand{\dxxn}[1]{\widetilde{D}_{\rm xx}^{#1}}
\newcommand{\lz}{\mathcal{L}_{0}}
\newcommand{\lzb}[1]{\mathcal{L}_{0}[#1]}
\newcommand{\lu}{\mathcal{L}}
\newcommand{\lub}[1]{\mathcal{L}[#1]}
\newcommand{\pe}[1]{\Theta(#1, \rpe)}
\newcommand{\me}[1]{\Theta(#1, \rme)}
\newcommand{\pes}[1]{\Theta(#1, \rpes)}
\newcommand{\mes}[1]{\Theta(#1, \rmes)}
\newcommand{\pep}[1]{\Theta(#1, \rpep)}
\newcommand{\mep}[1]{\Theta(#1, \rmep)}
\newcommand{\mpi}[0]{\texttt{MPI}}
\newcommand{\omp}[0]{\texttt{OpenMP}}

\newcommand{\exxm}[0]{\texttt{exx}\xspace}
\newcommand{\qe}[0]{\texttt{Quantum ESPRESSO} }
\newcommand{\ie}[0]{\textit{i.e.}, }
\newcommand{\eg}[0]{\textit{e.g.}, }
\newcommand{\via}[0]{\textit{via} }
\newcommand{\cf}[0]{\textit{cf}. }
\newcommand{\bl}[1]{\textcolor{blue}{#1}}

\usetikzlibrary{calc}
\usetikzlibrary{positioning}
\usetikzlibrary{shapes,arrows}
\usetikzlibrary{arrows.meta}
\tikzstyle{if} = [diamond, draw, fill=blue!20, text width=4.5em, text badly centered, node distance=3cm, inner sep=0pt]
\tikzstyle{prg} = [rectangle, draw, fill=blue!40, text width=5em, text centered, rounded corners, minimum height=4em]
\tikzstyle{prgN} = [rectangle, draw, fill=brown!50, text width=5em, text centered, rounded corners, minimum height=4em]
\tikzstyle{sbrt} = [rectangle, draw, fill=blue!0, text width=15em, text centered, rounded corners, minimum height=5em]
\tikzstyle{sbrtR} = [rectangle, draw, red, fill=blue!0, text width=15em, text centered, rounded corners, minimum height=5em]
\tikzstyle{sbrtN} = [rectangle, draw, fill=brown!30, text width=6.2em, text centered, rounded corners, minimum height=4em]
\tikzstyle{mod} = [rectangle, draw, fill=yellow!20, text width=5em, text centered, rounded corners, minimum height=4em]
\tikzstyle{inp} = [circle, draw, fill=purple!40, text width=3.5em, text centered, rounded corners, minimum height=3em]
\tikzstyle{inpN} = [circle, draw, fill=red!40, text width=3.5em, text centered, rounded corners, minimum height=3em]
\tikzstyle{outp} = [circle, draw, fill=yellow!40, text width=3.5em, text centered, rounded corners, minimum height=3em]
\tikzstyle{line} = [draw, -latex']
\tikzstyle{set} = [draw, ellipse,fill=red!20, node distance=3cm, minimum height=2em]
\tikzstyle{var} = [draw, circle,fill=green!0, node distance=3cm,text width=5em,text centered, minimum height=2em] 
\tikzstyle{vary} = [draw, circle,fill=yellow!30, node distance=3cm,text width=5em,text centered, minimum height=2em] 
\tikzstyle{varg} = [draw, circle,fill=violet!30, node distance=3cm,text width=5em,text centered, minimum height=2em] 
\tikzstyle{varb} = [draw, circle,fill=brown!30, node distance=3cm,text width=5em,text centered, minimum height=2em] 
\tikzstyle{varN} = [draw, circle,fill=green!40, node distance=3cm,text width=3.5em,text centered, minimum height=2em] 
\tikzstyle{dummy} = [] 
\tikzstyle{comm} = [rectangle, draw, fill=cyan!20, text width=5em, text centered, rounded corners, minimum height=4em]

\begin{document}
\title{Enabling Large-Scale Condensed-Phase Hybrid Density Functional Theory Based \\ \textit{Ab Initio} Molecular Dynamics I: Theory, Algorithm, and Performance}
\author{Hsin-Yu Ko}
\affiliation{Department of Chemistry and Chemical Biology, Cornell University, Ithaca, NY 14853, USA}
\affiliation{Department of Chemistry, Princeton University, Princeton, NJ 08544, USA}
\author{Junteng Jia}
\affiliation{Department of Chemistry and Chemical Biology, Cornell University, Ithaca, NY 14853, USA}
\author{Biswajit Santra}
\affiliation{Department of Chemistry, Princeton University, Princeton, NJ 08544, USA}
\affiliation{Department of Physics, Temple University, Philadelphia, PA 19122, USA}
\author{Xifan Wu}
\affiliation{Department of Physics, Temple University, Philadelphia, PA 19122, USA}
\author{Roberto Car}
\affiliation{Department of Chemistry, Princeton University, Princeton, NJ 08544, USA}
\affiliation{Department of Physics, Princeton University, Princeton, NJ 08544, USA}
\author{Robert A. DiStasio Jr.}
\email{distasio@cornell.edu} 
\affiliation{Department of Chemistry and Chemical Biology, Cornell University, Ithaca, NY 14853, USA}

\date{\today}

\begin{abstract}
By including a fraction of exact exchange (EXX), hybrid functionals reduce the self-interaction error in semi-local density functional theory (DFT), and thereby furnish a more accurate and reliable description of the underlying electronic structure in systems throughout biology, chemistry, physics, and materials science.
However, the high computational cost associated with the evaluation of all required EXX quantities has limited the applicability of hybrid DFT in the treatment of large molecules and complex condensed-phase materials.
To overcome this limitation, we describe a linear-scaling approach that utilizes a local representation of the occupied orbitals (\eg maximally localized Wannier functions (MLWFs)) to exploit the sparsity in the real-space evaluation of the quantum mechanical exchange interaction in finite-gap systems.
In this work, we present a detailed description of the theoretical and algorithmic advances required to perform MLWF-based \textit{ab initio} molecular dynamics (AIMD) simulations of large-scale condensed-phase systems of interest at the hybrid DFT level.
We focus our theoretical discussion on the integration of this approach into the framework of Car-Parrinello AIMD, and highlight the central role played by the MLWF-product potential (\ie the solution of Poisson's equation for each corresponding MLWF-product density) in the evaluation of the EXX energy and wavefunction forces.
We then provide a comprehensive description of the \exxm algorithm implemented in the open-source \qe program, which employs a hybrid \mpi{}/\omp{} parallelization scheme to efficiently utilize the high-performance computing (HPC) resources available on current- and next-generation supercomputer architectures.
This is followed by a critical assessment of the accuracy and parallel performance (\eg strong and weak scaling) of this approach when performing AIMD simulations of liquid water in the canonical ($NVT$) ensemble.
With access to HPC resources, we demonstrate that \exxm enables hybrid DFT based AIMD simulations of condensed-phase systems containing $500\mathrm{-}1000$~atoms (\eg \ce{(H2O)256}) with a walltime cost that is comparable to semi-local DFT.
In doing so, \exxm takes us one step closer to routinely performing AIMD simulations of complex and large-scale condensed-phase systems for sufficiently long timescales at the hybrid DFT level of theory.
\end{abstract}

\maketitle

\section{Introduction \label{sec:intro}} 
In view of its quite favorable balance of accuracy and computational cost, Kohn-Sham (KS) density functional theory~\cite{hohenberg_inhomogeneous_1964,kohn_self-consistent_1965,jones_density_1989,parr_density-functional_1989} (DFT) has become the most widely used electronic structure method for~\textit{ab initio} molecular dynamics (AIMD) simulations of large molecules and complex condensed-phase materials~\cite{burke_perspective_2012,marx_ab_2009,iftimie_ab_2005}.
Within the framework of KS-DFT, the total ground-state energy ($E$) is given as the sum of the following contributions:
\begin{equation}
  E = E_{\rm kin} + E_{\rm ext} + E_{\rm H} + E_{\rm xc} ,
  \label{eq:totE_ctr}
\end{equation}
in which $E_{\rm kin}$ is the KS kinetic energy, $E_{\rm ext}$ is the external potential which accounts for the nuclear-electronic and nuclear-nuclear potential energies (as well as any external fields), $E_{\rm H}$ is the Hartree energy, \ie the average (classical) Coulomb interaction energy of the electrons, and $E_{\rm xc}$ is the electronic exchange-correlation (xc) energy. Explicit forms for all of the energy contributions in Eq.~\eqref{eq:totE_ctr} are known except $E_{\rm xc}$, the approximation of which is still the subject of active research to date.

Functional approximations to $E_{\rm xc}$ are often described as the rungs of ``Jacob's Ladder,'' which connect the Hartree world to the exact solution of the time-independent Schr{\"o}dinger equation~\cite{manual_perdew_jacobs_2001}.
In this hierarchical classification of DFT, the first rung is given by the local (spin) density approximation (LDA)~\cite{ceperley_ground_1980,parr_density-functional_1989}, in which the form of $E_{\rm xc}^{\rm LDA}$ is obtained from the solution to the homogeneous electron gas.
As such, LDA works particularly well for systems with a (nearly) uniform electron density ($\rho(\bm r)$), \eg the valence electrons in metallic solids. The next rung includes xc functionals based on the semi-local generalized gradient approximation (GGA)~\cite{becke_density-functional_1988,lee_development_1988,perdew_generalized_1996}, which utilize the gradient of the electron density ($\bm \nabla \rho(\bm r)$) to correct the LDA description of systems with spatially varying $\rho(\bm r)$, \eg molecules and heterogeneous materials. At the current time, GGAs such as the non-empirical Perdew-Burke-Ernzerhof (PBE) xc functional~\cite{perdew_generalized_1996} are the computational workhorses for AIMD simulations of condensed-phase systems containing 100s--1000s of atoms. In this size regime, GGA-based approaches provide a favorable compromise between accuracy and computational cost, and have been quite successful in qualitatively (and sometimes even quantitatively) describing a number of systems and processes of interest throughout chemistry, physics, and materials science. 

Despite such widespread success, GGA functionals are unable to account for non-local electron correlation effects, which are responsible for the ubiquitous class of dispersion (or van der Waals) interactions. As such, several approaches have been devised to incorporate these long-range forces into the framework of DFT~\cite{klimes_perspective:_2012,grimme_dispersion-corrected_2016,hermann_first-principles_2017,berland_van_2015}, and include effective pairwise models~\cite{becke_exchange-hole_2007,tkatchenko_accurate_2009,grimme_consistent_2010,ferri_electronic_2015,caldeweyher_extension_2017}, methods that account for many-body dispersion interactions~\cite{tkatchenko_accurate_2012,distasio_jr._collective_2012,distasio_jr._many-body_2014,ambrosetti_long-range_2014,blood-forsythe_analytical_2016}, as well as non-local xc functionals~\cite{dion_van_2004,vydrov_nonlocal_2009,lee_higher-accuracy_2010}. We note in passing that third-rung meta-GGA functionals, which incorporate second-derivative information \via the Laplacian ($\nabla^2 \rho(\bm r)$) or the kinetic-energy density ($\tau(\bm r)$), are able to account for intermediate-range correlation effects~\cite{ghosh_phase-space_1986,becke_exchange_1989,tao_climbing_2003,zhao_new_2006,sun_semilocal_2013,sun_strongly_2015,yu_perspective:_2016}.
As such, these approaches have experienced a resurgence with the recent introduction of the SCAN functional~\cite{sun_strongly_2015}, which has shown promising results for bulk water systems~\cite{sun_accurate_2016,chen_ab_2017,zheng_structural_2018,lacount_ensemble_2019,xu_first-principles_2019} and interfacial water~\cite{calegari_andrade_structure_2018,calegari_free_2020}.

Another significant shortcoming associated with GGA (as well as meta-GGA) functionals is their propensity to suffer from self-interaction error (SIE), an artifact in approximate xc functionals that manifests as a spurious interaction between an electron and itself~\cite{perdew_self-interaction_1981,cohen_insights_2008}.
In the presence of SIE, $\rho(\bm r)$ is too delocalized, which in turn often leads to deleterious effects such as inadequate descriptions of transition states and charge transfer complexes~\cite{grafenstein_impact_2003,lundberg_quantifying_2005,leblanc_pervasive_2018}, underestimation of band gaps~\cite{janesko_screened_2009}, overestimation of lattice parameters in a wide variety of solids~\cite{marsman_hybrid_2008}, as well as excessive proton delocalization in liquid water~\cite{zhang_first_2011,zhang_structural_2011,gaiduk_first-principles_2018}, to name a few.
While SIE can be largely eliminated by self-interaction correction (SIC) based methods,~\cite{perdew_self-interaction_1981,manual_perdew_size-consisteny_1990,manual_pederson_self-interaction_2012,pederson_communication:_2014} the most commonly adopted approach for ameliorating the SIE present in semi-local KS-DFT is through the admixture of a fraction of exact exchange (EXX) in the underlying GGA (or meta-GGA) xc functional~\cite{becke_densityfunctional_1993}. These so-called hybrid (or hyper-GGA) xc functionals constitute the fourth rung in the DFT hierarchy, and can be written as (shown here as a correction to a GGA xc functional):
\begin{equation}
  E_{\rm xc}^{\rm hybrid} = a_x E_{\rm xx} + \left( 1 - a_x \right) E_{\rm x}^{\rm GGA} + E_{\rm c}^{\rm GGA} ,
  \label{eq:hyb}
\end{equation}
in which $E_{\rm xx}$ is the EXX energy, $E_{\rm x}^{\rm GGA}$ is the GGA exchange energy, and $E_{\rm c}^{\rm GGA}$ is the GGA correlation energy. The mixing parameter ($a_x$) in this expression depends on the hybrid xc functional approximation~\cite{perdew_rationale_1996,becke_new_1993,becke_densityfunctional_1993}, the optimal value of which (for a given system) can be determined from a self-consistent GW calculation~\cite{skone_self-consistent_2014}. By reducing the SIE, hybrid xc functionals are typically more accurate than GGA (or meta-GGA) approaches, in particular for the prediction of lattice parameters~\cite{marsman_hybrid_2008}, reaction energy barriers~\cite{grafenstein_impact_2003,lundberg_quantifying_2005,leblanc_pervasive_2018}, and band gaps~\cite{garza_predicting_2016}.
In this work, we limit our focus to the non-empirical PBE0~\cite{perdew_rationale_1996} hybrid xc functional, in which $a_x = \sfrac{1}{4}$ and the PBE GGA functional~\cite{perdew_generalized_1996} is used for $E_{\rm x}^{\rm GGA}$ and $E_{\rm c}^{\rm GGA}$.
Application of our approach (which is described below) to other popular hybrid xc functionals such as B3LYP~\cite{becke_densityfunctional_1993,lee_development_1988} is straightforward.

For a closed-shell system with $N_{o}$ doubly occupied orbitals (bands), $E_{\rm xx}$ can be written as:
\begin{equation}
  E_{\rm xx}=-\sum_{ij} \int \dd \bm{r} \int \dd \bm{r}' \, \frac{\phi_{i}^{*}(\bm{r})\phi_{j}^{*}(\bm{r}')\phi_{j}(\bm{r})\phi_{i}(\bm{r}')}{\left| \bm{r}-\bm{r}' \right|} ,
  \label{eq:exxGen0}
\end{equation}
in which $\phi_i$ and $\phi_j$ represent the occupied KS orbitals and the sum extends over all $N_o$ states. Defining the orbital-product density as 
\begin{equation}
  \rho_{ij}(\bm{r}) \equiv \phi_{i}^{*}(\bm{r}) \phi_{j}(\bm{r}) ,
  \label{eq:rho_from_phi}
\end{equation}
and the corresponding orbital-product potential (\ie the Coulomb potential felt by a test charge located at $\bm{r}$ originating from the $\rho_{ij}(\bm{r}')$ charge distribution) as
\begin{equation}
  v_{ij}(\bm{r}) \equiv \int \dd\bm{r}' \, \frac{\rho_{ij}(\bm{r}')}{|\bm{r}-\bm{r}'|} ,
  \label{eq:vxxGen}
\end{equation}
allows one to express Eq.~\eqref{eq:exxGen0} in the following compact form:
\begin{equation}
  E_{\rm xx} = -\sum_{ij} \int \dd \bm{r} \, \rho_{ij}(\bm{r}) v_{ji}(\bm{r}) .
  \label{eq:exxGen}
\end{equation}
Evaluation of $v_{ji}(\bm{r})$ is therefore of central importance in EXX calculations. For periodic systems, this quantity is usually computed through the convolution theorem (shown here at the $\Gamma$-point only),
\begin{align}
  \rho_{ji}(\bm{r}) &\xrightarrow{\texttt{fwdFFT}} \rho_{ji}(\bm G) \nonumber \\
  v_{ji}(\bm{G}) = 4\pi \frac{\rho_{ji}(\bm{G})}{\left|\bm{G}\right|^2}  &\xrightarrow{\texttt{invFFT}} v_{ji}(\bm r) ,
  \label{eq:convol}
\end{align}
in which $\rho_{ji}(\bm{G})$ and $v_{ji}(\bm{G})$ are the Fourier coefficients of $\rho_{ji}(\bm{r})$ and $v_{ji}(\bm{r})$, respectively.
We note in passing that the divergence of $v_{ji}(\bm{G})$ when $\bm{G}=\bm 0$ needs to be treated with care when evaluating $\exx$ using reciprocal-space methods.
In the real-space algorithm described herein, we sidestep this divergence as $v_{ji}(\bm{G}=\bm 0)$ is implicitly determined by the boundary conditions imposed during the solution of Poisson's equation (\ie $v_{ji}(r \rightarrow \infty) = 0$).
The computational scaling associated with both the forward (\texttt{fwdFFT}) and inverse (\texttt{invFFT}) fast Fourier transforms is $\mathcal{O}(N_{\rm FFT} \log N_{\rm FFT})$, where $N_{\rm FFT}$ is the size of the reciprocal space (planewave) grid, which grows linearly with system size.
Since the evaluation of $E_{\rm xx}$ in Eq.~\eqref{eq:exxGen} requires a sum over the contributions from all $N_o(N_o+1)/2$ unique pairs of occupied orbitals, the overall computational scaling becomes $\mathcal{O}(N_o^{2} N_{\rm FFT} \log N_{\rm FFT})$.
Neglecting the logarithmic dependence, the resulting cubic-scaling cost makes this reciprocal-space EXX algorithm quite computationally demanding and limits routine performance of hybrid DFT based AIMD simulations on large-scale condensed-phase systems.
Hence, most condensed-phase calculations with hybrid DFT still remain limited to predicting energetic and structural properties in the absence of thermal effects.

Significant progress has been made to accelerate condensed-phase EXX calculations by employing the following theoretical and numerical techniques: range-separation~\cite{heyd_hybrid_2003} or truncation~\cite{guidon_robust_2009} of the underlying Coulomb operator, implementation of massively parallel algorithms,~\cite{duchemin_scalable_2010,bylaska_parallel_2011,barnes_improved_2017,varini_enhancement_2013} employment of auxiliary atom-centered (localized) basis sets,~\cite{guidon_auxiliary_2010,hu_interpolative_2017,dong_interpolative_2018} adaptive compression (low-rank decomposition) of the EXX operator (ACE),~\cite{lin_adaptively_2016,jia_fast_2019,lin_convergence_2019} use of the projected commutator direct inversion of the iterative subspace (PC-DIIS) method to reduce the number of self-consistent field (SCF) iterations,~\cite{hu_projected_2017} utilization of sparsity through localization methods (\eg maximally localized Wannier functions (MLWFs),~\cite{marzari_maximally_1997,wu_order-n_2009,marzari_maximally_2012,distasio_jr._individual_2014} recursive subspace bisection (RSB),~\cite{gygi_compact_2009,gygi_efficient_2013} selected columns of the density matrix (SCDM),~\cite{damle_compressed_2015,damle_computing_2017,damle_scdm-k:_2017} and other localized representations~\cite{mountjoy_exact_2017}), as well as combinations thereof.~\cite{izmaylov_efficient_2006,guidon_ab_2008,guidon_robust_2009,guidon_auxiliary_2010,carnimeo_fast_2018}

To enable large-scale hybrid DFT based AIMD simulations in the condensed phase, the most promising methods for reducing the intrinsic computational cost and scaling associated with EXX exploit sparsity \via localized representations of the occupied space or density matrix.
For example, the RSB method of Gygi and coworkers~\cite{gygi_compact_2009,gygi_efficient_2013} uses a non-iterative algebraic decomposition of the wavefunction coefficients, which provides a transformation from the occupied KS eigenstates to a set of localized orbitals that are contained within prescribed domains in real space.
This method has already enabled a number of AIMD simulations using hybrid xc functionals (\eg computational investigations into the density of ice at finite temperature~\cite{gaiduk_density_2015}, ion solvation~\cite{zhang_communication:_2013,gaiduk_structural_2014}, as well as the structural and vibrational properties of liquid water~\cite{zhang_first_2011,zhang_structural_2011,gaiduk_first-principles_2018}) and is particularly convenient for simulating heterogeneous systems such as solid-liquid interfaces~\cite{dawson_performance_2015} due to the ease of selecting the prescribed localization domains. The SCDM method by Damle, Lin, and Ying exploits the sparsity of the off-diagonal elements of the density matrix,~\cite{damle_compressed_2015,damle_computing_2017,damle_scdm-k:_2017} and does not rely on an initial guess to iteratively localize the occupied space. As such, this approach sidesteps issues related to gauge invariance and can furnish more robust (\ie non-iterative) localized orbitals than other optimization-based schemes.~\cite{damle_disentanglement_2018} The MLWF formalism introduced by Marzari and Vanderbilt~\cite{marzari_maximally_1997} uses an iterative scheme to obtain a localized representation of the occupied KS orbitals by minimizing the total spread functional (\eg the sum of the spreads of the individual localized orbitals) and therefore extends the well-known Boys orbital localization scheme~\cite{boys_construction_1960} used in quantum chemistry into the condensed phase. MLWFs have shown great promise as both qualitative and quantitative analysis tools due to their similarity to the orbitals encountered in molecular orbital (MO) theory (\ie bonding and lone pairs) and the fact that they allow one to obtain molecular multipole moments~\cite{resta_quantum-mechanical_1998,silvestrelli_water_1999,souza_polarization_2000}, partition the charge density~\cite{kirchner_solvent_2004} and/or electrostatic potential~\cite{sagui_ab_2004}, and even compute non-bonded dispersion interactions~\cite{silvestrelli_van_2008} in complex condensed-phase environments. Numerous algorithms (such as \texttt{wannier90}~\cite{mostofi_wannier90:_2008}) for obtaining MLWFs have been incorporated into a number of existing community codes such as \texttt{Quantum ESPRESSO (QE)}~\cite{giannozzi_quantum_2009,giannozzi_advanced_2017}, \texttt{SIESTA}~\cite{soler_siesta_2002}, \texttt{ABINIT}~\cite{gonze_abinit:_2009}, \texttt{NWChem}~\cite{valiev_nwchem:_2010}, \texttt{GPAW}~\cite{enkovaara_electronic_2010}, \texttt{CP2K}~\cite{hutter_cp2k:_2013}, and \texttt{VASP}~\cite{kresse_ultrasoft_1999}, which makes this localization scheme readily available and quite practical for \textit{a posteriori} analyses of DFT-based calculations and AIMD simulations.
Furthermore, the MLWF localization scheme is particularly suitable for large-scale hybrid DFT based AIMD simulations since a Car-Parrinello-like propagation of the MLWFs has already been demonstrated~\cite{sharma_ab_2003,iftimie_--fly_2004,thomas_field_2004}, making the computational cost associated with orbital localization negligible between AIMD steps.
In light of this computationally efficient orbital localization scheme, the wide availability of MLWFs, and the promise of a robust tool for on-the-fly analytics, we will now focus our discussion on the development and implementation of a linear-scaling (order($N$)) MLWF-based EXX algorithm which can be used to perform large-scale condensed-phase AIMD simulations at the hybrid DFT level of theory.

In this work, we will focus on Car-Parrinello molecular dynamics (CPMD)~\cite{car_unified_1985} simulations of sufficiently large and finite-gap condensed-phase systems such that the first Brillouin zone can be accurately sampled at the $\Gamma$-point. Extensions to Born-Oppenheimer molecular dynamics (BOMD) and metallic systems~\cite{cornean_localised_2019,damle_variational_2018} are possible and will be discussed in future work. Working at the $\Gamma$-point allows us to consider real-valued orbitals only, \ie $\phi_i(\bm{r})=\phi_i^{*}(\bm{r})$, from which it follows that $\rho_{ij}(\bm{r})=\rho_{ji}(\bm{r})$ and $v_{ij}(\bm{r})=v_{ji}(\bm{r})$ in Eqs.~\eqref{eq:rho_from_phi}--\eqref{eq:exxGen}. Without loss of generality, we will also assume that the total wavefunction is closed shell (spin-unpolarized). Under these conditions, one can show that the set of MLWFs, which are obtained \via an orthogonal (unitary) transformation of the occupied KS eigenstates, \ie
\begin{equation}
  \widetilde{\phi}_i (\bm r) = \sum_j U_{ij} \phi_j (\bm r) ,
  \label{eq:ks_to_mlwf}
\end{equation}
have a significantly smaller support (or compact domain) than the entire simulation cell, and are in fact exponentially localized in real space~\cite{kohn_analytic_1959,des_cloizeaux_analytical_1964,nenciu_existence_1983,marzari_maximally_1997,niu_theory_1991,panati_bloch_2013}. These features of the MLWF representation of the occupied space provide a theoretical and computational framework for exploiting the natural sparsity in the real-space evaluation of the EXX energy (and wavefunction forces) that we will explore in this work.

To demonstrate that the use of MLWFs leads to a linear-scaling EXX approach, consider the expression for $\exx$ in Eq.~\eqref{eq:exxGen}. Since this quantity is invariant to orthogonal transformations of the occupied orbitals (see Sec.~\ref{sec:RealSpaceEXX}), evaluation of $\exx$ can be performed exactly within the MLWF representation. The first level of computational savings originates from the fact that a given MLWF only appreciably overlaps with a subset of neighboring MLWFs. This makes the number of non-vanishing EXX pair interactions \textit{per orbital} independent of the system size, and thereby reduces the total number of orbital pairs required in the summation over $i$ and $j$ in Eq.~\eqref{eq:exxGen}. In addition, one can further exploit the fact that a numerically exact evaluation of $\exx$ only requires that the spatial integral in Eq.~\eqref{eq:exxGen} be performed on the support of the orbital-product density. Since this quantity is sparse in the MLWF representation, this integration can be restricted to a real-space domain that is also independent of the system size. Taken together, these observations can be leveraged to construct a computationally efficient and linear-scaling MLWF-based algorithm for computing $\exx$. For a more detailed description of the theoretical underpinnings of this approach and the associated algorithmic implementation, see Secs.~\ref{sec:RealSpaceEXX} and \ref{sec:Implementation}, respectively. 

The initial concept and several pilot algorithms for this MLWF-based EXX approach~\cite{wu_order-n_2009,distasio_jr._individual_2014} have already been successfully used to enable a number of large-scale hybrid DFT based applications, \eg computational investigations into the electronic structure of semi-conducting solids~\cite{wu_hybrid_2009,chen_electronic_2011}, the structural properties of ambient liquid water~\cite{distasio_jr._individual_2014,santra_local_2015,ko_isotope_2019}, the structure and dynamics of aqueous ionic solutions~\cite{bankura_systematic_2015,chen_hydroxide_2018}, as well as the thermal properties of the pyridine-I molecular crystal~\cite{ko_thermal_2018}. In this manuscript, we build upon our earlier work by presenting a detailed description of the theoretical and algorithmic advances that are required to perform accurate and efficient MLWF-based AIMD simulations of large-scale condensed-phase systems at the hybrid DFT level. We focus our theoretical discussion on the integration of this approach into the CPMD framework by providing a detailed derivation of the EXX contributions to the equations of motion underlying fixed-cell CPMD simulations in the microcanonical ($NVE$) and canonical ($NVT$) ensembles. In particular, we include an in-depth discussion of a dual-level strategy which describes how the use of localized orbitals (like MLWFs) can lead to a linear-scaling EXX algorithm by exploiting the underlying sparsity in the real-space evaluation of the exchange interaction. Influenced by the work of Gygi and co-workers,~\cite{gygi_compact_2009,gygi_efficient_2013,dawson_performance_2015} we also introduce the concepts of MLWF-orbital and MLWF-product domains, which can be used to design an algorithmic framework that has the potential to enable accurate and efficient hybrid DFT simulations of condensed-phase systems with widely varying MLWF spreads.

In addition to this theoretical discussion, we also provide a comprehensive description of a massively parallel algorithm that extends well beyond our earlier pilot algorithms and uses this MLWF-based approach to compute all of the EXX contributions needed during hybrid DFT simulations of large-scale condensed-phase systems in the $NVE$ and $NVT$ ensembles. Recently implemented in the pseudopotential- and planewave-based open-source \texttt{QE} package~\cite{giannozzi_advanced_2017}, the so-called \exxm module exploits this dual-level linear-scaling strategy and employs a hybrid message-passing interface (\mpi{}) and open multi-processing (\omp{}) parallelization scheme to efficiently utilize high-performance computing (HPC) resources. Compared with earlier pilot versions, \exxm significantly improves the applicability of our MLWF-based approach to large-scale AIMD (\ie the strong-scaling limit) by introducing a hybrid parallelization scheme (which allows users to exploit both internode and intranode computational resources), a completely revised algorithm (which balances computation and communication, and reduces the overall memory footprint), as well as a more flexible and general-purpose implementation (which accommodates a wide range of users, including those with limited computational resources as well as those working at the massively parallel HPC limit).

This is followed by a critical assessment of the accuracy and parallel performance (\eg strong and weak scaling) of our implementation when performing AIMD simulations of liquid water in the $NVT$ ensemble on multiple different HPC architectures. In doing so, we demonstrate that \exxm enables hybrid DFT based AIMD simulations of \ce{(H2O)256}---a condensed-phase system containing $> 750$~atoms---with a walltime cost that is comparable to semi-local DFT and minimal errors in the EXX contribution to the total energy, wavefunction forces, ionic forces, and binding energetics. As such, the work described herein will further enable us to utilize the fourth rung of DFT in the study of the structure, properties, and dynamics of a number of important condensed-phase systems, as well as perform hybrid DFT based AIMD simulations across extended length and time scales which have been prohibitively difficult to access to date.

Although the current version of \exxm is restricted to condensed-phase systems in fixed orthorhombic simulation cells, an extension of this approach that treats general Bravais lattices and allows for hybrid DFT based AIMD simulations in the isobaric-isoenthalpic ($NpH$) and isobaric-isothermal ($NpT$) ensembles will be discussed in the next paper in this series. Since \exxm is quite modular, this algorithm can also be incorporated into any planewave-based DFT code; when combined with linear-scaling GGA codes such as \texttt{PARSEC}~\cite{kronik_parsec_2006,saad_numerical_2010}, \texttt{BigDFT}~\cite{mohr_accurate_2015}, \texttt{ONETEP}~\cite{skylaris_introducing_2005}, or \texttt{CONQUEST}~\cite{bowler_recent_2006}, \exxm could also be leveraged to achieve a fully (overall) linear-scaling hybrid DFT approach. We note in passing that the MLWF-based EXX approach described herein also sets the stage for performing large-scale condensed-phase AIMD simulations based on quantum chemical (\ie wavefunction theory) methodologies. Since a majority of the theoretical and algorithmic developments presented in this work are directly applicable to the iterative solution of the Hartree-Fock (HF) equations, this approach can be extended to enable a hierarchy of post-HF local electron correlation methods. Additional directions also include range-separated hybrids (RSH)~\cite{gerber_hybrid_2005,vydrov_assessment_2006,baer_tuned_2010,kronik_excitation_2012,karolewski_using_2013} as well as fifth-rung xc functionals (\eg MLWF-based GW approaches~\cite{chen_x-ray_2010,swartz_ab_2013}).

The remainder of the paper is organized as follows. In Sec.~\ref{sec:Theory}, we describe the theoretical framework for performing CPMD simulations at the hybrid DFT level of theory within the MLWF representation. Sec.~\ref{sec:Implementation} contains details of our massively parallel algorithmic implementation in the open-source \texttt{QE} package. This is followed by a detailed systematic analysis of the accuracy and computational performance of the current implementation in Sec.~\ref{sec:Perf}. The paper is then completed in Sec.~\ref{sec:conclusion}, which provides some brief conclusions as well as the future outlook of AIMD simulations using hybrid DFT.

\section{Theory \label{sec:Theory}}

In this section, we describe the theory behind our real-space MLWF-based framework for performing large-scale AIMD simulations of finite-gap condensed-phase systems at the hybrid DFT level of theory.
We will focus the discussion below on the equations of motion underlying fixed-cell CPMD simulations in the $NVE$ and $NVT$ ensembles.
Extension to constant-pressure CPMD simulations will be discussed in a forthcoming paper in this series.
Although we limit our scope here to CPMD, which provides a computationally efficient localized orbital propagation scheme~\cite{sharma_ab_2003,iftimie_--fly_2004,thomas_field_2004}, a cost-effective and competitive extension to BOMD has been achieved by our group and will also be addressed in another paper in this series.

\subsection{Index Conventions \label{sec:notation}}

We will utilize the following conventions for the various indices encountered in this work:
\begin{itemize}
  \item $i$, $j$, $k$: indices for the $N_o$ occupied orbitals (or MLWFs)
  \item $a$, $b$, $c$: indices corresponding to the Cartesian directions ${\bm x}$, ${\bm y}$, and ${\bm z}$
  \item $I$, $J$, $K$: indices for the $N_A$ ions
  \item $q$: index for points on the real-space grid
  \item $l,m$: indices for spherical harmonics
\end{itemize}

\subsection{EXX-Based CPMD in the $NVE$ Ensemble \label{sec:hybrid_aimd}}
\subsubsection{Equations of Motion \label{sec:eom}}

In CPMD simulations, fictitious dynamics are introduced on the $N_o$ occupied KS orbitals $\left\{ \phi_i \left( \bm r \right) \right\}$ \via artificial (fictitious) masses $\mu$.
Hence, CPMD simulations in the $NVE$ ensemble are governed by the following equations of motion for the electronic and ionic degrees of freedom:~\cite{marx_ab_2009} 
\begin{align}
  \mu \ddot{\phi}_i (\bm r) &= -\left( \frac{\delta E}{\delta \phi^*_i (\bm r)} \right) + \sum_{j} \Lambda_{ij} \phi_j (\bm r) \label{eq:cpE} \\
  M_I \ddot{\bm R}_I        &= - \left( \nabla_{\bm R_I} E \right) , \label{eq:cpI}
\end{align}
in which Newton's dot notation is used to indicate time derivatives, $E$ is the total ground-state DFT energy in Eq.~\eqref{eq:totE_ctr}, $-(\delta E/\delta \phi^*_i (\bm r))$ is the force acting on the $i$-th occupied KS wavefunction, $\Lambda_{ij}$ is a Lagrange multiplier enforcing orthonormality in $\left\{ \phi_i (\bm r) \right\}$, and $-\nabla_{\bm R_I} E$ is the force acting on the $I$-th ion (which is located at $\bm R_I$ with mass $M_I$).
At the hybrid DFT level, the equations of motion in Eqs.~\eqref{eq:cpE} and~\eqref{eq:cpI} will only depend on $E_{\rm xx}$ \via the wavefunction forces, $-(\delta E/\delta \phi^*_i (\bm r))$, which are discussed in detail below.

\subsubsection{EXX Contribution to the Wavefunction Forces \label{sec:exx_wff}}

In KS-DFT, $E_{\rm xc}$ is a functional of the electron density, which is given by $\rho(\bm r) = 2 \sum_i \phi_i^*(\bm r) \phi_i(\bm r)$.
As such, one can write the $E_{\rm xc}$ contribution to the (negative of the) wavefunction force for the $i$-th KS orbital as the action of the so-called xc potential, $v_{\rm xc}(\bm r) \equiv(\delta E_{\rm xc}/\delta \rho (\bm r))$, on the orbital itself, \ie
\begin{align}
  \left( \frac{\delta E_{\rm xc}}{\delta \phi^*_i (\bm r)} \right) = \left( \frac{\delta E_{\rm xc}}{\delta \rho (\bm r)} \right) \! \left( \frac{\delta \rho (\bm r)}{\delta \phi^*_i (\bm r)} \right) = 2 v_{\rm xc}(\bm r) \phi_i (\bm r) .
  \label{eq:vxc}
\end{align}
Since the explicit functional dependence of $E_{\rm xx}$ (in $E_{\rm xc}^{\rm hybrid}$) on $\rho(\bm r)$ is unknown, one needs special procedures such as the optimized effective potential (OEP) method~\cite{kummel_orbital-dependent_2008} to derive the EXX contribution to the wavefunction forces within a strict KS-DFT scheme.
In this work, we adopt a generalized KS-DFT scheme (\ie by allowing for an orbital-dependent $v_{\rm xc}(\bm r)$), which requires significantly less computational effort and yields the same ground-state energies as the OEP formalism.
In this approach (which is currently the standard practice in the field), we compute the corresponding orbital-dependent EXX wavefunction forces, $D^i_{\rm xx}(\bm r)=-(\partial E_{\rm xx}/\partial \phi_i^*(\bm r))$, by taking the functional derivative of $E_{\rm xx}$ in Eq.~\eqref{eq:exxGen} with respect to $\phi_i^*(\bm r)$, yielding:
\begin{align}
  D^i_{\rm xx}(\bm r) = \sum_{j} v_{ij}(\bm r) \phi_j(\bm r) \equiv \sum_{j} D_{\rm xx}^{ij}(\bm r) .
  \label{eq:Dxx}
\end{align}
To derive this expression, we have used Eqs.~\eqref{eq:rho_from_phi} and \eqref{eq:vxxGen} for the orbital-product density and potential, $\rho_{ij}(\bm r)$ and $v_{ij}(\bm r)$, and defined $D_{\rm xx}^{ij}(\bm r)$ as the action of $v_{ij}(\bm r)$ on $\phi_j(\bm r)$. From Eq.~\eqref{eq:Dxx}, it is again clear that the evaluation of the orbital-product potential, $v_{ij}(\bm{r})$, is of central importance to the calculation of $D^i_{\rm xx}(\bm{r})$.

\subsection{Real-Space EXX Calculations: Linear Scaling \textit{via} Orbital Localization \label{sec:RealSpaceEXX}}

The efficient evaluation of $v_{ij}(\bm r)$---which is a required ingredient for computing $E_{\rm xx}$ and $D_{\rm xx}^{i}(\bm r)$---is key to enabling large-scale condensed-phase AIMD simulations at the hybrid DFT level of theory.
In this section, we will describe a linear-scaling EXX method that exploits the natural sparsity of the quantum mechanical exchange interaction in real space \via the use of a localized (MLWF) representation of the occupied orbitals.
Within this framework, $\tv{ij}$ (which is the MLWF analog of $v_{ij}(\bm r)$ in Eq.~\eqref{eq:vxxGen}) only needs to be computed for \textit{overlapping pairs} of MLWFs on a real-space domain that is independent of the system size, thereby paving the way to a linear-scaling EXX method in the condensed phase (see Sec.~\ref{sec:Implementation} for algorithmic details).
As such, the cornerstone of our method is the efficient real-space evaluation of $\tv{ij}$, which is accomplished herein \via the solution of Poisson's equation on a system-size independent real-space domain for each overlapping MLWF pair.

For a finite-gap condensed-phase system, the occupied KS orbitals (or bands) can be mapped \via an orthogonal transformation onto a unique set of MLWFs (see Eq.~\eqref{eq:ks_to_mlwf}), that are exponentially localized in real space~\cite{kohn_analytic_1959,des_cloizeaux_analytical_1964,nenciu_existence_1983,marzari_maximally_1997,niu_theory_1991,panati_bloch_2013} and have a significantly smaller support than the entire simulation cell.
As such, the MLWF representation of the occupied space allows one to exploit the underlying sparsity in the quantum mechanical exchange interaction, and provides a theoretical and computational framework for substantially reducing the computational scaling and cost associated with EXX-based approaches.

To see how MLWFs can be leveraged to attain a linear-scaling EXX algorithm, we first transform the canonical $\exx$ expression in Eq.~\eqref{eq:exxGen0} into the MLWF representation.
Since $\phi_j (\bm{r}) = \sum_i (U^{-1})_{ji} \tphi{i}$ (\cf Eq.~\eqref{eq:ks_to_mlwf}), $\exx$ can be written as follows:
\begin{align}
  \exx &= -\sum_{ij} \sum_{\substack{kk' \\ k''k'''}} \int\dd\bm{r} \int\dd\bm{r}' \, \frac{\widetilde{\phi}_{k}(\bm{r})\widetilde{\phi}_{k'}(\bm{r}')\widetilde{\phi}_{k''}(\bm{r})\widetilde{\phi}_{k'''}(\bm{r}')}{\left| \bm{r}-\bm{r}' \right|} \nonumber \\
  &\times (U^{-1})_{ik} (U^{-1})_{jk'} (U^{-1})_{j{k''}}(U^{-1})_{i{k'''}} .
\end{align}
Utilizing the fact that $\bm{U}\bm{U}^{T}=\bm{U}\bm{U}^{-1}=\bm{I}$ for an orthogonal matrix, summation over $i$ and $j$ in this expression leads to $\sum_{ij} (U^{-1})_{ik} (U^{-1})_{jk'} (U^{-1})_{j{k''}}(U^{-1})_{i{k'''}} = \sum_{i} U_{ki} (U^{-1})_{i{k'''}} \sum_{j} U_{k'j} (U^{-1})_{j{k''}} = \delta_{k{k'''}} \, \delta_{k'{k''}}$, from which we see that 
\begin{align}
  \exx &= -\sum_{ij} \int\dd\bm{r} \int\dd\bm{r}' \, \frac{\widetilde{\phi}_{i}(\bm{r})\widetilde{\phi}_{j}(\bm{r}')\widetilde{\phi}_{j}(\bm{r})\widetilde{\phi}_{i}(\bm{r}')}{\left| \bm{r}-\bm{r}' \right|}
  \label{eq:exxGen0_mlwf}
\end{align}
upon dummy variable substitutions of $k \rightarrow i$ and $k' \rightarrow j$.
This proof demonstrates that the expression for evaluating $E_{\rm xx}$ is invariant to the orthogonal transformation between the KS and MLWF representations.
In fact, this invariance property of $\exx$ also holds for any arbitrary orbital representation $\left\{ \psi_{i}(\bm r) \right\}$ that is derived from an orthogonal rotation $\bm U'$ within the occupied KS subspace (\ie $\psi_{i} (\bm r) = \sum_j U'_{ij} \phi_j (\bm r)$).
In analogy to Eq.~\eqref{eq:exxGen}, the MLWF expression for $\exx$ in Eq.~\eqref{eq:exxGen0_mlwf} can also be written in the following compact form:
\begin{align}
  \exx = -\sum_{ij} \int \dd\bm{r} \, \trho{ij}\tv{ij} ,
  \label{eq:exx}
\end{align}
in terms of the MLWF-product density,
\begin{equation}
  \trho{ij} \equiv \tphi{i} \tphi{j} = \trho{ji} ,
  \label{eq:rho_from_phi_mlwf}
\end{equation}
and the corresponding MLWF-product potential,
\begin{equation}
  \tv{ij} \equiv \int \dd\bm{r}' \, \frac{\widetilde{\rho}_{ij}(\bm{r}')}{|\bm{r}-\bm{r}'|} =\tv{ji}.
  \label{eq:vxxGen_mlwf}
\end{equation}
We note in passing that while $\exx$ is invariant to any orthogonal transformation, the values of $\trho{ij}$ and $\tv{ij}$---despite the fact that they have the same expression as that given in Eqs.~\eqref{eq:rho_from_phi} and \eqref{eq:vxxGen}---do in fact depend on the employed representation.
It is this freedom in the choice of the orthogonal transformation that allows one to select an appropriate localized orbital representation (\eg MLWF) for exploiting the underlying sparsity in the EXX interaction.
Throughout this work, we will dress each of the MLWF-specific quantities with a tilde to distinguish them from their analogous expressions in the canonical KS representation. 

Given the expression for $\exx$ in the MLWF representation (\cf Eqs.~\eqref{eq:exxGen0_mlwf}--\eqref{eq:exx}), the corresponding EXX contributions to the wavefunction forces that are required to propagate the CPMD equations of motion (Eqs.~\eqref{eq:cpE} and \eqref{eq:cpI}) can be derived following the same procedure given above in Sec.~\ref{sec:exx_wff}.
In this regard, the wavefunction force on the $i$-th MLWF, $\dxx{i}=-(\delta \exx/\delta \widetilde{\phi}_i^*(\bm r))$, can be obtained from Eqs.~\eqref{eq:exxGen0_mlwf}--\eqref{eq:vxxGen_mlwf}, yielding:
\begin{align}
  \dxx{i} = \sum_{j} \tv{ij}\tphi{j} \equiv \sum_{j} \dxx{ij} ,
  \label{eq:Dxx_mlwf}
\end{align}
where $\dxx{ij}$ has been defined as the action of $\tv{ij}$ on $\tphi{j}$.
Here, $\dxx{i}$ and $\dxx{ij}$ also depend on the MLWF representation and therefore take on different values when compared to their KS analogs in Eq.~\eqref{eq:Dxx}.
From Eqs.~\eqref{eq:exx} and \eqref{eq:Dxx_mlwf}, it is again clear that the evaluation of the MLWF-product potential, $\tv{ij}$, is the cornerstone of our MLWF-based EXX approach.

With the expressions required for the evaluation of $\exx$ and $\dxx{i}$ in hand, we will now discuss in detail how MLWFs lead to a linear-scaling EXX algorithm by exploiting the underlying sparsity in the exchange interaction.
Since the set of MLWFs are exponentially localized in real space and therefore have a significantly smaller support than the entire simulation cell, this allows us to exploit two levels of sparsity during the computational evaluation of all required EXX-related quantities.
The first level of computational savings originates from the fact that a given MLWF, $\tphi{i}$, will only appreciably overlap with a number, $\widetilde{n}_{i}$, of neighboring MLWFs.
For all other MLWFs, the product density, $\trho{ij} = \tphi{i} \tphi{j}$ (and hence the corresponding product potential, $\tv{ij}$\bl), will be vanishingly small.
In these cases, the contributions to $\exx$ and $\dxx{i}$ are numerically zero, and this directly reduces the number of terms that are required in the summation over $j$ in Eqs.~\eqref{eq:exx} and \eqref{eq:Dxx_mlwf}.
As such, the number of EXX pair interactions \textit{per orbital} becomes independent of system size (assuming a fixed system density), which reduces the total number of orbital pairs, $N_{\rm pair}$, from $\mathcal{O}(N_o^2)$ to $\mathcal{O}(N_o)$, \ie $N_{\rm pair} = N_o(N_o+1)/2 \rightarrow \widetilde{n}N_o$.
In this last expression, $\widetilde{n} = \max_i \{ \widetilde{n}_i \} < N_o$ is independent of the system size, hence $\widetilde{n}N_o$ represents an upper bound to the number of EXX pair interactions in our approach.
Since the contributions from the omitted MLWF pairs to $\exx$ and $\dxx{i}$ are vanishingly small, this reduction in $N_{\rm pair}$ still allows for a numerically exact evaluation of all EXX-related quantities.
Although this leads to significant computational savings, the overall scaling associated with evaluating these quantities is still formally quadratic as the real-space domain associated with the simulation cell, $\Omega$, grows linearly with the size of the system. 

\begin{figure}[ht!]
  \includegraphics[width=\linewidth]{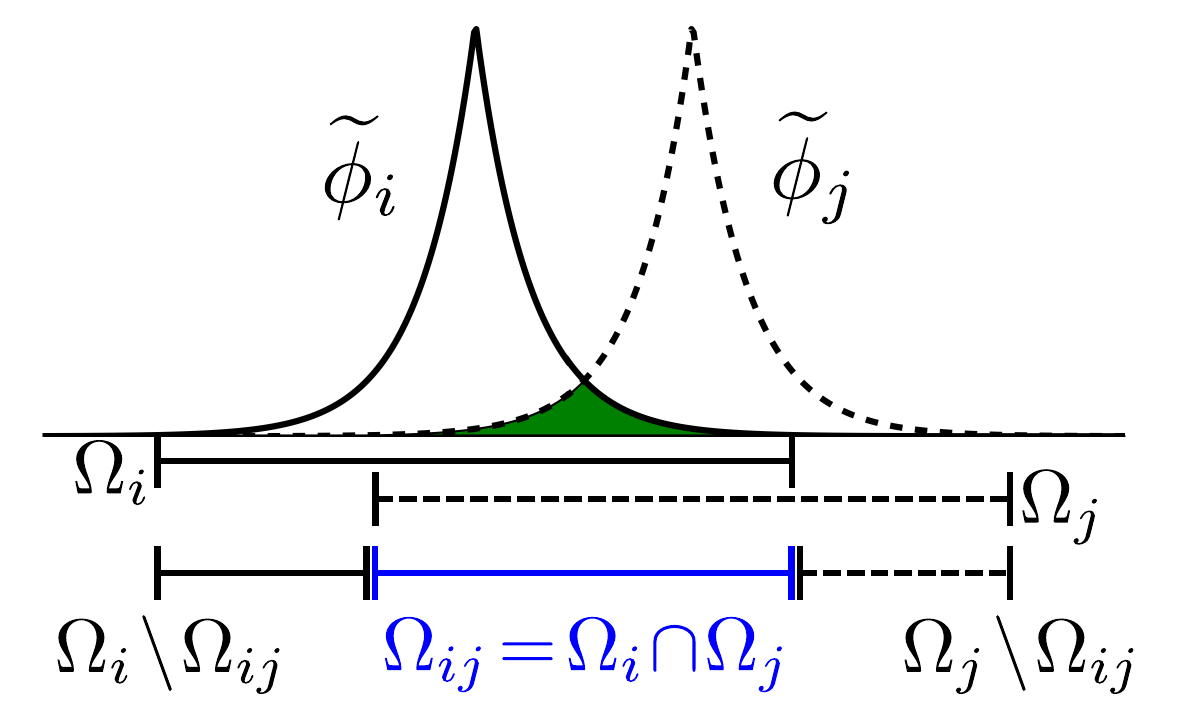}
  \caption{
  Graphical depiction of an overlapping pair of exponentially localized MLWFs, $\tphi{i}$ (solid) and $\tphi{j}$ (dashed).
  As such, the supports for $\tphi{i}$ and $\tphi{j}$ are compact and labelled by $\Omega_i$ and $\Omega_j$, respectively.
  The support of the corresponding MLWF-product density, $\trho{ij} = \tphi{i}\tphi{j}$ (green), is also compact and labelled by $\Omega_{ij} = \Omega_{i} \cap \Omega_{j}$.
  As described in the text, a numerically exact evaluation of the EXX contribution to the energy ($\exx$) requires the solution to Poisson's equation for the near-field potential ($\tv{ij}$) on the $\Omega_{ij}$ domain (see Eqs.~\eqref{eq:exx_omega} and \eqref{eq:pe}), while a numerically exact evaluation of the EXX contribution to the wave function forces ($\dxx{ij}$ and $\dxx{ji}$) requires a converged multipole expansion for the far-field potential ($\tv{ij}$) on the $\Omega_{j}\setminus\Omega_{ij}$ and $\Omega_{i}\setminus\Omega_{ij}$ domains (see Eqs.~\eqref{eq:Dxx_mlwf_omega} and \eqref{eq:me}).
  }
  \label{fig:1d_cover}
\end{figure}

To achieve linear scaling with system size, one can further exploit the fact that the set of exponentially localized MLWFs have a substantially smaller support than $\Omega$.
This allows us to employ real-space domains that are independent of system size and still maintain a numerically exact evaluation of $\exx$ and $\dxx{i}$.
To harness this second level of computational savings, we
follow the work of Gygi and co-workers~\cite{gygi_compact_2009,gygi_efficient_2013,dawson_performance_2015} by defining an MLWF-orbital domain as $\Omega_i = \{ \bm r \in \Omega \,\bm{\mid}\, | \tphi{i} | > \epsilon \}$, in which $\epsilon$ is a small (but finite) positive threshold. For small enough values of $\epsilon$, $\Omega_i$ will encompass the support of $\tphi{i}$ in real space (see Fig.~\ref{fig:1d_cover}).
This domain is focused around the so-called MLWF center, $\tC{i}$, which is given by the expectation value (or first moment) of $\bm r$, \ie $\tC{i} = \braket{\tphin{i}|\bm r|\tphin{i}} = \int \dd\bm{r} \, \bm{r} \trho{ii}$.
In analogy, we also define an MLWF-product domain as $\Omega_{ij} \equiv \Omega_i \cap \Omega_j$, which (for sufficiently small $\epsilon$ values) encompasses the support of $\trho{ij}$ (see Fig.~\ref{fig:1d_cover}).
Since $\Omega_{ij}$ corresponds to the points in real space where $\tphi{i}$ and $\tphi{j}$ \textit{are both non-negligible}, this domain is even more sparse than $\Omega_{i}$ or $\Omega_{j}$.
When $i = j$, one can straightforwardly compute the corresponding center of charge for $\trho{ii}$ as $\tC{ii} = \int \dd\bm{r} \, \bm{r} \trho{ii} \,/ \int \dd\bm{r} \, \trho{ii}$.
Since $\trho{ii}$ integrates to unity, $\tC{ii} = \tC{i}$, which is simply the center of the $i$-th MLWF given above.
When $i \neq j$, $\trho{ij}$ now corresponds to a localized charge distribution with a \textit{vanishing monopole} due to the orthogonality of the MLWFs; hence, the center of this charge distribution cannot be analogously defined as $\int \dd\bm r \, \bm r \trho{ij} \,/ \int \dd\bm r \, \trho{ij}$.
As such, we utilize an analog of the standard gauge in molecular quantum mechanics for an electrically neutral system, wherein the ``center of charge'' is taken as the position at which the nuclear (ionic) dipole moment vanishes. This allows us to define $\tC{ij} = \int \dd\bm{r} \, \bm{r} |\trho{ij}| \,/ \int \dd\bm{r} \, |\trho{ij}|$ as the corresponding center for $\trho{ij}$.
By making all sectors of this charge distribution positive, $|\trho{ij}|$ now has a sizable monopole and a well-defined center of charge given by $\tC{ij}$.
By construction, this choice of gauge recovers the correct center of charge when $i = j$, \ie $\tC{ii} = \tC{i}$, and is therefore consistent with the expression used above for $\trho{ii}$.

Within this framework, both $\Omega_i$ and $\Omega_{ij}$ are system-size independent and substantially smaller than $\Omega$.
Furthermore, since $\Omega_{ij}$ is defined as the overlapping region between two exponentially decaying MLWFs, $\tphi{i}$ and $\tphi{j}$, the extent of this domain is smaller than both $\Omega_i$ and $\Omega_j$, which holds true even when $i=j$.
From Eqs.~\eqref{eq:exx} and \eqref{eq:rho_from_phi_mlwf}, one sees that a numerically exact evaluation of $\exx$ (neglecting self-consistency effects, \textit{vide infra}) only requires 
summation over overlapping $ij$ pairs (denoted by $\braket{ij}$) and spatial integration over 
$\Omega_{ij}$, \ie
\begin{align}
  \exx = - \sum_{\braket{ij}} \int_{\Omega_{ij}} \dd \bm{r} \, \trho{ij} \tv{ij} .
  \label{eq:exx_omega}
\end{align}
In the same breath, Eq.~\eqref{eq:Dxx_mlwf} shows that a numerically exact evaluation of $\dxx{ij}$ only requires the action of $\tv{ij}$ over $\Omega_j$, \ie
\begin{align}
  \dxx{ij} = \tv{ij}\tphi{j} \qquad \bm r \in \Omega_{j} .
  \label{eq:Dxx_mlwf_omega}
\end{align}
This implies that one only needs $\tv{ij}$ on $\Omega_{ij}$ for a numerically exact evaluation of $\exx$, and $\tv{ij}$ on $\Omega_{j}$ for a numerically exact evaluation of $\dxx{ij}$.
As such, the evaluation of $\tv{ij}$ can also be restricted to system-size independent real-space domains, despite the fact that this quantity is formally non-zero across $\Omega$, and asymptotically goes as $1/r$ for $i = j$ (due to the non-vanishing monopole associated with $\trho{ii}$) and $1/r^2$ (or higher order) for $i \neq j$ (due to the vanishing monopole associated with $\trho{ij}$).
This leads to even further computational savings as $\tv{ij}$ can be obtained exactly by solving Poisson's equation (PE) over $\Omega_{ij}$ in the near field,
\begin{equation}
  \nabla^2\tv{ij}=-4\pi\trho{ij} \qquad \bm r \in \Omega_{ij} ,
  \label{eq:pe}
\end{equation}
subject to Dirichlet boundary conditions given by an appropriately converged multipole expansion (ME) of $\trho{ij}$ in the far field, \ie
\begin{equation}
  \tv{ij} = 4\pi \sum_{lm} \frac{Q_{lm}}{(2l+1)} \frac{Y_{lm}(\theta,\varphi)}{r^{l+1}} \qquad \bm r \notin \Omega_{ij} .
  \label{eq:me}
\end{equation}
In this expression, $\tC{ij}$ is taken as the origin, $\bm r = (r,\theta,\varphi)$ is given in spherical polar coordinates, $Y_{lm} (\theta,\varphi)$ are the spherical harmonics, and
\begin{equation}
  Q_{lm} = \int_{\Omega_{ij}} \dd\bm{r} \, Y_{lm}^{*}(\theta,\varphi) r^l \widetilde{\rho}_{ij}(\bm{r}) ,
  \label{eq:mepole}
\end{equation}
are the multipole moments corresponding to $\trho{ij}$. For typical systems, a ME with a maximum value of $l = 6$ is sufficiently converged.~\cite{wu_order-n_2009,distasio_jr._individual_2014}
We note in passing that the ME in Eq.~\eqref{eq:me} serves a dual purpose and will also be employed during the evaluation of $\dxx{ij}$, which requires $\tv{ij}$ on $\Omega_j$.
In other words, $\dxx{ij}$ is computed with $\tv{ij}$ on $\Omega_{ij}$ \via the solution to the PE in Eq.~\eqref{eq:pe}, and $\tv{ij}$ on the $\Omega_j \setminus \Omega_{ij}$ domain, \ie for all points in $\Omega_j$ that are not contained in $\Omega_{ij}$, \via the ME in Eq.~\eqref{eq:me}.

This discussion again clearly highlights that an efficient real-space evaluation of $\tv{ij}$---on compact and system-size independent domains---is the cornerstone of our linear-scaling MLWF-based EXX approach.
In the next section, we will focus our discussion on the algorithmic implementation of this approach, which can be used to perform large-scale condensed-phase AIMD simulations with hybrid DFT.

\section{Implementation and Algorithmic Details \label{sec:Implementation}}

In this section, we describe the implementation of our linear-scaling MLWF-based EXX algorithm in the \texttt{CP} module of \texttt{QE}~\cite{giannozzi_quantum_2009, giannozzi_advanced_2017}.
This algorithm has been implemented as a standalone module named \exxm, which has been integrated with the MLWF-enabled semi-local DFT routines in \texttt{QE} \via a portable input/output interface (see flowchart in Fig.~\ref{fig:flowchart}).
During each CPMD step, the main input required for \exxm includes the current set of MLWFs, $\{ \tphi{i} \}$, while the output produced by this module includes $\exx$ and $\{\dxx{i}\}$.
As such, adaptation of the \exxm module to other periodic DFT codes should be straightforward, as long as the capability to produce MLWFs ``on-the-fly'' during CPMD simulations is available (\textit{vide infra}).
In fact, the current \exxm module only requires that the input orbitals are sufficiently local and form an orthonormal set, and can therefore accommodate (with appropriate modifications) other orbital localization schemes such as RSB~\cite{gygi_compact_2009,gygi_efficient_2013} and SCDM~\cite{damle_compressed_2015,damle_computing_2017,damle_scdm-k:_2017}.
To enable large-scale EXX-based AIMD using this approach, we employ a hybrid \mpi{} and \omp{} parallelization scheme that allows us to differentially exploit both internode and intranode computational resources provided by massively parallel supercomputer architectures.

\subsection{MLWF-based EXX-CPMD: Prerequisites \label{subsec:MLWFpropagation}}

To start a CPMD simulation, one needs to reach the electronic ground state for a given initial configuration of the system \via a SCF calculation.
In the \texttt{CP} module of \texttt{QE}, the iterative solution of the non-linear KS equations is accomplished using either conjugate gradient (CG) or second-order damped dynamics (SODD) to minimize the fictitious kinetic energy associated with the electronic degrees of freedom (while keeping the ions fixed)~\cite{tassone_acceleration_1994}.
During the SODD minimization, the proto-KS orbitals are evolved according to the following equations of motion (which are equivalent to Eq.~\eqref{eq:cpE} with an additional damping term):
\begin{equation}
  \mu \ddot{\varphi}_{i} (\bm r) =  D_i(\bm r) - 2 \mu \gamma \dot{\varphi}_i (\bm r) ,
  \label{eq:sodd_eom}
\end{equation}
in which $\{\varphi_{i} (\bm r)\}$ are the proto-KS orbitals during the SCF calculation, $D_i(\bm r) \equiv -(\delta E/\delta \varphi^\ast_{i} (\bm r)) + \sum_j \Lambda_{ij} \varphi_j (\bm r)$ is the force acting on the $i$-th orbital, and $\gamma$ is a damping parameter.
To evolve the proto-KS orbitals, Eq.~\eqref{eq:sodd_eom} can be integrated to yield:~\cite{tassone_acceleration_1994}
\begin{align}
  \varphi_{i} (\bm r, \tau+\Delta\tau) &= \frac{2}{1+\Gamma }\varphi_{i} (\bm r, \tau) -\frac{1-\Gamma }{1+\Gamma} \varphi_{i} (\bm r, \tau-\Delta\tau) \nonumber \\
  &+ \frac{\Delta\tau^2 }{1+\Gamma}\frac{D_i(\bm r, \tau)}{\mu} ,
  \label{eq:sodd_eom_discrete}
\end{align}
in which $\Delta \tau$ is the time step for the fictitious proto-KS dynamics and $\Gamma \equiv \gamma \Delta \tau$~\cite{note_typical_choice}.
Upon convergence of the SODD procedure, $\{\varphi_{i} (\bm r, \tau)\}$ becomes a set of ground-state KS orbitals, which is chosen as the initial condition for the AIMD simulation (\ie $\{\phi_{i} (\bm r, t = 0)\}$).
In doing so, cubic-scaling matrix operations such as diagonalization of the Fock (or effective Hamiltonian) matrix are completely sidestepped during the SCF procedure; as such, this approach does not require (nor produce) unoccupied/virtual states, and provides a solid foundation upon which one can build a fully linear-scaling DFT (or HF) code base.

In fact, this CP-like approach to the SCF solution of the KS equations can be combined with the MLWF localization procedure by performing a nested SODD optimization of the Marzari-Vanderbilt~\cite{marzari_maximally_1997, marzari_maximally_2012} functional to incrementally localize the proto-KS orbitals between each SCF step~\cite{sharma_ab_2003}.
In \texttt{CP}, this is accomplished by splitting Eq.~\eqref{eq:sodd_eom_discrete} into an extrapolation step,
\begin{align}
  \chi_{j} (\bm r, \tau+\Delta \tau) &=  \frac{2}{1+\Gamma }\widetilde{\varphi}_{j} (\bm r, \tau) -\frac{1-\Gamma }{1+\Gamma} \widetilde{\varphi}_{j} (\bm r, \tau-\Delta \tau) \nonumber\\
  &+ \frac{\Delta \tau^2 }{1+\Gamma}\frac{\widetilde{D}_{j}(\bm r, \tau)}{\mu} ,
  \label{eq:sodd_mlwf_eom_discrete_1} 
\end{align}
followed by a localization step,
\begin{align}
  \widetilde{\varphi}_{i} (\bm r, \tau+\Delta \tau) =& \sum_j U_{ij}(\tau+\Delta \tau) \chi_{j} (\bm r, \tau+\Delta \tau) .
  \label{eq:sodd_mlwf_eom_discrete_2}
\end{align}
In the extrapolation step, an intermediary set of orbitals, $\{\chi_{j} (\bm r, \tau + \Delta \tau)\}$, is formed \via SODD evolution of the proto-MLWF orbitals, $\{\widetilde{\varphi}_{j} (\bm r)\}$, according to $\widetilde{D}_j(\bm r, \tau) \equiv -(\delta E/\delta \widetilde{\varphi}^\ast_{j} (\bm r, \tau)) + \sum_k \widetilde{\Lambda}_{jk}(\tau) \widetilde{\varphi}_{k} (\bm r, \tau)$, the force acting on the $j$-th proto-MLWF orbital (which includes $\{ \widetilde{\Lambda}_{jk}(\tau) \}$, the set of Lagrange multipliers needed to preserve orthonormality).
In the localization step, the proto-MLWFs, $\{\widetilde{\varphi}_{j} (\bm r, \tau+\Delta \tau)\}$, are incrementally localized \via a unitary (orthogonal) transformation over $\{\chi_{i} (\bm r, \tau + \Delta \tau)\}$, the intermediary set of orbitals obtained during the extrapolation step in Eq.~\eqref{eq:sodd_mlwf_eom_discrete_1}.
This unitary transformation is accomplished \via $\bm U(\tau+\Delta \tau)$, a matrix which is generated from a nested SODD optimization~\cite{sharma_ab_2003} of the Marzari-Vanderbilt functional~\cite{marzari_maximally_1997,marzari_maximally_2012}.
During the SCF procedure, it is computationally unfavorable (and \textit{not} necessary) to converge the nested SODD minimization for $\bm U$ at each step, as the current (unconverged) orbitals do not yet represent the ground electronic state.
In this regard, only a few nested SODD steps (\eg a maximum of $20$ during the SCF procedure for liquid water) were used to incrementally localize the proto-MLWF orbitals.
Upon convergence of this combined SCF and localization procedure, the set of proto-MLWF orbitals, $\{\widetilde{\varphi}_{i} (\bm r, \tau)\}$, becomes the set of ground-state MLWF orbitals, which can now be chosen as the initial condition for an MLWF-based AIMD simulation (\ie $\{\widetilde{\phi}_{i} (\bm r, t = 0)\}$).
As such, this approach provides a cost-effective alternative to the standard \textit{a posteriori} procedure of localizing the canonical (Bloch) occupied orbitals from a fully converged SCF calculation.

At the hybrid DFT level, we adopt this CP-like approach to incrementally localize the occupied orbitals \textit{during} the EXX-based SCF procedure, thereby avoiding a preliminary EXX calculation in reciprocal space.
Since the incrementally localized proto-MLWF orbitals are not equivalent to the final set of MLWFs at a given SCF step, the orbital-dependent EXX contributions to $v_{\rm xc}(\bm r)$ are approximately evaluated \via Eq.~\eqref{eq:Dxx_mlwf_omega}; however, the resulting errors are inconsequential as the incremental refinement of the localized orbitals (and therefore $\dxx{i}$) at each step leads to the desired set of MLWFs upon SCF convergence.
Since our approach is based on an incremental ``on-the-fly'' refinement of the proto-MLWF orbitals during the SCF procedure, it is therefore unsuitable for standard Fock matrix diagonalization routines, in which global rotations between the occupied and virtual orbitals during each diagonalization step would lead to marked delocalization of the occupied states.
Such delocalization would require substantial effort (essentially from scratch) to relocalize the orbitals after each diagonalization step, and would thereby nullify the computational savings obtained from a sparse evaluation of all EXX-related quantities.

\begin{figure}[t!]
  \centering
  \begin{tikzpicture}[scale=0.60, node distance = 2cm, auto, every node/.style={transform shape}]
    \node [sbrt]                                   (s1)   {\Large \texttt{Input MLWFs}\\};
    \node [sbrt,  below of=s1, node distance=12em] (s2)   {\Large \texttt{Redistribution}\\ \texttt{of MLWFs}\\};
    \node [sbrt,  below of=s2, node distance= 8em] (s3)   {\Large \texttt{Construction of}\\ \texttt{Pair List and}\\ \texttt{Proto-Subdomains}\\};
    \node [sbrt,  below of=s3, node distance= 8em] (s4)   {\Large \texttt{Communication}\\ \texttt{of MLWFs}\\};
    \node [sbrt,  below of=s4, node distance= 8em] (s5)   {\Large \texttt{Solution of Poisson's}\\ \texttt{Equation}\\};
    \node [sbrt,  below of=s5, node distance=10em] (s6)   {\Large \texttt{Computation of}\\ \texttt{Energy and Forces}\\}; 
    \node [sbrt,  below of=s6, node distance=10em] (s7)   {\Large \texttt{Redistribution}\\ \texttt{of Wavefunction}\\ \texttt{Forces}\\};
    \node [sbrt,  below of=s7, node distance=12em] (s8)   {\Large \texttt{Output EXX}\\ \texttt{Energy and Forces}\\};
    \node [left of=s2,  node distance=13.5em] (l2)    {\huge \texttt{I}};
    \node [left of=s3,  node distance=13.5em] (l3)    {\huge \texttt{II}};
    \node [left of=s4,  node distance=13.5em] (l4)    {\huge \texttt{III}};
    \node [left of=s5,  node distance=13.5em] (l5)    {\huge \texttt{IV}};
    \node [left of=s6,  node distance=13.5em] (l6)    {\huge \texttt{V}};
    \node [left of=s7,  node distance=13.5em] (l7)    {\huge \texttt{VI}};
    \node [varg,  below right = 4em and 10em of s1] (d1)  {\LARGE $\widetilde{\phi}_{i}(\bm r)$};
    \node [varb,  below right = 0em and 10em of s2] (d2)  {\LARGE $\widetilde{\phi}_{i}(\bm r)$};
    \node [varb,  below right = 0em and 10em of s3] (d3)  {\LARGE $\widetilde{\phi}_{i}(\overline{\bm r})$};
    \node [varb,  below right = 0em and 10em of s4] (d4)  {\LARGE $\widetilde{\rho}_{ij}(\overline{\bm r})$};
    \node [varb,  below right = 0em and 10em of s5] (d5)  {\LARGE $\widetilde{v}_{ij}(\overline{\bm r})$};
    \node [varb,  below right = 2em and 6em of s6]  (d6)  {\LARGE $\widetilde{D}^{i}_{\rm xx}(\bm r)$};
    \node [varg,  below of=d6, node distance= 8em]  (d7)  {\LARGE $\widetilde{D}^{i}_{\rm xx}(\bm r)$};
    \node [vary,  right of=d6, node distance=8em]   (d8)  {\LARGE $E_{\rm xx}$};
    \node [dummy,  right of=d7, node distance=8em]   (d9)  {};
    \node [dummy, above of = d1,  node distance= 8.5em] (e1)  {};
    \node [dummy, above of = d6,  node distance= 6.5em] (e2)  {}; 
    \node [dummy, above of = d8,  node distance= 6.5em] (e3)  {}; 
    \node [dummy, below of = d7,  node distance= 7.5em] (e4)  {}; 
    \node [dummy, below of = d9,  node distance= 7.5em] (e5)  {}; 
    \node [dummy,  left of = s4,  node distance=9.60em] (e6) {};
    \node [dummy,  left of = s6,  node distance=9.60em] (e7) {};
    \draw [-{Stealth[scale=0.8,angle'=50]},semithick]   (s2) -- (s3);
    \draw [-{Stealth[scale=0.8,angle'=50]},semithick]   (s3) -- (s4);
    \draw [-{Stealth[scale=0.8,angle'=50]},semithick]   (s4) -- (s5);
    \draw [-{Stealth[scale=0.8,angle'=50]},semithick]   (s5) -- (s6);
    \draw [-{Stealth[scale=0.8,angle'=50]},semithick]   (s6) -- (s7); 
    \draw [-{Stealth[scale=0.8,angle'=50]},semithick,dashed] (e1) -- (d1);
    \draw [-{Stealth[scale=0.8,angle'=50]},semithick,dashed] (d1) -- (s2);
    \draw [-{Stealth[scale=0.8,angle'=50]},semithick,dashed] (s2) -- (d2);
    \draw [-{Stealth[scale=0.8,angle'=50]},semithick,dashed] (d2) -- (s3);
    \draw [-{Stealth[scale=0.8,angle'=50]},semithick,dashed] (s3) -- (d3);
    \draw [-{Stealth[scale=0.8,angle'=50]},semithick,dashed] (d3) -- (s4);
    \draw [-{Stealth[scale=0.8,angle'=50]},semithick,dashed] (s4) -- (d4);
    \draw [-{Stealth[scale=0.8,angle'=50]},semithick,dashed] (d4) -- (s5);
    \draw [-{Stealth[scale=0.8,angle'=50]},semithick,dashed] (s5) -- (d5);
    \draw [-{Stealth[scale=0.8,angle'=50]},semithick,dashed] (d5) -- (s6);
    \draw [-{Stealth[scale=0.8,angle'=50]},semithick,dashed] (d6) -- (s7);
    \draw [-{Stealth[scale=0.8,angle'=50]},semithick,dashed] (s7) -- (d7);
    \draw [-{Stealth[scale=0.8,angle'=50]},semithick,dashed] (e2) -- (d6);
    \draw [-{Stealth[scale=0.8,angle'=50]},semithick,dashed] (e3) -- (d8);
    \draw [-{Stealth[scale=0.8,angle'=50]},semithick,dashed] (e5) -- (s8);
    \draw [dotted,semithick]      (s1) -- (s2); 
    \draw [dashed,semithick]      (s1) -- (e1);
    \draw [dashed,semithick]      (s6) -- (e3);
    \draw [dashed,semithick]      (d9) -- (d8);
    \draw [dashed,semithick]      (e4) -- (d7);
    \draw [dashed,semithick]      (e5) -- (d9);
    \draw [dotted,semithick]      (s7) -- (s8); 
    \draw[dashed, rounded corners, very thick, color=dartmouthgreen] ($(s2.north west)+(-3.0em,6.0em)$) rectangle ($(s8.south east)+(20.0em,+6.0em)$);
  \end{tikzpicture}
  \caption{
  Flowchart of the \exxm module (dashed green box) in \texttt{CP}.
  As described in the main text, the input required by this module includes the current set of MLWFs, $\{ \tphi{i} \}$, at each CPMD step.
  The output produced by \exxm includes the EXX energy ($\exx$) and the EXX contribution to the wavefunction forces ($\{ \dxx{i} \}$).
  Purple (brown) circles indicate that a given quantity is represented according to the \texttt{GRID} (\texttt{ORBITAL}) data distribution scheme (see Fig.~\ref{fig:redist} and Sec.~\ref{subsec:impl_distribution}), while the pale yellow circles represent data that are globally broadcasted.
  The $\overline{\bm r}$ notation indicates local (relative) Cartesian coordinates in a given subdomain.
  For a detailed description of each step in the \exxm module, see Secs.~\ref{subsubsec:impl_redist_phi}--\ref{subsubsec:impl_redist_d}. 
  }
  \label{fig:flowchart}
\end{figure}
In practice, EXX-based SCF calculations in \texttt{CP} take advantage of the incremental nature of the aforementioned MLWF refinement process by starting with a relatively inexpensive semi-local xc functional (\eg PBE~\cite{perdew_generalized_1996} for PBE0~\cite{perdew_rationale_1996, adamo_toward_1999}), which stabilizes $\rho(\bm r)$ and initiates the orbital localization procedure.
Once the semi-local DFT iterations reach $\approx 10\times$ the target SCF convergence threshold, the orbitals are typically quite localized and closely resemble the final set of MLWFs corresponding to the chosen semi-local xc functional.
At this point, the \exxm module is activated to perform the remaining steps required to reach SCF convergence at the hybrid DFT level, upon which one obtains the final set of MLWFs corresponding to the chosen hybrid xc functional.
For all systems tested, this approach has a significantly reduced computational cost when compared to the alternative procedure of: (\textit{i}) performing a standard (canonical) PBE calculation, (\textit{ii}) localizing the converged PBE orbitals from scratch, and (\textit{iii}) using these localized PBE orbitals as input into the incremental localization procedure described above to perform a PBE0 SCF calculation to convergence.
For insulating systems with small band gaps (\eg \ce{InN}), one should exercise caution when using GGA orbitals for the initial guess in this procedure, as the GGA xc functional might incorrectly predict a metallic system with the wrong band-index ordering.~\cite{wu_hybrid_2009}
With that caveat in mind, we also note that the computational cost associated with this initial SCF procedure is completely negligible when compared to the overall CPMD simulation, which is the focus of this work.
In future work, we hope to further optimize this incremental SCF procedure to enable high-throughput hybrid DFT-based single-point energy evaluations on large-scale condensed-phase systems (as well as linear-scaling EXX-based BOMD).

In analogy to the incremental localization procedure described above for EXX-based SCF calculations (\textit{cf}. Eqs.~\eqref{eq:sodd_mlwf_eom_discrete_1}--\eqref{eq:sodd_mlwf_eom_discrete_2}), we have also introduced this nested SODD determination of $\bm U$ into the propagation of the CPMD equations of motion.
This is accomplished by a CPMD propagation step, 
\begin{align}
  \chi_{j} (\bm r, t+\Delta t) &=  2\widetilde{\phi}_{j} (\bm r, t) - \widetilde{\phi}_{j} (\bm r, t-\Delta t) \nonumber\\
  & +\frac{\Delta t^2}{\mu}\widetilde{D}_{j}(\bm r, t) ,
  \label{eq:cpE_verlet}
\end{align}
in which an intermediary set of orbitals, $\{\chi_{j} (\bm r, t + \Delta t)\}$, is formed \via CPMD evolution (with time step $\Delta t$) of the MLWF orbitals, $\{\widetilde{\phi}_{j} (\bm r)\}$. 
During the CPMD propagation step, these intermediary orbitals become slightly more delocalized than the set of MLWFs (yet remain on the ground state potential energy surface), and are therefore refined by a subsequent localization step,
\begin{align}
  \widetilde{\phi}_{i} (\bm r, t+\Delta t) =& \sum_j U_{ij}(t+\Delta t) \chi_{j} (\bm r, t+\Delta t) ,
  \label{eq:cpE_relocalize}
\end{align}
in which the unitary transformation $\bm U$ is generated by \textit{tightly converging} the nested SODD optimization of the Marzari-Vanderbilt functional.
By doing so after every CPMD propagation step, we ensure that the resulting $\{\widetilde{\phi}_{i} (\bm r, t)\}$ and $\{\widetilde{D}_{i} (\bm r, t)\}$ are indeed the MLWFs and the forces acting on them.
We note in passing that the need to perform the additional localization step in Eq.~\eqref{eq:cpE_relocalize} reflects the lack of gauge invariance in the electronic CPMD equations of motion within the MLWF representation~\cite{iftimie_--fly_2004,thomas_field_2004}.
Nevertheless, the intermediary orbitals generated by Eq.~\eqref{eq:cpE_verlet} are typically good approximations to the MLWFs, and thereby provide a rather good initial guess to the SODD localization procedure~\cite{sharma_ab_2003}.
As a result, the localization procedure typically converges with a small number of nested SODD iterations (\eg $3$--$4$ iterations for the liquid water systems in Sec.~\ref{subsubsec:c-g_para}), which results in minimal computational overhead when compared to the cost of the EXX calculation.
Moving forward, this incremental localization scheme could be avoided using the field-theoretic approach proposed by Tuckerman and coworkers~\cite{thomas_field_2004}, which introduces additional fictitious dynamics on a set of gauge fields to enable ``on-the-fly'' propagation of the MLWF transformation ($\bm U$) matrix.

\subsection{MLWF-based EXX-CPMD: Data Distribution Schemes \label{subsec:impl_distribution}}

\begin{figure*}[t!]
  \begin{center}
    \includegraphics[width=1.0\linewidth]{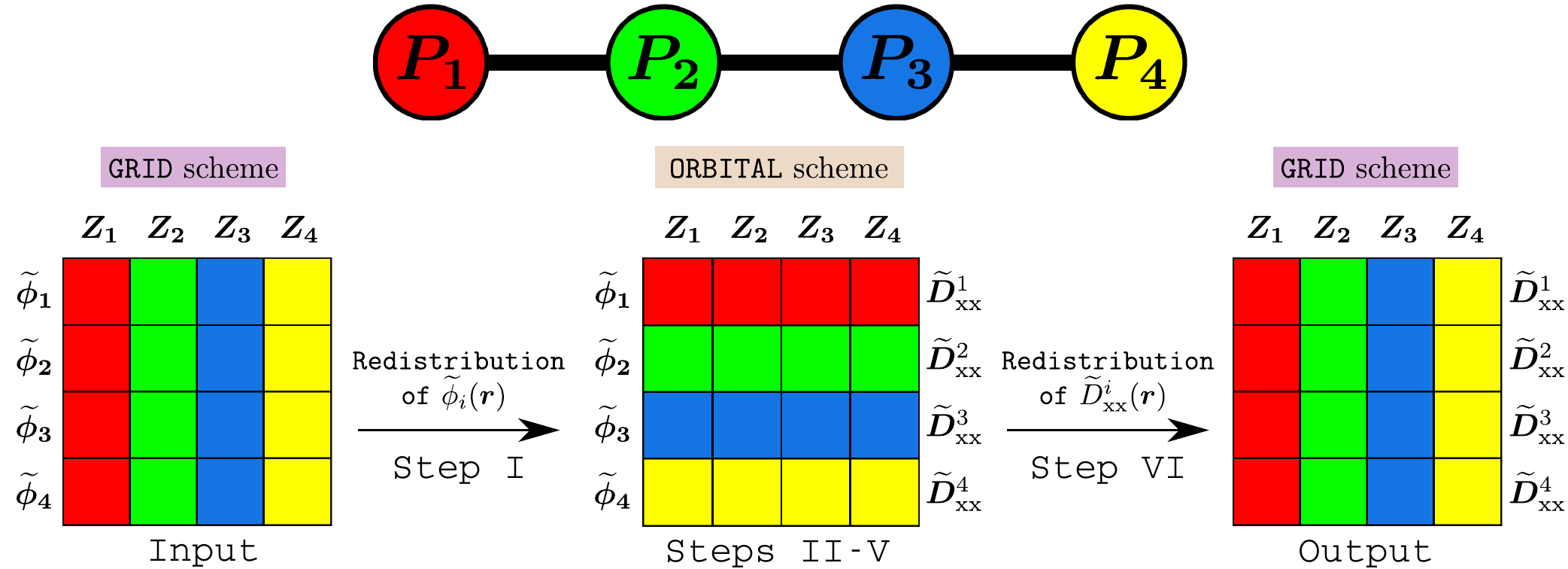}
  \end{center}
  \caption{
  Schematic illustration of the \texttt{GRID} and \texttt{ORBITAL} data distribution schemes in \texttt{QE}.
  For simplicity, we consider a system consisting of a single water molecule with $N_{o}=4$ MLWFs ($\tphi{i}$), a simulation cell consisting of a real-space simple-cubic grid that has been partitioned into $N_{\rm slab}=4$ slabs along the $z$-direction ($Z_{i}$), and a pool of $N_{\rm proc}=4$ \mpi{} processes ($P_{i}$).
  As depicted at the top of the figure, each of these \mpi{} processes (and the corresponding data it holds in local memory) is assigned a color: $P_1$ (red), $P_2$ (green), $P_3$ (blue), and $P_4$ (yellow). 
  As input into the \exxm module, the $\tphi{i}$ are provided in the \texttt{GRID} scheme, in which a given \mpi{} process, $P_{i}$, holds the data corresponding to \textit{all} $N_{o}$ MLWFs on \textit{one} slab, $Z_{i}$, of the real-space grid.
  During Step~I of the \exxm module (Sec.~\ref{subsubsec:impl_redist_phi}), the $\tphi{i}$ are redistributed according to the \texttt{ORBITAL} scheme, in which a given \mpi{} process, $P_{i}$, holds the data corresponding to only \textit{one} MLWF, $\tphi{i}$, across \textit{all} $N_{\rm slab}$ slabs of the real-space grid.
  As described in Secs.~\ref{subsubsec:impl_spdomain}--\ref{subsubsec:impl_compute_esf}, Steps~II--V involve selective communication of the $\tphi{i}$ between \mpi{} processes and computation of all EXX-related quantities ($\exx$ and $\{\dxx{i}\}$).
  At the end of Step~V, the $\{\dxx{i}\}$ are stored according to the \texttt{ORBITAL} scheme, and are redistributed back to the \texttt{GRID} scheme during Step~VI (Sec.~\ref{subsubsec:impl_redist_d}), the final step of the \texttt{exx} module.
  }
  \label{fig:redist}
\end{figure*}
As mentioned above, we employ a hybrid \mpi{}/\omp{} parallelization scheme to enable large-scale EXX-based AIMD on massively parallel supercomputer architectures containing 1000s of nodes.
Our algorithm, which is described in Sec.~\ref{subsec:impl_algorithm} below, is primarily based upon the \mpi{} distributed-memory paradigm, which requires specific data distribution schemes to minimize communication overhead and maximize computational efficiency.
During a GGA-based CPMD simulation in \texttt{QE}, the orbitals, charge density, and potential are constantly transformed between real- and reciprocal-space \via the \texttt{fwdFFT} and \texttt{invFFT} operations.
With all real-space quantities numerically represented on a grid (mesh) that is discretized along the corresponding lattice vectors, \texttt{QE} employs the \texttt{GRID} data distribution scheme to scatter these quantities across $N_{\rm proc}$ \mpi{} processes (ranks).
In the \texttt{GRID} data distribution scheme (see Fig.~\ref{fig:redist}), the real-space grid is partitioned into $N_{\rm slab}$ slabs along the $z$ axis.
Assuming $N_{\rm proc} = N_{\rm slab}$ for simplicity, each \mpi{} process will hold the data corresponding to \textit{all} distributed real-space quantities on a \textit{single} slab of the real-space grid.
In doing so, this data distribution scheme facilitates efficient parallel FFT by dividing the 3D FFT into a set of 2D FFTs (each of which can be executed by a given \mpi{} process within a given slab) followed by a 1D FFT along the direction of the slab partition.

As depicted in Fig.~\ref{fig:flowchart}, the input to the \exxm module in \texttt{QE} includes the current set of MLWFs, $\{ \tphi{i} \}$, at each CPMD step.
These MLWFs are distributed across \mpi{} processes according to the \texttt{GRID} data distribution scheme, in which a given process holds the data corresponding to all MLWFs on a given slab of the real-space grid.
Although the \texttt{GRID} scheme is convenient for efficient parallel FFT, this data distribution model is far from ideal for an efficient massively parallel implementation of our MLWF-based EXX approach.
As such, we have introduced an alternative \texttt{ORBITAL} data distribution scheme in \texttt{QE} (see Fig.~\ref{fig:redist}), in which a given \mpi{} process now holds quantities like $\tphi{i}$ and $\dxx{i}$ for a \textit{single} MLWF across the \textit{entire} real-space grid (for the case in which $N_{\rm proc} = N_o$; for other cases, see the discussion below in Sec.~\ref{subsubsec:impl_redist_phi}).
The details behind the transformation between the \texttt{GRID} and \texttt{ORBITAL} data distribution schemes are provided below in Secs.~\ref{subsubsec:impl_redist_phi} and \ref{subsubsec:impl_redist_d}.

The \texttt{ORBITAL} data distribution scheme is particularly suited for our real-space MLWF-based EXX algorithm, since this approach is centered around orbital sparsity and the efficient evaluation of $\tv{ij}$.
For one, the \texttt{ORBITAL} scheme allows us to utilize a significantly larger number of \mpi{} processes ($N_{\rm proc} \gg N_{\rm slab}$), as the number of MLWFs or overlapping MLWF pairs (both of which grow linearly with system size) quickly exceeds $N_{\rm slab}$ (which grows with the cubic root of the system size).
The \texttt{ORBITAL} scheme also allows us to exploit intranode parallelization with $N_{\rm thread}$ \omp{} threads during the most computationally intensive steps in our algorithm, \eg solving the PE to obtain $\tv{ij}$ (see Sec.~\ref{sec:impl_PS}).
As a result, this hybrid \mpi{}/\omp{} parallelization scheme not only provides us with access to even more computational resources during EXX-based simulations, but also allows us to sidestep the prohibitively large data communication overhead associated with an \mpi{}-based solution to the PE.

\subsection{MLWF-based EXX-CPMD: Algorithm \label{subsec:impl_algorithm}}

In this section, we provide a detailed description for each of the steps inside the \exxm module in \texttt{QE}.
Our discussion will follow the flowchart depicted in Fig.~\ref{fig:flowchart}, in which the current set of MLWFs in real space ($\{ \tphi{i} \}$, distributed according to the \texttt{GRID} scheme) are provided as input into the \exxm module.
Subsequent output of \exxm includes the EXX energy ($\exx$) as well as the EXX contribution to the wavefunction forces ($\{ \dxx{i} \}$, which are again distributed according to the \texttt{GRID} scheme).
This preserves compatibility with the rest of \texttt{CP}, and allows for a modular \exxm codebase.

\subsubsection{Step I: Redistribution of MLWFs \label{subsubsec:impl_redist_phi}}

In the \exxm algorithm, the assignment of MLWFs to a given \mpi{} process is based on $\zeta \equiv N_{\rm proc} / N_{o}$, \ie the ratio of available \mpi{} processes to the number of MLWFs.
When $\zeta = 1$, there is one \mpi{} process per MLWF, and each process, $P_{i}$, is assigned a unique MLWF, $\tphin{i}$.
With limited computational resources ($N_{\rm proc} < N_{o}$), $\zeta < 1$ and multiple MLWFs are assigned to each process; as such, a balanced distribution of MLWFs across \mpi{} processes is only possible when $N_{\rm proc}$ is a divisor of $N_{o}$.
In the strong-scaling limit, our \exxm algorithm allows for $\zeta$ to take on integer values greater than one, in which a given MLWF is assigned to multiple \mpi{} processes.
Unless otherwise specified, we will assume that $\zeta = 1$ throughout the remainder of Sec.~\ref{subsec:impl_algorithm}.

Given the current set of MLWFs in real space, $\{ \tphi{i} \}$, which are distributed among the available $N_{\rm proc}$ \mpi{} processes according to the \texttt{GRID} scheme, the first step in the \exxm module is the forward redistribution of these quantities into the \texttt{ORBITAL} data distribution scheme.
For this purpose, each \mpi{} process collects an assigned $\tphi{i}$ across the entire real-space grid \via an \texttt{ALL-TO-ALL} internode communication step, as shown in Fig.~\ref{fig:redist}.
This \texttt{ALL-TO-ALL} communication is performed twice per CPMD step: once here in the forward redistribution of $\{ \tphi{i} \}$ from the \texttt{GRID} to the \texttt{ORBITAL} scheme, and once in Step~VI in the inverse redistribution of $\{ \dxx{i} \}$ from the \texttt{ORBITAL} to the \texttt{GRID} scheme (see Sec.~\ref{subsubsec:impl_redist_d}).
As discussed in Sec.~\ref{sec:Perf}, this communication overhead represents a substantial fraction of the total cost associated with the \exxm module, and can be significantly reduced by a more sophisticated communication scheme over select subsets of the \mpi{} process pool, \ie those containing the regions of real space containing the relevant MLWF-orbital domain, $\Omega_i$ (see Sec.~\ref{sec:RealSpaceEXX}).
This algorithmic improvement is currently underway and will be described in future work.

\subsubsection{Step II: Construction of Pair List and Proto-Subdomains \label{subsubsec:impl_spdomain}}

With the MLWFs distributed among \mpi{} processes according to the \texttt{ORBITAL} scheme, we now explain how the \exxm module exploits the sparsity of the MLWFs, and utilizes system-size independent subdomains of $\Omega$ during the computation of all EXX-related quantities.
To accomplish this goal, we will first describe the construction of the so-called unique MLWF-pair list, $\lu$, which not only contains the relevant set of overlapping MLWF pairs, but also determines how the computational workload associated with these pairs is distributed among the pool of available \mpi{} processes.
This is followed by a detailed description of the set of ``proto-subdomains'' employed in the \exxm module, which represent computationally efficient alternatives to the formal $\Omega_{i}$ and $\Omega_{ij}$ subdomains introduced above in Sec.~\ref{sec:RealSpaceEXX}.

\begin{center}
  \textit{Construction of the MLWF-Pair List}
\end{center}

To exploit the first level of computational savings, which originates from the fact that MLWFs are exponentially localized and only overlap with a limited number of neighbors, two MLWFs, $\tphi{i}$ and $\tphi{j}$, are considered an overlapping pair if $| \tC{i} - \tC{j} | < \rpr$.
A judicious choice for $\rpr$ is required for accurately calculating all EXX-related quantities, and an analysis of the convergence of $\exx$ with respect to $\rpr$ will be provided in Sec.~\ref{sec:ecr}.

At the current point in the algorithm, each $\tphi{i}$ is stored according to the \texttt{ORBITAL} data distribution scheme on one (or more) \mpi{} processes (depending on the value of $\zeta$ employed during runtime, see Sec.~\ref{subsubsec:impl_redist_phi}).
For simplicity, we will discuss the $\zeta = 1$ case first, in which there is only one \mpi{} process per MLWF, and each process, $P_{i}$, is assigned a unique MLWF, $\tphi{i}$.
As such, $P_{i}$ lacks direct access to $\tphi{j}$ for $j \neq i$, which is required for the construction of $\trho{ij}$ and the subsequent computation of $\tv{ij}$ for evaluating the $\braket{ij}$-pair contribution to $\exx$, $\dxx{ij}$, and $\dxx{ji}$.
Although $\trho{ii}$ can be constructed \textit{locally} on $P_{i}$ to evaluate the $\braket{ii}$-pair (self pair) contribution to each of these quantities, all of the other $\braket{ij}$-pair contributions will require communication between \mpi{} processes.

To design an MLWF-based EXX algorithm that achieves a minimal time to solution while efficiently utilizing all parallel computational resources, one needs to (\textit{i}) minimize the total computational workload, (\textit{ii}) minimize the number of interprocess communication events, and (\textit{iii}) maintain a balanced workload among the pool of available \mpi{} processes.
To accomplish this goal, we now describe the procedure employed in the \texttt{exx} module to construct the so-called unique MLWF-pair list, $\lu$, which defines the computation and communication protocol in our algorithm.
To construct $\lu$, the indices corresponding to \textit{all} overlapping MLWF pairs (as determined by the aforementioned criteria based on $\rpr$) are first assembled into the non-unique MLWF-pair list, $\lz$, which contains all possible permutations $ij$ and $ji$ of these overlapping MLWF pairs.
Since the $\braket{ij}$- and $\braket{ji}$-pair contributions to $\exx$ are equivalent (\textit{cf}. Eqs.~\eqref{eq:exx}--\eqref{eq:vxxGen_mlwf}), it is clear that $\lz$ is redundant and contains twice as many pairs as needed.

Before discussing the procedure used to determine $\lu$, we first demonstrate that exploiting such redundancy within the more parallelizable \texttt{ORBITAL} data distribution scheme leads to the requirement for two interprocess communication events per unique MLWF pair.
To see this more clearly, one only needs to consider a minimalistic system which contains a single $\braket{ij}$-pair of overlapping MLWFs.
Throughout this example, spatial arguments will be suppressed (\eg $\tphin{i}$ will be used instead of $\tphi{i}$), since all computation and communication events will be performed using system-size independent subdomains (\textit{vide infra}).
With $\tphin{i}$ located on $P_{i}$ and $\tphin{j}$ located on $P_{j}$, we will first consider what happens if the inherent pair redundancy is \textit{not} exploited.
In this case, $\tphin{i}$ ($\tphin{j}$) is first communicated to $P_{j}$ ($P_{i}$) for a total of two interprocess communication events.
At this point, each \mpi{} process constructs the corresponding MLWF-product density, $\trhon{ij} = \trhon{ji}$, and proceeds to compute $\tvn{ij} = \tvn{ji}$ by solving two equivalent PEs.
Since the solution of the PE is the dominant computational step in our EXX algorithm, this will count for a total of two computation events.
With $\tvn{ij} = \tvn{ji}$ available on both $P_{i}$ and $P_{j}$, each process is now in a position to compute the $\braket{ij}$- and $\braket{ji}$-pair contributions to $\exx$ \via Eq.~\eqref{eq:exx}, as well as $\dxxn{ij} = \tvn{ij}\tphin{j}$ and $\dxxn{ji} = \tvn{ji}\tphin{i}$ \via Eq.~\eqref{eq:Dxx_mlwf}.
As depicted in Fig.~\ref{fig:flowchart}, the $\{\dxxn{i}\}$ are needed in the \texttt{ORBITAL} data distribution scheme before these quantities are finally redistributed back to the \texttt{GRID} scheme to ensure compatibility with the other modules in \texttt{QE} (see Fig.~\ref{fig:redist}).
As such, the local evaluation of $\dxxn{ij}$ on $P_{i}$ and $\dxxn{ji}$ on $P_{j}$ directly provides these quantities in the requisite \texttt{ORBITAL} data distribution scheme without the need for any additional communication.
Hence, the total cost per unique MLWF pair amounts to two units of communication followed by two units of computation, when pair redundancy is not exploited.

Since the removal of all MLWF-pair redundancy is crucial for minimizing the total number of computational events (and hence the overall time to solution), we now consider the case where this inherent pair redundancy is exploited.
In this case, only $\tphin{j}$ would be sent to $P_{i}$ with an associated cost of one unit of communication, and $P_{i}$ will therefore be solely responsible for computing all EXX-related quantities.
Since the $\braket{ij}$- and $\braket{ji}$-pair contributions to $\exx$ are equivalent, these quantities can be computed on $P_{i}$ \via a single computation event (\ie the solution to the corresponding PE), and then sent to any other process with minimal communication (\ie one double-precision number for each pair contribution to $\exx$).
Although $\dxxn{ij} \neq \dxxn{ji}$, this also poses no problem as $P_{i}$ has direct access to $\tphin{i}$ and $\tphin{j}$, and hence both $\dxxn{ij}$ and $\dxxn{ji}$ can be computed \textit{locally}.
With the requirement that the $\{\dxxn{i}\}$ are stored in the \texttt{ORBITAL} data distribution scheme, this will incur an additional communication event as $\dxxn{ji}$ is shipped back to $P_{j}$.
Hence, exploiting the inherent MLWF-pair redundancy reduces the computational workload by half (as expected), but it does not change the requirement for two communication events per unique MLWF pair.

\begin{figure*}[t!]
  \includegraphics[width=\linewidth]{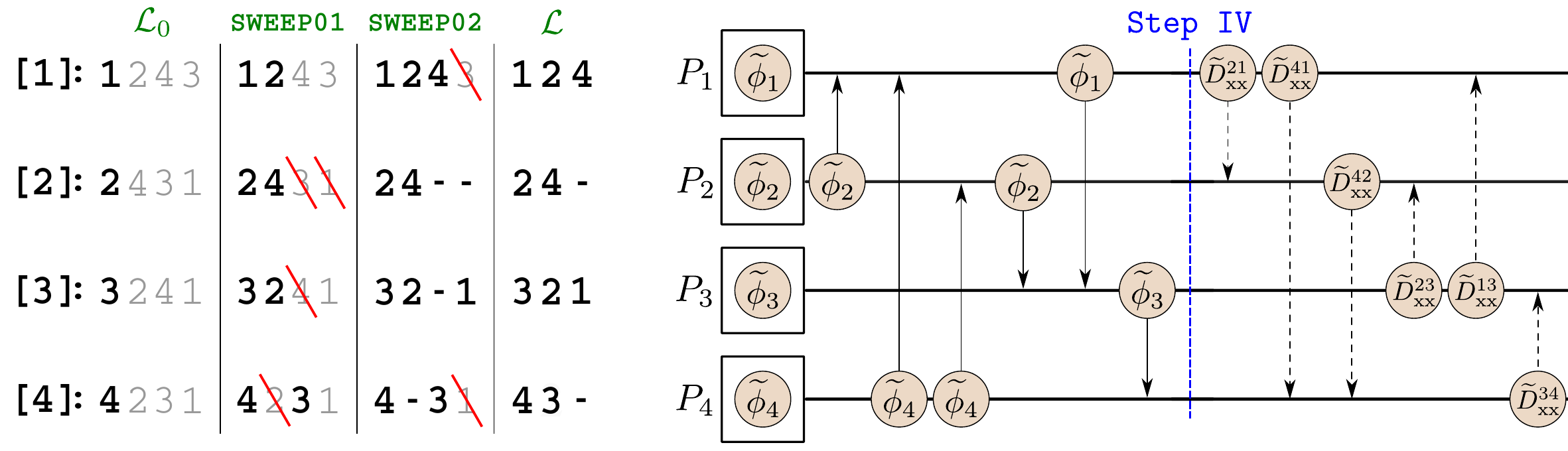}
  \caption{
  Graphical depiction of the unique MLWF-pair list construction process and corresponding MLWF communication scheme in the \exxm module.
  For simplicity, we will again consider a single water molecule with $N_{o}=4$ MLWFs ($\tphin{i}$) and a pool of $N_{\rm proc}=4$ \mpi{} processes ($P_{i}$), \ie $\zeta = 1$.
  Starting with the non-unique MLWF-pair list, $\lz$, which contains all possible permutations of overlapping MLWF pairs, the step-by-step procedure employed to transform $\lz$ into the unique MLWF-pair list, $\lu$, is depicted in the \textit{left} panel.
  Since all of the MLWFs are mutually overlapping in a single water molecule, $\lzb{i}$ contains $\{j\}$, the indices corresponding to all MLWFs (including $j = i$), which have been sorted according to $|\tC{i}-\tC{j}|$.
  During the process of reducing $\lz$ to $\lu$, $i$ is always selected (bold black font) to remain in $\lzb{i}$, while the remaining non-unique indices are shown in gray.
  In the first sweep (\texttt{SWEEP01}) of Algorithm~\ref{alg:upl}, the next element $j\in\lzb{i}$ is selected to remain in $\lzb{i}$, while the now redundant index $i$ is removed (red slash) from $\lzb{j}$.
  During the first sweep in this example, the indices $2,4,2,3$ were selected to remain in $\lzb{1},\lzb{2},\lzb{3},\lzb{4}$, while the corresponding redundant indices, $1,2,3,4$, were removed from $\lzb{2},\lzb{4},\lzb{2},\lzb{3}$.
  This process is repeated until all of the MLWF-pair redundancy is removed from $\lz$, upon which one is left with the final $\lu$.
  For a single water molecule, only two sweeps are required to reach this stage; at that point, all of the unique MLWF pairs have been assigned to a given $\lub{i}$, and no redundant indices remain.
  With each MLWF stored according to the \texttt{ORBITAL} data distribution scheme (in which $\tphin{i}$ is assigned to $P_{i}$), the final $\lu$ determines how the computational workload will be distributed among the pool of available \mpi{} processes.
  Even in this simple example, there exists a mismatch in the number of pairs assigned to each process, with three MLWF pairs assigned to $P_1$ and $P_3$, and only two MLWF pairs assigned to $P_2$ and $P_4$.
  Such discrepancies are expected (even for the most homogeneous systems) and lead to an imbalance in the computational workload.
  By virtue of the \texttt{ORBITAL} data distribution scheme, $\lu$ also determines the corresponding MLWF communication protocol, which is depicted in the \textit{right} panel for the single water molecule.
  After a given $\tphin{j}$ is directly communicated to $P_{i}$, this \mpi{} process forms $\trhon{ij}$, solves the corresponding PE for $\tvn{ij}$ (depicted by the blue dashed line), and computes the $\braket{ij}$-pair contribution to $\exx$, $\dxxn{ij}$, and $\dxxn{ji}$ (which is sent back to $P_{j}$). 
  As an example, consider $\lub{2}$, which contains two indices ($2,4$).
  Since $P_{2}$ already holds $\tphin{2}$, $\trhon{22}$ (required for computing $\tvn{22}$) can be constructed locally without the need for any interprocess communication.
  To construct $\trhon{24}$ (and hence $\tvn{24}$), $\tphin{4}$ is sent from $P_{4}$ to $P_{2}$ (solid arrow).
  After the corresponding $\dxxn{42}$ is formed, it is shipped back to $P_{4}$ (dashed arrow).
  Besides this straightforward dependency between receiving an MLWF and computing the corresponding EXX contributions to the wavefunction forces, the communication in the \exxm module has no discernible time axis in the figure above.
  }
  \label{fig:pair_list}
\end{figure*}

During this non-redundant evaluation of the $\braket{ij}$- and $\braket{ji}$-pair contributions, the fact that $P_{j}$ was idle while $P_{i}$ performed all of the required computations creates an imbalance in the computational workload assigned to each \mpi{} process.
With the freedom to assign the computational workload associated with the $\braket{ij}$-pair to either $P_{i}$ or $P_{j}$, the \exxm module is now tasked with determining how the total computational workload will be distributed among the pool of available \mpi{} processes.
Armed with knowledge of the total number of non-unique MLWF pairs in the system (\via $\lz$) as well as the use of system-size independent subdomains to regularize the computational cost associated with the solution to each PE (\textit{vide infra}), the process for doing so involves a static load-balancing algorithm which seeks to minimize the overall time to solution by reducing the imbalances present in the computational workload, and hence the number of idle processes.
Although it is certainly possible in the current version of the algorithm, we chose not to involve a third process, $P_{k}$, in the evaluation of the $\braket{ij}$-pair contribution, as this would introduce two additional communication events, \ie $\tphin{i}$ to $P_{k}$ and $\dxxn{ij}$ back to $P_{i}$ (in addition to $\tphin{j}$ to $P_{k}$ and $\dxxn{ji}$ back to $P_{j}$).
In this regard, the local computation of $\dxxn{ij}$ on $P_{i}$ not only avoids additional unnecessary communication events, but also allows for reduced storage requirements as this quantity can be cumulatively incremented (over multiple $j$) within a single array corresponding to $\xmega{i}$.

This static load-balancing algorithm can be represented by the so-called unique MLWF-pair list, $\lu$, the construction of which is described in the \textit{left} panel of Fig.~\ref{fig:pair_list} (for the illustrative case of a single water molecule) as well as Algorithm~\ref{alg:upl} (for the general case).
We start with the $\lz$ array, which contains all possible permutations of overlapping MLWF pairs, \ie $\lzb{i}$ (the $i$-th row of $\lz$) is populated with a list of indices, $\{j\}$, corresponding to all $\tphin{j}$ that overlap with $\tphin{i}$.
For each $i$, the indices $j\in\lzb{i}$ are sorted into ascending order based on their vicinity to $\tphin{i}$ \via $|\tC{i}-\tC{j}|$.
By construction, each $\lzb{i}$ also contains $i$ (self pair) and will retain this non-redundant index throughout the refinement of $\lz$ to $\lu$ in Algorithm~\ref{alg:upl}.
While there are still redundant pairs in $\lz$, this algorithm will consecutively sweep over MLWFs to locate redundant pairs such as $\braket{ij}$ and $\braket{ji}$; in our approach, this is tantamount to finding both $j\in\lzb{i}$ and $i\in\lzb{j}$.
Once located, the algorithm eliminates this redundancy from $\lz$ by removing the index $i$ from $\lzb{j}$.
At the end of these sweeps, all of the redundancies in $\lz$ are removed, and we are left with $\lu$, the unique MLWF-pair list.
This list contains the minimum number of computational tasks required to evaluate all EXX-related quantities, and dictates how this computational workload will be distributed among the pool of available \mpi{} processes.
By virtue of the \texttt{ORBITAL} data distribution scheme, $\lu$ also encodes the communication protocol that will be followed throughout the remainder of the \exxm module (see Sec.~\ref{subsubsec:impl_commphi}).
With $\zeta=1$, this amounts to sending $\xto{\tphin{j}}{P_{i}}$ and $\xto{\dxxn{ji}}{P_{j}}$ for each unique $\braket{ij}$ pair, as depicted in the \textit{right} panel of Fig.~\ref{fig:pair_list}.
\begin{algorithm}[H]
  \begin{algorithmic}
    \State \texttt{any\_removal} $\gets$ \texttt{TRUE}
    \While {\texttt{any\_removal}}
    \State \texttt{any\_removal} $\gets$ \texttt{FALSE}
    \For{$i=1, N_{o}$}
    \For{$j\neq i \in \lzb{i}$}
    \If {$i\in\lzb{j}$}
    \State $\lzb{j} \gets \lzb{j} \setminus \{i\}$
    \State \texttt{any\_removal} $\gets$ \texttt{TRUE}
    \State \Break
  \EndIf
\EndFor
    \EndFor
  \EndWhile
  \State $\lu \gets \lz$
  \caption{Refinement of $\lz$ to $\lu$
  }
  \label{alg:upl}
\end{algorithmic}
\end{algorithm}

By construction, static load-balancing algorithms (such as Algorithm~\ref{alg:upl}) yield fairly well-balanced workload distributions by mitigating potential imbalances during the refinement of $\lz$ to $\lu$.
Here, we note that the distance-based sorting of the indices in each row of $\lz$ is crucial for avoiding severe workload imbalances due to sequential index ordering.
In this regard, an equally effective load-balancing algorithm would be possible by performing random sweeps over row indices (and completely avoiding the initial distance-based sorting procedure). 
Since the number of overlapping pairs per MLWF will often change throughout an AIMD simulation, this static load-balancing algorithm is performed during each MD step in an attempt to determine an optimal workload balance.
For a detailed discussion regarding the performance of this static load-balancing algorithm during CPMD simulations of liquid water, as well as future possible improvements of this approach, see Sec.~\ref{subsubsec:c-g_para}.

When computational resources are limited, the \exxm module can utilize less \mpi{} processes during runtime (\ie $\zeta<1$).
In this case, multiple MLWFs are contiguously assigned to each process, and a balanced distribution of the workload (within the framework defined by Algorithm~\ref{alg:upl}) is more likely when $N_{\rm proc}$ is a divisor of $N_{o}$; as such, this is the current recommended setting whenever applicable (see Sec.~\ref{subsubsec:impl_redist_phi}).
In the isolated water molecule example in Fig.~\ref{fig:pair_list}, the use of $\zeta=1/2$ would start with $P_{1}$ holding $\tphin{1}$ and $\tphin{2}$, and $P_{2}$ holding $\tphin{3}$ and $\tphin{4}$.
After running Algorithm~\ref{alg:upl} to generate $\lu$, the workload associated with a given MLWF is mapped onto the process holding this orbital.
This results in five units of computation assigned to each \mpi{} process: two self pairs, one local pair (in which both MLWFs are held on the same process), and two non-local pairs (in which one of the MLWFs is held on a different process), \eg $P_{1}$ would be responsible for $\braket{ii} = \braket{11}$ and $\braket{ii} = \braket{22}$ (two self pairs), $\braket{ij} = \braket{12}$ (one local pair), and $\braket{ij} = \braket{14}$ and $\braket{ij} = \braket{24}$ (two non-local pairs).
In this case, the workload is optimally balanced and the maximum number of computation events per process is only $5/3 \times$ (instead of $2 \times$) larger than $\zeta=1$.
This allows for a more computationally efficient means to performing an EXX calculation, albeit with a longer time to solution.

With access to massively parallel resources ($\zeta > 1 \in \mathbb{N}$), each $\tphin{i}$ is now replicated and stored in memory on the $P_{i}, P_{i+N_{o}}, \ldots, P_{i+(\zeta-1) N_{o}}$ processes.
After running Algorithm~\ref{alg:upl} to generate $\lu$, the workload associated with a given MLWF is split into $\zeta$ parts, each of which is assigned to one of the processes holding this orbital.
For the isolated water molecule, the use of $\zeta=2$ ($\zeta=1$) results in processes assigned with $1\mathrm{-}2$ ($2\mathrm{-}3$) computational tasks.
This reduces the maximum number of computation events per process from $3$ to $2$, and hence lowers the overall time to solution.
However, this gain comes at the expense of increasing the workload imbalance from $1/3$ (\ie processes with the lightest workload idling for $\approx 1/3$ of the time) to $1/2$, and is therefore a less efficient use of the available computational resources.

\begin{center}
  \textit{Construction of Proto-Subdomains}
\end{center}

As discussed throughout this work, the efficient evaluation of $\tv{ij}$ is the cornerstone of our MLWF-based EXX approach.
To exploit the sparsity of the MLWFs and still retain a numerically exact evaluation of $\tv{ij}$, this quantity is computed \via the solution to the corresponding PE for all points in $\Omega$ that are contained in $\Omega_{ij}$ (see Eq.~\eqref{eq:pe}).
Since the PE is a boundary-value problem, the required boundary conditions are provided by the ME of $\trho{ij}$ about $\tC{ij}$ on the thin shell of the real-space grid surrounding $\Omega_{ij}$ (see Eqs.~\eqref{eq:me}--\eqref{eq:mepole}).
By computing $\tv{ij}$ for all $\bm r \in \Omega_{ij}$, $\exx$ can be computed in a numerically exact fashion (\textit{cf}. Eq.~\eqref{eq:exx_omega}).
However, the numerically exact evaluation of the EXX contribution to the wavefunction forces, $\dxx{ij}$ and $\dxx{ji}$, requires $\tv{ij}$ for all $\bm r \in \Omega_{j}$ and $\bm r \in \Omega_{i}$, respectively (see Eq.~\eqref{eq:Dxx_mlwf_omega}).
Since $\Omega_{ij} \subset \Omega_{j}$ and $\Omega_{ij} \subset \Omega_{i}$, $\dxx{ij}$ and $\dxx{ji}$ are evaluated with $\tv{ij}$ from the solution to the PE for all $\bm r \in \Omega_{ij}$. 
For $\bm r \in \Omega_{j} \setminus \Omega_{ij}$ and $\bm r \in \Omega_{i} \setminus \Omega_{ij}$, $\tv{ij}$ can be conveniently and accurately supplied by a sufficiently converged ME of $\trho{ij}$.
As such, the ME serves the dual purpose of providing the necessary boundary conditions for the PE as well as the far-field $\tv{ij}$ required for a numerically exact computation of both $\dxx{ij}$ and $\dxx{ji}$.
\begin{figure}[ht!]
  \includegraphics[width=\linewidth]{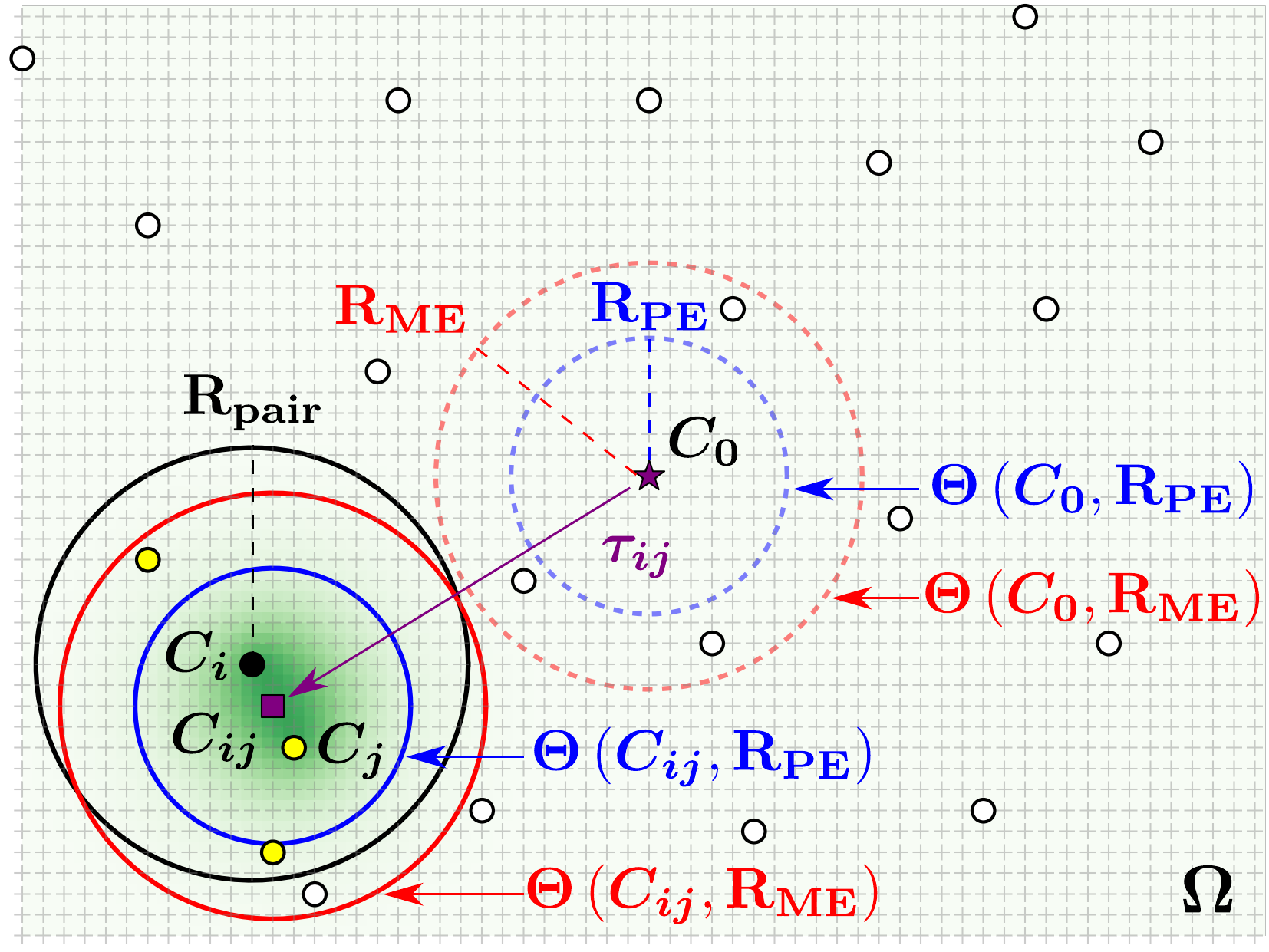}
  \caption{
  Graphical depiction of the proto-subdomains used in the \texttt{exx} module.
  Dots are used to denote the MLWF centers, $\tC{i}$, which are approximated by the closest points, $\gC{i}$, on the real-space grid, $\Omega$.
  The dashed blue and red circles bound the two concentric spherical proto-subdomains, $\pe{\gC{0}}$ and $\me{\gC{0}}$, which are assembled around $\gC{0}$ (purple star) with radii $\rpe$ and $\rme$, respectively.
  Pair-exchange interactions involving $\tphi{i}$ include all overlapping MLWFs (yellow dots), whose centers are located within a distance, $\rpr$, of $\tC{i} \approx \gC{i}$ (black dot) that is large enough to account for all $\tphi{k}$ with $\Omega_{ik} \neq \emptyset$.
  For the overlapping $\braket{ij}$ pair, the $\pe{\gC{0}}$ and $\me{\gC{0}}$ proto-subdomains are translated across $\Omega$ \via a rigid grid offset, $\tau_{ij}$, to form $\pe{\gC{ij}}$ and $\me{\gC{ij}}$, which are centered at $\gC{ij} \approx \tC{ij}$ (purple square).
  To evaluate the $\braket{ij}$ contribution to all EXX-related quantities, the corresponding PE, $\nabla^2 \tv{ij} = -4\pi\trho{ij}$, is solved for $\tv{ij}$ on the $\pe{\gC{ij}}$ subdomain, which encompasses $\Omega_{ij}$, the support of $\trho{ij}$ (shaded in dark green).
  Boundary conditions for the PE (as well as the far-field $\tv{ij}$) are computed \via a ME of $\trho{ij}$ on the $\me{\gC{ij}} \setminus \pe{\gC{ij}}$ shell surrounding $\pe{\gC{ij}}$.
  }
  \label{fig:spheres}
\end{figure}

To exploit this second level of computational savings, which originates from the fact that a numerically exact evaluation of all EXX-related quantities can be restricted to real-space domains that are system-size independent and significantly smaller than $\Omega$, the \exxm module introduces an alternative formulation of the $\Omega_{i}$ and $\Omega_{ij}$ subdomains described above in Sec.~\ref{sec:RealSpaceEXX}.
To begin, we first note that subdomains like $\xmega{i}$ and $\xmega{ij} = \xmega{i} \cap \xmega{j}$ are formally defined as the points in $\Omega$ for which $\tphi{i}$ and $\trho{ij}$ are non-negligible (\ie larger than some predetermined numerical cutoff).
As such, both of these subdomains can have irregular and even disjoint shapes.
However, this is a cumbersome and computationally demanding definition that would require screening substantial sectors of $\Omega$ for each pair of MLWFs during every CPMD step.
To combat this issue and still maintain a numerically exact evaluation of all required quantities, one could simply utilize two concentric \textit{spherical} subdomains per $\braket{ij}$ pair, \ie $\Theta(\tC{ij},\rpe^{ij})$ and $\Theta(\tC{ij},\rme^{ij})$, which are spheres centered at $\tC{ij}$ with radii $\rpe^{ij}$ and $\rme^{ij}$ chosen to be large enough to encompass $\Omega_{ij}$ and $\Omega_{i} \cup \Omega_{j}$, respectively.
In doing so, the corresponding PE, $\nabla^2 \tv{ij} = -4\pi\trho{ij}$, could then be solved without any domain truncation error on $\Theta(\tC{ij},\rpe^{ij})$, which is significantly smaller than $\Omega$.
Computing the ME of $\trho{ij}$ on the $\Theta(\tC{ij},\rme^{ij}) \setminus \Theta(\tC{ij},\rpe^{ij})$ shell would again provide the necessary boundary conditions for the PE as well as the far-field $\tv{ij}$ needed for evaluating both $\dxx{ij}$ on $\xmega{j}$ and $\dxx{ji}$  on $\xmega{i}$.
Since both of these subdomains are contained in $\Theta(\tC{ij},\rme^{ij})$, $\dxx{ij}$ and $\dxx{ji}$ can also be computed in a numerically exact fashion on a subset of points contained in $\Omega$.

To efficiently utilize this concept of concentric spherical subdomains in the \exxm module, we assemble two fixed-size proto-subdomains, $\pe{\gC{0}}$ and $\me{\gC{0}}$, centered around a predetermined origin, $\gC{0}$, which is chosen to be one of the grid points in $\Omega$.
When dealing with all computations involving a given $\braket{ij}$ pair, these proto-subdomains are simply translated to $\tC{ij}$, which will be approximated (with no discernible error) by $\gC{ij}$, the closest grid point in $\Omega$ (see Fig.~\ref{fig:spheres} and Sec.~\ref{subsubsec:impl_commphi}).
Since these fixed-size proto-subdomains will be used for all $\braket{ij}$ pairs, their radii should be chosen such that $\rpe = \max_{ij} \{ \rpe^{ij} \}$ and $\rme = \max_{ij} \{ \rme^{ij} \}$.
With judicious choices for $\rpe$ and $\rme$ (see Secs.~\ref{sec:ecr}--\ref{sec:pcr}), these proto-subdomains allow for an accurate evaluation of all EXX-related quantities, and have several algorithmic advantages that will be described below.

With $\rpe$ and $\rme$ in hand, we now describe the construction of these proto-subdomains around $\gC{0}$, an arbitrary center that is coincident with a grid point in $\Omega$.
In this work, the grid point closest to the center of $\Omega$ was chosen as the reference $\gC{0}$, since this allows us to avoid the use of both the minimum image convention and wrap-around (periodic) boundary conditions during grid point screening.
Assembly of the proto-subdomains begins by looping over grid points, $\bm r \in \Omega$, and determining whether or not a given grid point is contained within $\pe{\gC{0}}$ or $\me{\gC{0}} \setminus \pe{\gC{0}}$.
For each grid point contained in either proto-subdomain, we increment the corresponding counter ($q'$ or $q''$) and store its relative (\textit{local}) Cartesian coordinates (in $\overline{\bm r}_{\rm PE}$ or $\overline{\bm r}_{\rm ME}$) and (\textit{global}) grid point indices (in $\bm g^0_{\rm PE}$ or $\bm g^0_{\rm ME}$), as depicted in Algorithm~\ref{alg:subdomain}.
\begin{algorithm}[H]
  \begin{algorithmic}
    \State $q' \gets 0$; $q'' \gets 0$
    \ForAll{$\bm r\in \Omega$}
    \If {$| \bm r - \bm  C_{0} | \le R_{\rm PE}$}
    \State $q' \gets q' + 1$
    \State $\overline{\bm r}_{\rm PE}[q'] \gets \bm r - \bm C_{0}$
    \State $\bm g^0_{\rm PE}[q'] \gets \texttt{NINT} \left[ N_{{\rm grid},a}  r_a / |\bm L_a|  \right], \quad a = 1,2,3$
    \ElsIf {$ R_{\rm PE} < | \bm r - \bm  C_{0} | \le R_{\rm ME}$}
    \State $q'' \gets q'' + 1$
    \State $\overline{\bm r}_{\rm ME}[q''] \gets \bm r - \gC{0}$
    \State $\bm g^0_{\rm ME}[q''] \gets \texttt{NINT} \left[ N_{{\rm grid},a}  r_a / |\bm L_a| \right], \quad a = 1,2,3$
  \EndIf
\EndFor
\State $N_{\rm PE} \gets q'$
\State $N_{\rm ME} \gets q'+q''$
\caption{Proto-Subdomain Construction}
\label{alg:subdomain}
  \end{algorithmic}
\end{algorithm}

Incrementing these counters throughout the loop over $\bm r \in \Omega$ yields $N_{\rm PE}$ and $N_{\rm ME}$, the (fixed) number of points in $\pe{\gC{0}}$ and $\me{\gC{0}}$.
By storing the relative Cartesian coordinates, $\bm r - \gC{0}$, we now have a set of \textit{local} coordinates that are invariant to rigid translations of $\pe{\gC{0}}$ and $\me{\gC{0}}$, thereby avoiding the need to recompute these coordinates for every $\braket{ij}$ pair.
This also provides a convenient platform for precomputing a number of quantities (\eg $\bm r$ in spherical polar coordinates, the set of spherical harmonics, etc.) that are required during the ME of $\trho{ij}$ (\textit{cf}. Eqs.~\eqref{eq:me}--\eqref{eq:mepole}).
For each point in the proto-subdomains, we also store its \textit{global} grid point indices, which are given by three integer values, $(g_1^0, g_2^0, g_3^0)$, representing the position of a given grid point along the cell (lattice) vectors, $\bm L_1$, $\bm L_2$, and $\bm L_3$ (which are aligned with the Cartesian directions for the orthorhombic cells considered in this work).
For an orthogonal grid, which has $N_{{\rm grid},a}$ equispaced grid points along each of the $\bm L_a$ lattice vectors (with grid spacing $\delta\xi_{a}=|\bm L_a|/N_{{\rm grid},a}$), the global grid index along $\bm L_a$ is given by $g^0_a = r_{a} / \delta\xi_{a} = N_{{\rm grid},a} r_{a}/|\bm L_a|$.
Since $\bm r$ is always coincident with a grid point in $\Omega$, $\{ g^0_a \}$ is formally an array of integers; this is enforced in a floating-point environment using the nearest integer function, $\texttt{NINT}$.

This accumulated data is then concatenated to form two $3 \times N_{\rm ME}$ arrays as follows: the local coordinates are stored in a double-precision array,
\begin{align}
  \hspace{-0.075in}\overline{\bm r}[q] &= \left.
  \begin{cases}
    \overline{\bm r}_{\rm PE}[q] & q = 1,\ldots,N_{\rm PE} \\
    \overline{\bm r}_{\rm ME}[q-N_{\rm PE}] & q = N_{\rm PE}+1,\ldots,N_{\rm ME}
  \end{cases}
  \right\} ,
  \label{eq:rbar}
\end{align}
while the global grid indices are stored in an integer array,
\begin{align}
  \hspace{-0.09in}\bm g^0[q] &= \hspace{-0.01in}\left.
  \begin{cases}
    \bm g^0_{\rm PE}[q] & q = 1,\ldots,N_{\rm PE} \\
    \bm g^0_{\rm ME}[q-N_{\rm PE}] & q = N_{\rm PE}+1,\ldots,N_{\rm ME}
  \end{cases}
  \right\} .
  \label{eq:g0}
\end{align}
By storing all of the subdomain data in this scheme, only a single local index, $q$, is required for labeling the elements in these arrays.
This still maintains access to the $\pe{\gC{0}}$ and $\me{\gC{0}}$ proto-subdomains (as well as the $\me{\gC{0}}\setminus\pe{\gC{0}}$ shell) through knowledge of $N_{\rm PE}$ and $N_{\rm ME}$, the number of elements in each proto-subdomain.
As such, this scheme provides us with a compact representation for the sparse quantities required in our EXX algorithm, as well as a convenient mapping between data stored in the proto-subdomain representation and the real-space grid ($\Omega$) representation.
This is crucial for loading and off-loading data to and from $\Omega$, as it only requires communication of the relevant sectors of $\Omega$ for sparse quantities like $\trho{ij}$ and $\tv{ij}$.

\subsubsection{Step III: Communication of MLWFs \label{subsubsec:impl_commphi}}

By virtue of the \texttt{ORBITAL} data distribution scheme, the unique MLWF-pair list, $\lu$, not only determines the computational workload associated with each \mpi{} process, but also encodes the communication protocol that will be followed throughout the remainder of the \exxm module (see Fig.~\ref{fig:pair_list}).
With a support that is significantly smaller than $\Omega$, the communication of any given MLWF on the entire real-space grid is clearly neither efficient nor necessary in our EXX algorithm.
To reduce the communication overhead associated with each overlapping $\braket{ij}$ pair, the \exxm module employs the proto-subdomains ($\pe{\gC{0}}$ and $\me{\gC{0}}$) introduced in Sec.~\ref{subsubsec:impl_spdomain}.
As discussed above, these system-size-independent proto-subdomains provide a compact data representation for the storage and communication of sparse quantities such as $\tphin{i}$, $\trhon{ij}$, $\tvn{ij}$, and $\dxxn{ij}$ (or $\dxxn{ji}$).
To utilize $\pe{\gC{0}}$ and $\me{\gC{0}}$ in practice, these proto-subdomins must be translated across $\Omega$ to form the subdomains, $\pe{\gC{ij}}$ and $\me{\gC{ij}}$, required for evaluating all quantities associated with a given $\braket{ij}$ pair, as shown in Fig.~\ref{fig:spheres}.

Before describing the translation of these proto-subdomains, we now discuss the employed convention used for $\tC{ij}$, and remind the reader of the flexibility one has in defining this quantity for neutral charge distributions like $\trho{ij}$ (see Sec.~\ref{sec:RealSpaceEXX}).
In the \exxm module, $\tC{ij}$ is defined as the midpoint between the $i$-th and $j$-th MLWF centers, \ie $\tC{ij}=(\tC{i} + \tC{j})/2$, which represents an excellent approximation to the aforementioned gauge used in molecular quantum mechanics and an algorithmically convenient choice.
By utilizing the MLWF centers, this definition for $\tC{ij}$ accounts for the spatial distribution of each MLWF through its first moment, and becomes equivalent to the conventional definition, $\tC{ij} = \int \dd\bm{r} \, \bm{r} |\trho{ij}| \,/ \int \dd\bm{r} \, |\trho{ij}|$, for a number of different symmetric cases (\eg when both $\tphi{i}$ and $\tphi{j}$ have the same spread and are spherically symmetric with respect to $\tC{i}$ and $\tC{j}$, when $\trho{ij}$ is centrosymmetric with respect to the midpoint, etc.).
Furthermore, this choice for $\tC{ij}$ recovers the correct center of charge for $\trho{ii}$, \ie $\tC{ij} \rightarrow \tC{ii} = \tC{i}$.
Algorithmically speaking, this convention for $\tC{ij}$ is also quite useful, as it only requires knowledge of the MLWF centers, which are available throughout a CPMD simulation.

As mentioned above, the \exxm module employs one additional simplification when dealing with MLWF and MLWF-pair centers: these quantities are approximated by the closest grid points in $\Omega$ and denoted throughout by either $\gC{i}$ or $\gC{ij}$.
This algorithmic simplification leads to no appreciable error during evaluation of $\exx$ and $\{ \dxxn{i} \}$, and allows us to rigidly translate the proto-subdomains to the appropriate center, $\gC{ij}$, as needed.
For an orthogonal grid, the component of the required grid translation vector, $\bm \tau^{ij}$, along a given lattice vector, $\bm L_a$, is given by:
\begin{align}
  \tau_{a}^{ij} &= \texttt{NINT} \left[ N_{{\rm grid},a}  \left( \gC{ij} - \gC{0} \right)_{a}/|\bm L_a|  \right] \nonumber \\
  &= \texttt{NINT} \left[ \frac{  \left( \gC{ij} - \gC{0}\right)_{a}}{\delta\xi_{a}} \right] .
\end{align}
Application of $\bm \tau^{ij}$ to a given proto-subdomain leaves the radius and local Cartesian coordinates untouched, and simply offsets the global grid indices as follows:
\begin{align}
  g^{ij}_{a}[q] = \texttt{MOD} \left[ g^{0}_{a}[q] + \tau^{ij}_{a}, \, N_{{\rm grid},a} \right] ,
  \label{eq:l2g}
\end{align}
thereby resulting in a subdomain centered at $\gC{ij}$.
The use of the remainder function, \texttt{MOD}, in Eq.~\eqref{eq:l2g} enforces the appropriate wrap-around boundary conditions; as such, this equation is specific to the grid convention used in \texttt{QE}, in which the grid points (along $\bm L_a$) are numbered from $0,1,\ldots,N_{{\rm grid},a}-1$.
For codes that number these grid points from $1,2,\ldots,N_{{\rm grid},a}$, Eq.~\eqref{eq:l2g} should be modified as follows: $g^{ij}_{a}[q] = \texttt{MOD} [\,g^{0}_{a}[q] + \tau^{ij}_{a} - 1, \, N_{{\rm grid},a}\,] + 1$.

With each MLWF stored according to the \texttt{ORBITAL} data distribution scheme, the \mpi{} process ($\zeta=1$) or processes ($\zeta>1$) that are currently storing $\tphi{i}$ on $\Omega$ are responsible for sending this MLWF to another \mpi{} process (or processes) according to the computation and communication protocol outlined by $\lu$.
In order to do so, $\tphi{i}$ is off-loaded onto the appropriately translated subdomains, $\me{\gC{ij}}$, corresponding to the overlapping $\braket{ij}$ pairs that will be handled remotely (\ie on other processes); all of the information required to do so is provided by local access to $\lu$ and $\{ \gC{ij} \}$, as both of these arrays have been broadcast to all processes.
For each of these overlapping $\braket{ij}$ pairs, a sparse quantity like $\tphi{i}$ is now stored in the more compact $\me{\gC{ij}}$ representation \via the use of three relatively small arrays: $\overline{\bm r}$, $\bm g^{ij}$, and $\tphib{i} \equiv \widetilde{\phi}_{i}(\overline{\bm r}+\gC{ij})$, with associated sizes (types) of $3 \times \nme$ (double-precision), $3 \times \nme$ (integer), and $1 \times \nme$ (double-precision), respectively.
Here, we remind the reader that all of the data on the $\pe{\gC{ij}}$ subdomain and $\me{\gC{ij}} \setminus\pe{\gC{ij}}$ shell are contained within $\me{\gC{ij}}$, and can easily be accessed using the local grid indexing scheme outlined in Eqs.~\eqref{eq:rbar}--\eqref{eq:g0}.
As mentioned above, the local Cartesian coordinates stored in the $\overline{\bm r}$ array are invariant to translations of the proto-subdomains; as such, this information does not need to be recomputed for each translated subdomain and can be broadcast across all processes.
Communication of the MLWFs on these compact subdomains then proceeds according to $\lu$ among the pool of available \mpi{} processes.
Once $\tphib{i}$ is received by a given process, $\trhob{ij}$ is assembled on the $\pe{\gC{ij}}$ subdomain and the \exxm module begins the process of solving the corresponding PE.

\subsubsection{Step IV: Solution of Poisson's Equation \label{sec:impl_PS}}

Based on $\lu$, each \mpi{} process, $P_{i}$, now holds an assigned MLWF-product density $\trhob{ij}$ as well as the relevant quantities that map the $\pe{\gC{ij}}$ and $\me{\gC{ij}}$ subdomains onto $\Omega$ (\ie $\npe$, $\nme$, $\{ \rbar \}$, and $\{ \bm g^{ij} \}$).
As such, $P_{i}$ has all of the required information to compute $\tvb{ij}$ on the $\pe{\gC{ij}}$ and $\me{\gC{ij}} \setminus \pe{\gC{ij}}$ subdomains.
On $\pe{\gC{ij}}$, $\tvb{ij}$ is obtained \via the solution of the PE in Eq.~\eqref{eq:pe}.
On $\me{\gC{ij}}\setminus\pe{\gC{ij}}$, $\tvb{ij}$ is obtained \via the ME in Eqs.~\eqref{eq:me}--\eqref{eq:mepole}, which provides the appropriate boundary conditions for the PE as well as the far-field potential.

While the ME of $\trhob{ij}$ (about $\gC{ij}$) can be straightforwardly computed using the local coordinates,  $\{ \rbar \}$, the PE requires a discrete representation of the Laplacian operator for computing numerical second derivatives on these subdomains.
Since the subdomains employed in the \exxm module are coincident with the underlying real-space grid (taken here to be orthogonal), the Laplacian operator can be expressed as a sum of second partial derivatives along each of the lattice directions, $\nabla^2 = \sum_a \frac{\partial^2 }{\partial \xi_a^2}$, in which $\xi_a$ is a coordinate of $\widehat{\bm L}_{a}\equiv \bm L_a / | \bm L_a |$.
At a given grid point, $\bm \xi_{0}$, the second partial derivative of a function, $f(\bm{\xi})$, along $\bm L_a$ was discretized \via the standard central-difference approach~\cite{fornberg_generation_1988}:
\begin{align}
  \left.\frac{\partial^2 f(\bm \xi)}{\partial \xi_a^2} \right|_{\bm \xi = \bm \xi_{0}} = \sum_{q=-n}^{n} w_{q}\frac{  f(\bm \xi_{0} + q \, \delta \xi_a \widehat{\bm L}_{a})}{ \delta \xi_a^2} .
  \label{eq:d2_1}
\end{align}
In this expression, the associated discretization error depends on the number, $n$, of (equispaced) neighboring grid points on each side of $\bm{\xi}_0$, and $w_{q}$ is the central-difference coefficient for the $q$-th grid point.
We note in passing that only $w_{|q|}$ is required due to the central symmetry ($w_{q}=w_{-q}$) of the equispaced finite-difference stencil.
Discretization of this derivative results in a ($2n+1$)-point stencil along each grid direction, $\bm{L}_{a}$; as such, the discrete 3D Laplacian operator corresponds to a finite-difference stencil covering $3 \times 2n+1=6n+1$ grid points.
We note in passing that the choice of $n=3$ (with corresponding central-difference coefficients~\cite{fornberg_generation_1988} given by $w_{0} = -49/18$, $w_{1} = +3/2 = w_{-1}$, $w_{2} = -3/20 = w_{-2}$, and $w_{3} = +1/90 = w_{-3}$) yields a second derivative with 
an associated discretization error of $\mathcal{O}\left( \delta\xi^6 \right)$; this choice is the default option in the \exxm module as it yields a well-converged value for $\exx$~\cite{wu_order-n_2009,distasio_jr._individual_2014}.

With this discrete representation of the Laplacian, we can express the PE in Eq.~\eqref{eq:pe}, $\nabla^2 \tvb{ij} = -4\pi \trhob{ij}$, as the following set of sparse linear equations on the $\pe{\gC{ij}}$ subdomain:
\begin{equation}
  \bm\nabla^2_{\rm {PE}} \tvnb{ij} = -4\pi \left( \trhonb{ij}-\terhonb{ij} \right) .
  \label{eq:pe_mat_red}
\end{equation}
In this expression, $\bm{\nabla}^2_{\rm {PE}}$ is a sparse $\npe\times\npe$ matrix containing the discretized Laplacian (whose stencil coverage has been restricted to $\pe{\gC{ij}}$), $\tvnb{ij}$ is a $\npe \times 1$ vector representing the (currently unknown) MLWF-product potential, and $\trhonb{ij}$ is a $\npe \times 1$ vector containing the MLWF-product density.
The final term on the right-hand side, $\terhonb{ij} \equiv -(1/4\pi) (\bm{\nabla}^2 - \bm{\nabla}^2_{\rm PE}) \tvnb{ij}$, is the so-called boundary charge, which accounts for the part(s) of the Laplacian stencil that extend outside of $\pe{\gC{ij}}$ (and into the $\me{\gC{ij}}\setminus \pe{\gC{ij}}$ shell) and have been truncated in the $\bm \nabla^2_{\rm PE}$ representation of this operator.
In doing so, $\terhonb{ij}$ accounts for the Dirichlet boundary conditions provided by the ME form of the potential on the $\me{\gC{ij}} \setminus \pe{\gC{ij}}$ shell surrounding $\pe{\gC{ij}}$ (see Eq.~\eqref{eq:me}), and therefore allows for a numerically exact solution of the PE using $\bm\nabla^2_{\rm {PE}}$, a Laplacian whose stencil coverage has been restricted to the $\pe{\gC{ij}}$ subdomain.
This restricts the PE to the support of $\trhon{ij}$ and substantially reduces the dimensionality and associated computational cost of solving the PE for each overlapping MLWF pair.

The system of sparse linear equations in Eq.~\eqref{eq:pe_mat_red} is then solved (for $\tvnb{ij}$) using an iterative conjugate-gradient (CG) approach.
Since the solution of the PE is notoriously difficult to parallelize efficiently over \mpi{} tasks, the CG-based PE solver in the \texttt{exx} module is largely parallelized over \omp{} threads to allow for an efficient real-space evaluation of $\tvnb{ij}$.
The efficient solution of the PE for each overlapping MLWF pair is the cornerstone of our MLWF-based EXX algorithm, and the performance of the CG-based PE solver will be discussed in Sec.~\ref{subsubsec:f-g_para}.
During CPMD simulations, the number of CG iterations required to solve a given PE can be substantially reduced by using a polynomial extrapolation~\cite{press_numerical_1992} of the potential from the previous CPMD steps as the initial guess.
More detailed considerations of this extrapolation scheme as well as extensions to BOMD will be discussed in future work.

\subsubsection{Step V: Computation of Energy and Forces \label{subsubsec:impl_compute_esf}}

After a process, $P_{i}$, arrives at the solution to the PE for one of its assigned pairs (\ie for a given $j \in \lu[i]$), this process now holds the corresponding MLWF-product potential $\tvb{ij}$ on the entire $\me{\gC{ij}}$ subdomain.
With $\tvb{ij}$ on the $\pe{\gC{ij}}$ subdomain (\via the solution of the PE) and $\tvb{ij}$ on the $\me{\gC{ij}} \setminus \pe{\gC{ij}}$ shell (\via the ME of $\trhob{ij}$), $P_{i}$ is now ready to evaluate the $\braket{ij}$ contribution to the EXX energy ($\exx$) and wavefunction forces ($\dxxb{ij}$ and $\dxxb{ji}$).

The evaluation of $\exx$ is quite straightforward \via Eq.~\eqref{eq:exx_omega}.
Here, we remind the reader that a numerically exact evaluation of this quantity only requires integration on the $\pe{\gC{ij}}$ subdomain; this integration over the support of $\trho{ij}$ is also parallelized over \omp{} threads and is therefore quite computationally efficient.
Partial summations over the assigned $\braket{ij}$ pairs on each process are accumulated to form $\exx$ with minimal associated communication (\ie one double-precision number for $\exx$ per \mpi{} process).

With $\tvb{ij}$ in hand, $P_{i}$ is also in position to compute both $\dxxb{ij} = \tvb{ij} \tphib{j}$ and $\dxxb{ji} = \tvb{ij} \tphib{i}$ on the entire $\me{\gC{ij}}$ subdomain.
For each $\dxxb{ji}$ computed on $P_{i}$, this quantity is shipped back to $P_{j}$ (assuming $\zeta = 1$ here for simplicity), which requires communication of $\nme$ double-precision numbers for each $\dxxb{ji}$; this array is equivalent in size to $\tphib{j}$ and represents the necessary second communication event described in Sec.~\ref{subsubsec:impl_spdomain}.
Since $P_{i}$ has access to $\tvb{ij} \ \forall \ j\in\lu[i]$, this process accumulates $\dxxb{i} = \sum_{j} \dxxb{ij}$ into a local temporary array that is the size of the global real-space grid; as $P_{i}$ receives a given $\dxxb{ij}$ array from $P_{j}$, this quantity is also accumulated into this local temporary array.
When all $\dxxb{ij}$ contributions are accounted for, this temporary array on $P_{i}$ holds the final $\dxx{i}$ according to the \texttt{ORBITAL} data distribution scheme.

\subsubsection{Step VI: Redistribution of Wavefunction Forces \label{subsubsec:impl_redist_d}}

At this stage, all of the EXX-related quantities have been evaluated; $\exx$ has been accumulated and broadcast to all processes, while $\{\dxx{i}\}$ is stored in the \texttt{ORBITAL} data distribution scheme.
As such, the remaining task for the \exxm module is the transformation of $\{\dxx{i}\}$ to the \texttt{GRID} data distribution scheme for compliance with the core functions in \texttt{QE} (see Sec.~\ref{subsec:impl_distribution}).
This redistribution is essentially the reverse operation of the \texttt{GRID} $\rightarrow$ \texttt{ORBITAL} redistribution of the MLWFs described in Fig.~\ref{fig:redist} and  Sec.~\ref{subsubsec:impl_redist_phi}.

At this stage, the \texttt{QE} executable exits the \exxm module with $\exx$ and $\{\dxx{i}\}$ (in the \texttt{GRID} data distribution scheme) as output.
These EXX-related quantities are then added to their semi-local exchange analogs with the appropriate EXX fraction, $a_x$, given in Eq.~\eqref{eq:hyb}.
With the EXX contribution to the wavefunction (MLWF) forces, the CPMD equations of motion are now propagated forward \via Eqs.~\eqref{eq:cpE}--\eqref{eq:cpI}.

\section{Accuracy and Performance \label{sec:Perf}}

During the implementation of the \exxm module, we have introduced three parameters: $\rpr$, $\rpe$, and $\rme$ (see Sec.~\ref{subsubsec:impl_spdomain}).
$\rpr$ is used to determine whether or not two MLWFs, $\tphi{i}$ and $\tphi{j}$, are considered to be an overlapping $\braket{ij}$ pair \via $| \tC{i} - \tC{j} | < \rpr$.
For all overlapping $\braket{ij}$ pairs, $\rpe$ and $\rme$ are the radii of the spherical $\pe{\gC{ij}}$ and $\me{\gC{ij}}$ subdomains, which are centered at $\gC{ij}$ and chosen to cover the product density, $\trho{ij}$, and individual orbitals, $\tphi{i}$ and $\tphi{j}$, respectively.
In order to efficiently perform large-scale hybrid DFT based AIMD simulations, judicious choices for $\rpr$, $\rpe$, and $\rme$ are required to balance the performance and accuracy, and are therefore the focus of this section.
In Sec.~\ref{sec:Perf_cutoff}, we introduce a systematic selection of these parameters based on user-defined error thresholds for $\exx$ and $\{\dxx{i}\}$.
In Sec.~\ref{sec:Perf_scaling}, we discuss the intranode (\omp{}) and internode (\mpi{}) parallel scaling performance when simulating liquid water using our MLWF-based EXX approach on several different supercomputer architectures.

\subsection{Parameters and Convergence Criteria \label{sec:Perf_cutoff}}

In this section, a systematic determination of all required parameters will be demonstrated using a snapshot from a liquid water simulation containing \ce{(H2O)64} in a cubic cell with $L = 23.52$~Bohr.
We begin by performing a reference single-point energy evaluation in \texttt{QE} at the PBE0 level with a planewave kinetic energy cutoff of $85$~Ry.
In this reference calculation (which yields $\exxr$), we use the largest possible values for all three parameters such that:
(\textit{i}) all radially overlapping MLWF pairs are included with $| \tC{i} - \tC{j} | < \rpr = L/2 = 11.76$~Bohr and (\textit{ii}) both proto-subdomains ($\pe{\gC{0}}$ and $\me{\gC{0}}$) are contained within the simulation cell (\ie $\rme=L/2=11.76$~Bohr and $\rpe=L/2-n \max_{a}\{\delta\xi_{a}\} = 11.11$~Bohr).
Subtraction of $n \max_{a}\{\delta\xi_{a}\}$ (with $n=3$) from $\rpe$ allows us to retain a thin shell of the real-space grid surrounding $\pe{\gC{0}}$; this shell provides the boundary conditions for the PE and accommodates the part(s) of the discretized Laplacian stencil that extend beyond $\pe{\gC{0}}$.
We note in passing that the energetic contributions to $\exxr$ from MLWF pairs with $| \tC{i} - \tC{j} | > \rpr = L/2 = 11.76$~Bohr (within the minimum image convention) are completely negligible (\ie $2.0 \times 10^{-5}$~kcal/mol or $\approx 4.8\times 10^{-6}\%$).

As such, this reference calculation provides an approximately exact evaluation of $\exx$ using the \exxm module, albeit at an excessive computational cost.
More specifically, the computational cost associated with the CG solution of the PE in \exxm scales as $\mathcal{O}(\npe^{4/3})$, with $\npe$ asymptotically growing as $\mathcal{O}(\rpe^3)$.
In the same breath, the associated communication and memory footprint in \exxm scale as $\mathcal{O}(\nme)$, with $\nme$ asymptotically growing as $\mathcal{O}(\rme^3)$.
In addition, both computation and communication in \exxm scale as $\mathcal{O}(N_{\rm pair})$, with $N_{\rm pair}$ asymptotically growing as $\mathcal{O}(\rpr^3)$ (for homogeneous systems with constant densities).
As such, judicious choices for $\rpr$, $\rpe$, and $\rme$ are required to balance the performance and accuracy of our algorithm (see Secs.~\ref{sec:ecr}--\ref{sec:pcr}).

Since the number of self pairs (with $i=j$) is $N_o$ and the number of non-self pairs (with $i \neq j$) is $(\widetilde{n}-1)N_o$ (see Sec.~\ref{sec:RealSpaceEXX}), the computational cost associated with evaluating the contribution to $\exx$ from non-self pairs will dominate that from self pairs when performing MLWF-based EXX calculations.
However, the energetic contribution to $\exx$ from the self pairs tends to dominate the contribution from non-self pairs, which is primarily due to the fact that the overlap between an MLWF and itself is substantially larger than the overlap between an MLWF and any one of its neighbors.
In the reference calculation described above, for example, the energetic contribution to $\exxr$ is dominated ($\approx 85\%$) by the self pairs while the computational cost to evaluate $\exxr$ ($\approx 87\%$) mainly originates from the non-self pairs.
Taken together, these observations suggest that we can further balance the performance and accuracy of our algorithm by employing a larger value of $\rpe$ for the self pairs ($\rpes$, which defines the $\pes{\gC{0}}$ proto-subdomain) and a smaller value of $\rpe$ for the non-self pairs ($\rpep$, which defines the $\pep{\gC{0}}$ proto-subdomain).
Since $\rpes$ will in general be larger than $\rpep$, we will employ the same strategy by using a larger value of $\rme$ for the self pairs ($\rmes$) and a smaller value of $\rme$ for the non-self pairs ($\rmep$), which define the $\mes{\gC{0}}$ and $\mep{\gC{0}}$ proto-subdomains, respectively.
Physically speaking, the choice to use larger (smaller) $\rpe$ and $\rme$ values for the self (non-self) pairs is also justified by the fact that: (\textit{i}) self MLWF-product densities ($\trho{ii}$) have a larger support than non-self MLWF-product densities ($\trho{ij}$) due to the increased overlap between an MLWF and itself, and (\textit{ii}) self MLWF-product potentials ($\tv{ii}$) are generally longer-ranged than non-self MLWF-product potentials ($\tv{ij}$) due to the absence of a monopolar contribution in the non-self cases (see Sec.~\ref{sec:RealSpaceEXX}).

\subsubsection{Convergence of the EXX Contribution to the Energy \label{sec:ecr}}

\begin{figure}[t!]
  \includegraphics[width=\linewidth]{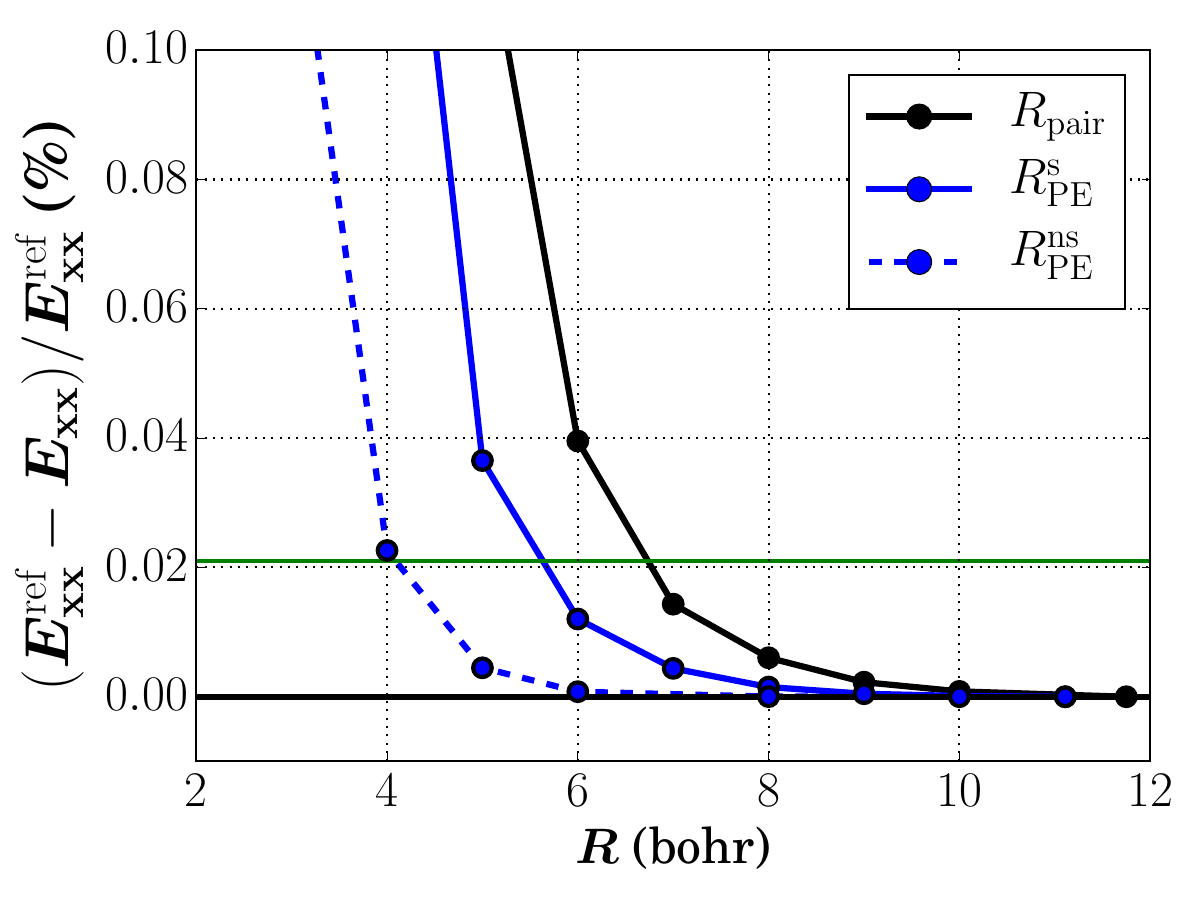}
  \caption{
  Convergence of $\exx$ as a function of $\rpr$, $\rpes$, and $\rpep$ on a snapshot of liquid water containing \ce{(H2O)64} in a cubic cell with $L = 23.52$~Bohr.
  Relative errors (in $\%$) with respect to $\exxr$ are evaluated individually by varying $\rpr$ (solid black line), $\rpes$ (solid blue line), and $\rpep$ (dashed blue line), while keeping all other parameters set to their maximum allowed values (see text for more details).
  An overall relative error of $\approx 0.02\%$ corresponds to the default parameter values in \texttt{QE} (\ie $\rpr = 8.0$~Bohr, $\rpes = 6.0$~Bohr, $\rpep = 5.0$~Bohr), and is depicted by the solid green line.
  }
  \label{fig:conv_exx}
\end{figure}

As an initial test, $\exx$ will be computed using the converged MLWFs obtained during the calculation of $\exxr$.
Since the effects of self-consistency are neglected in this test (and will be investigated below), we can quickly assess the convergence of $\exx$ with respect to $\exxr$ as a function of $\rpr$, $\rpes$, and $\rpep$ (without the need for modifying $\rmes$ and $\rmep$).
We do so by independently varying each of the $\rpr$, $\rpes$, and $\rpep$ parameters, while keeping all other parameters fixed at their largest possible values (see Fig.~\ref{fig:conv_exx}).
From this figure, one can immediately see that the relative (percent) error in $\exx$ rapidly decays as each of these parameter values is increased.
This observation can be justified by considering the fact that each MLWF (in finite-gap systems) is exponentially localized, and hence products of MLWFs (\ie $\trho{ii}$ or $\trho{ij}$) are also exponentially localized.
As such, increasing $\rpes$ (or $\rpep$) leads to spherical PE domains that increasingly cover the exponentially decaying tails of $\trho{ii}$ (or $\trho{ij}$); as seen in Eq.~\eqref{eq:exx_omega} (and the surrounding discussion), this results in rapid convergence to the reference value for $\exx$. 
Since $\exx$ converges more quickly with $\rpep$ (than $\rpes$), this finding confirms our physical intuition that $\rpes > \rpep$, and further justifies our use of separate self and non-self proto-subdomains as a means to improving the balance between performance and accuracy in this algorithm.
By increasing $\rpr$, the incremental contribution to $\exx$ from (more) distant MLWF pairs becomes negligible as $\trho{ij} \rightarrow 0 \, \forall \, \bm r$, which also results in rapid convergence to the reference value for $\exx$.

Based on this initial convergence test, we have chosen $\rpr = 8.0$~Bohr, $\rpes = 6.0$~Bohr, and $\rpep = 5.0$~Bohr as the default parameter set in \texttt{QE}; for EXX-based simulations of liquid water, the overall relative error is $\approx 0.02\%$, which is essentially additive (in each of these parameters) and typically rather stringent when obtaining ground-state energies, binding/cohesive energetics, and ionic forces in the condensed phase (\textit{vide infra}).
To increase the convergence of $\exx$, a general rule of thumb is to first increase $\rpes$, since the self-pair contribution is the dominant contribution to $\exx$, yet requires evaluation of less terms (and is therefore significantly cheaper to compute) than the contribution from non-self pairs.
In this example, one can further reduce the relative error in $\exx$ by an additional factor of two (\ie to $\approx 0.01\%$) by simply increasing $\rpes$ from $6.0$~Bohr to $7.0$~Bohr, with negligible ($< 1\%$) additional computational cost.

In this convergence test, we have neglected the effects of orbital self-consistency when determining $\exx$.
To quantify this effect, we first performed a fully self-consistent EXX calculation using the default parameter values for $\rpr$, $\rpes$, and $\rpep$ determined above (while keeping $\rmes$ and $\rmep$ set to the maximum reference value of $L/2 = 11.76$~Bohr).
In doing so, we found that the inclusion of orbital self-consistency leads to a negligible ($<0.01\%$) variation in $\exx$, which indicates that: (\textit{i}) there is excellent agreement between the PE and ME evaluations of the far-field (beyond $\rpes$ and $\rpep$) contribution to the wavefunction forces ($\dxx{i}$), and (\textit{ii}) the default value of $\rpr$ is sufficient to capture all relevant overlapping MLWF pairs in this system. 
In a non-self-consistent calculation, the ME only provides the boundary conditions for the PE, and therefore has no direct effect on $\exx$, provided that $\rmes$ ($\rmep$) is larger than $\rpes$ ($\rpep$) by the extent of the Laplacian stencil (\ie $n \max_{a}\{\delta\xi_{a}\}$, see Sec.~\ref{sec:Perf_cutoff}).
In a self-consistent calculation, however, $\rmes$ and $\rmep$ govern the accuracy and cost of obtaining $\exx$ \via the sparse evaluation of $\{\dxx{i}\}$ (see Eq.~\eqref{eq:Dxx_mlwf}, Eq.~\eqref{eq:Dxx_mlwf_omega}, and the surrounding discussion); in other words, larger values for 
$\rmes$ and $\rmep$ lead to more accurate $\exx$ values (\via the convergence of the orbitals which is self-consistently driven by $\{\dxx{i}\}$), although this is accompanied by a higher computational cost (as well as communication overhead and memory footprint) during the EXX calculation.
To quantify this effect (and determine the appropriate default values for $\rmes$ and $\rmep$), we now study the convergence of $\{\dxx{i}\}$ with respect to these parameters.

\subsubsection{Convergence of the EXX Contribution to the Wavefunction Forces \label{sec:pcr}}

\begin{figure}[t!]
  \includegraphics[width=\linewidth]{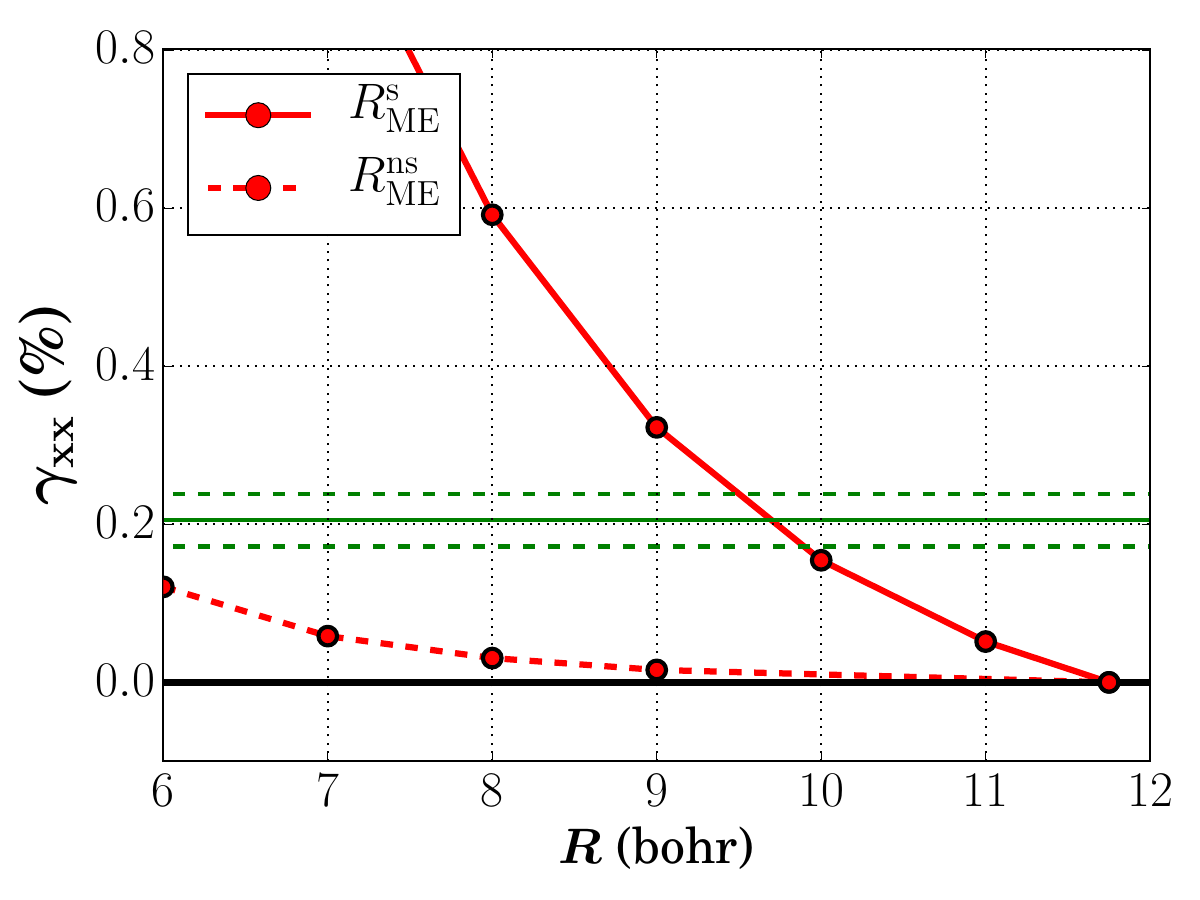}
  \caption{
  Convergence of $\{\dxx{i}\}$ (\via the $\gamma_{\rm xx}$ metric defined in Eqs.~\eqref{eq:gammaxx_i} and~\eqref{eq:gammaxx}) as a function of $\rmes$ and $\rmep$ on a snapshot of liquid water containing \ce{(H2O)64} in a cubic cell with $L = 23.52$~Bohr.
  Relative errors (in $\%$) with respect to $\{\dxxr{i}\}$ are evaluated by varying $\rmes$ (solid red line) and $\rmep$ (dashed red line), while keeping all other parameters ($\rpr$, $\rpes$, $\rpep$) set to their default values in \texttt{QE} (see text for more details).
  An overall relative error of $\approx 0.2\%$ (with a standard deviation of $\pm 0.03\%$) corresponds to the default parameter values in \texttt{QE} ($\rmes = 10.0$~Bohr, $\rmep = 7.0$~Bohr), and is depicted by the solid (dashed) green line.
  }
  \label{fig:conv_dxx}
\end{figure}

To study the convergence of $\{\dxx{i}\}$ and determine the default values for $\rmes$ and $\rmep$, use of the reference calculation performed above (in which all parameters were set to the largest possible values) is inconvenient as it lacks the flexibility to vary $\rmes$ and $\rmep$ (as these parameters must be larger than $\rpes$ and $\rpep$ to provide the boundary conditions for the PE).
Since the use of the default parameter values for $\rpr$, $\rpes$, and $\rpep$ (with $\rmes$ and $\rmep$ each set to the maximum reference value of $L/2 = 11.76$~Bohr) reproduces $\exxr$ with negligible error (\ie to within $0.02\%$), we will base our convergence tests on this as our new reference calculation.
As an initial test, $\{\dxx{i}\}$ will be computed using the converged MLWFs and $\{\dxxr{i}\}$ obtained during this \textit{new} reference calculation.
Since the effects of self-consistency are neglected in this test (and will be investigated below), we can quickly assess the convergence of $\{\dxx{i}\}$ with respect to $\{\dxxr{i}\}$  as a function of $\rmes$ and $\rmep$.
We do so by independently varying the $\rmes$ and $\rmep$ parameters, while keeping all other parameters ($\rpr$, $\rpes$, and $\rpep$) fixed at their default values (see Fig.~\ref{fig:conv_dxx}).
By employing this new reference calculation, $\rmes$ can now be varied from $7.65$~Bohr (\ie $\rpes + n \max_a \delta \xi_{a}$) to a maximum value of $L/2 = 11.76$~Bohr, and $\rmep$ can be varied from $5.65$~Bohr (\ie $\rpep + n \max_a \delta \xi_{a}$) to a maximum value of $L/2 = 11.76$~Bohr.
To quantify the error in $\{\dxx{i}\}$, we will utilize the relative $L^1$-norm for each $\dxx{i}$:
\begin{align}
  \gamma_{\rm xx}^{i} \equiv \frac{\int \dd \bm r \, \left| \dxxr{i} - \dxx{i} \right|}{\int \dd \bm r \, \left| \dxxr{i} \right|} ,
  \label{eq:gammaxx_i}
\end{align}
which can then be averaged over MLWFs to furnish the following convergence metric:
\begin{align}
  \gamma_{\rm xx} = \frac{1}{N_{o}}\sum_i \gamma_{\rm xx}^{i} .
  \label{eq:gammaxx}
\end{align}

From Fig.~\ref{fig:conv_dxx}, one can again see that the relative (percent) error in $\{\dxx{i}\}$ rapidly decays as each of these parameter values is increased.
Since $\gamma_{\rm xx}$ converges more quickly with $\rmep$ (than $\rmes$), this finding also confirms our physical intuition that $\rmes > \rmep$, and further justifies our use of separate self and non-self proto-subdomains as a means to improving the balance between performance and accuracy in this algorithm.
As discussed above in Secs.~\ref{sec:RealSpaceEXX} and~\ref{sec:Perf_cutoff}, this results from the fact that self MLWF-product potentials ($\tv{ii}$) are generally longer-ranged than non-self MLWF-product potentials ($\tv{ij}$) due to the absence of a monopolar contribution in the non-self cases.
Since the cost of our algorithm (\ie computation, communication, and memory) is dominated by the non-self contributions and scales cubically with $\rmep$, it is preferable to choose the smallest possible value for this parameter.
The plots depicted in Fig.~\ref{fig:conv_dxx} clearly suggest that $\rmep$ is converged for $R \gtrsim 7.0$~Bohr and (the noticeably slower) $\rmes$ only begins to plateau for $R \gtrsim 10.0$~Bohr.
When used in conjunction with the default parameter values for $\rpr$, $\rpes$, and $\rpep$ determined above, parameter values of $\rmes = 10.0$~Bohr and $\rmep = 7.0$~Bohr lead to an overall relative error of $\gamma_{\rm xx} \approx 0.2\%$ for this snapshot of liquid water.
To see how this $\gamma_{\rm xx} \approx 0.2\%$ error translates into the final value of $\exx$, we again performed a fully self-consistent EXX calculation on this \ce{(H2O)64} snapshot; in doing so, we found that the use of these parameter values leads to a completely negligible error ($\approx 10^{-7}$\%) in $\exx$.
We further note that a number of EXX-based CPMD simulations of solid and liquid aqueous systems have been performed by our group using these parameter values; in all cases, we have found that the appropriate constant of motion was reasonably maintained.
As such, we have set $\rmes = 10.0$~Bohr and $\rmep = 7.0$~Bohr (in addition to $\rpr=8.0$~Bohr, $\rpes=6.0$~Bohr, and $\rpep=5.0$~Bohr) as the default parameters used in \texttt{QE}.

As seen above for $\rpes$, one can further reduce $\gamma_{\rm xx}$ by an additional factor of two (\ie to $\approx 0.1\%$) by increasing $\rmes$ from $10.0$~Bohr to $11.0$~Bohr, with negligible ($< 1\%$) additional computational cost; for sufficiently large simulation cells, increasing $\rmes$ is therefore another efficient way to improve the accuracy of the EXX calculation.
Even with $\rmes = 11.0$~Bohr, however, one can still observe a finite slope in the tail of the $\rmes$ curve in Fig.~\ref{fig:conv_dxx}. This is an artifact of performing this convergence test on \ce{(H2O)64}, and a stricter convergence of $\gamma_{\rm xx}$ with $\rmes$ would be observed when using a larger simulation cell.
To estimate the effect of this artifact on $\gamma_{\rm xx}$, we performed an exponential fit (with $R^2 > 0.9999$) to the $\rmes$ curve, and found that the residual error in $\gamma_{\rm xx}$ (\ie that was caused by truncating $\rmes$ to $L/2 = 11.76$~Bohr) is $\approx 0.1\%$.
Based on the self-consistency test described above, we expect that this residual error in $\gamma_{\rm xx}$ would only lead to a negligible ($\approx 10^{-7}$\%) error in $\exx$ for our \ce{(H2O)64} test system; when treating similarly sized (or slightly smaller) systems, we recommend that users quantify this truncation error and potentially set $\rmes$ to the largest possible value (\ie $\rmes = L/2$).

\subsubsection{Transferability of the Default EXX Parameters \label{sec:transferability}}

To provide an alternative gauge for the $0.02$\% and $0.2$\% error thresholds in $\exx$ and $\{\dxx{i}\}$, we also considered the errors in the binding energy and ionic forces in this \ce{(H2O)64} snapshot by comparing the results obtained with the default and reference (\ie $\rpr=\rmes=\rmep=L/2=11.76$~Bohr and $\rpes=\rpep=11.11$~Bohr, see Sec.~\ref{sec:Perf_cutoff}) parameter sets.
For the binding energy (per \ce{H2O} molecule), we found an error of $0.04$~kcal/mol using the default parameters in \exxm, which is comparable to the typical pseudopotential error.
For the $3N$ ionic force components (with $N$ being the number of atoms), the default parameter set leads to a mean absolute error of $6.2\times 10^{-5}$~Ha/Bohr and a maximum absolute error of $2.5\times 10^{-4}$~Ha/Bohr, which is approximately half of the default convergence criteria used during structural/geometry optimizations in \texttt{QE}.
Taken together, all of these tests strongly indicate that the default parameter values in \exxm are more than adequate when performing EXX calculations on systems like liquid water.

We note in passing that our choice to treat self ($\braket{ii}$) and overlapping non-self ($\braket{ij}$) MLWF pairs differently using the $\pes{\gC{0}}$ and $\pep{\gC{0}}$ proto-subdomains is only the first step towards exploiting the concept of variable-size subdomains during MLWF-based EXX calculations.
By using a single set of $\rpes$ and $\rpep$ values, this choice is particularly well-suited for condensed-phase systems characterized by a narrow distribution of MLWF spreads (\eg liquid water, wherein each \ce{H2O} molecule has a set of four similarly localized MLWFs).
As such, we expect that the chosen default parameter values determined above for bulk liquid water will yield similar errors in $\exx$ and $\{\dxx{i}\}$ for systems with similarly large band gaps.
For systems with smaller gaps (and hence more diffuse MLWFs), one would need to use more stringent parameter values to obtain a similar level of accuracy; as such, a series of test calculations (in analogy to those above) should be run to determine the optimal $\rpes$ and $\rpep$ values prior to performing large-scale production AIMD simulations.
For systems with a wider distribution of MLWF spreads (due to a smaller band gap and/or a more heterogeneous environment), our current algorithm would be forced to sacrifice computational efficiency for accuracy (\textit{vide infra}), since $\rpes$ and $\rpep$ (as well as $\rmes$ and $\rmep$) would need to be large enough to provide sufficient cover for the most diffuse MLWF in the system (and would therefore be overkill for MLWFs with substantially smaller spreads).
As pointed out by Dawson and Gygi~\cite{dawson_performance_2015}, this issue is particularly important in small-gap heterogeneous condensed-phase systems, such as solvated semiconducting nanoparticles and water-semiconductor interfaces.

Although the current implementation of \exxm would sacrifice efficiency when applied to such challenging cases, we would still argue that the algorithmic framework of \exxm is general enough to perform accurate hybrid DFT based CPMD simulations of finite-gap systems.
While certain hacks can be used to ameliorate the efficiency degradation in these cases, we have chosen to focus on a comprehensive revision of our algorithm that will explicitly account for the MLWF-orbital ($\Omega_{i}$) and MLWF-product ($\Omega_{ij}$) domains introduced in Sec.~\ref{sec:RealSpaceEXX}.
Inspired by the work of Gygi and coworkers~\cite{gygi_compact_2009,gygi_efficient_2013,dawson_performance_2015}, we are currently working on a significantly more efficient (\textit{vide infra}) $\beta$-version of \exxm in which each MLWF will have a spread-dependent $\Omega_i$ and each overlapping $\braket{ij}$ pair will have a overlap-dependent $\Omega_{ij}$.
As such, MLWFs with larger spreads will automatically be treated more accurately without sacrificing computational efficiency for MLWFs with smaller spreads.
In addition, more distant MLWF pairs will have smaller MLWF-product domains by construction due to the overlap dependence in $\Omega_{ij} = \Omega_{i} \cap \Omega_{j}$.
In doing so, we expect that the $\beta$-version of \exxm will be a single EXX algorithm that is general enough to accurately and efficiently handle condensed-phase systems ranging from large-gap homogeneous systems like liquid water (with a narrow distribution of MLWF spreads) to small-gap heterogeneous systems like solvated semiconducting nanoparticles (with a wide distribution of MLWF spreads). 
When paired with an orbital localization scheme that can appropriately treat small-to-vanishing band gap systems~\cite{cornean_localised_2019,damle_variational_2018}, we expect that the $\beta$-version of this algorithm will also be an important step towards treating large-scale metallic systems with screened and/or range-separated exchange~\cite{heyd_hybrid_2003,gerber_hybrid_2005,vydrov_assessment_2006,janesko_screened_2009,baer_tuned_2010,kronik_excitation_2012,karolewski_using_2013}.

\subsubsection{Tight Convergence to the Electronic Ground State \label{sec:tight_conv}}

While \exxm is designed to perform efficient AIMD simulations, this module can also be adapted to evaluate precise ground state energetics (\eg to within an uncertainty of $\Delta E < 10^{-8}$~Ha), which are needed for property evaluations, numerical phonon calculations, etc.
In order to achieve tight convergence to the electronic ground state using \exxm, the default convergence criteria used during the CG solution to the PE (\texttt{exx\_poisson\_eps} = $10^{-6}$~au) and the nested SODD optimization of the Marzari-Vanderbilt functional (\texttt{tolw} = $10^{-8}$~au) must be tightened accordingly.
Doing so minimizes the noise in the force acting on $\{\tphi{i}\}$ ($\{\widetilde{D}_{i}(\bm r)\}$ in  Eq.~\eqref{eq:sodd_mlwf_eom_discrete_1}), and reduces oscillatory behavior in the energy profile during the SODD-based SCF procedure.

\subsection{Parallel Scaling and Performance \label{sec:Perf_scaling}}

As demonstrated above in Sec.~\ref{sec:Perf_cutoff}, judicious choices for the five parameter values in \exxm allow one to evaluate all EXX-related quantities with a high level of accuracy.
To enable large-scale EXX-based AIMD simulations using this approach, we employ a hybrid \mpi{}/\omp{} parallelization scheme that allows us to minimize the walltime cost (\ie time to solution) by exploiting both internode and intranode computational resources provided by massively parallel supercomputer architectures (see Sec.~\ref{sec:Implementation}).
Here, we remind the reader that our massively parallel implementation of \exxm seamlessly distributes the major computational workload across thousands of \mpi{} processes.
Within each \mpi{} process, the CG-based PE solver is further parallelized over \omp{} threads.
The scaling and performance of the internode (\mpi{}, first level) and intranode (\omp{}, second level) levels in our hybrid parallelization scheme were evaluated by performing large-scale simulations of liquid water with \exxm on the IBM Blue Gene/Q platform (\textit{Mira}), and will be discussed below in Secs.~\ref{subsubsec:c-g_para} and \ref{subsubsec:f-g_para}, respectively.
In Sec.~\ref{subsubsec:cori-performance}, the computational performance of \exxm will also be considered on the \textit{Cori} Haswell and KNL architectures.

\subsubsection{Internode Parallelization \via \mpi{} \label{subsubsec:c-g_para}}

For the first level of parallelization, the \exxm module employs internode \mpi{} communication to distribute the computational workload associated with a given EXX calculation across (many) thousands of compute nodes.
To critically assess the computational performance of this parallelization level, which is at the very heart of our massively parallel algorithm, we performed a strong-scaling analysis (\ie by varying the number of processing elements for a fixed problem size) and a weak-scaling analysis (\ie by varying the problem size for a fixed ratio of problem size to number of processing elements).
To investigate the strong and weak scaling of \exxm, we performed a series of $12$ different EXX-based CPMD simulations of liquid water, in which (\textit{i}) the problem (system) size was varied to include $N_{\rm water}=64,128,256$ water molecules (each with $N_{o} = 4 \times N_{\rm water}$ MLWFs), and (\textit{ii}) the number of processing elements ($N_{\rm proc}$ \mpi{} processes) was varied \via $\zeta = N_{\rm proc}/N_{o} = \sfrac{1}{2},1,2,4$.
In these calculations, we used one \mpi{} process per node on the \textit{Mira} IBM Blue Gene/Q platform, and used all of the $64$ hyperthreads available per node for the intranode \omp{} parallelization.

\begin{table*}[hbt!]
  \centering
  \caption{
  Computational timings profile for CPMD simulations of liquid water at the hybrid PBE0 level on the \textit{Mira} IBM Blue Gene/Q platform using the \exxm module in \texttt{QE}.
  Parameters include: (\textit{i}) the system size, which was varied to include $N_{\rm water}=64,128,256$ water molecules (each with $N_{o} = 4 \times N_{\rm water}$ MLWFs); (\textit{ii}) the number of \mpi{} processes ($N_{\rm proc}$), which was varied to cover $\zeta = N_{\rm proc}/N_{o} = \sfrac{1}{2},1,2,4$; and (\textit{iii}) the level of task-group parallelization in \texttt{QE}, which was varied to include $N_{\rm tg}=1,2,4,8,16$ task groups (see below and text for more details). 
  All timings have been averaged over $50$ CPMD steps, and are reported (in s/step) for the following \texttt{QE} modules: $\braket{t_{\rm GGA}}$, the walltime associated with the underlying GGA calculation; $\braket{t_{\rm MLWF}}$, the walltime associated with optimizing the Marzari-Vanderbilt functional (\ie to localize the MLWFs between CPMD steps as shown in Eq.~\eqref{eq:cpE_relocalize}); $\braket{t_{\exxm}}$, the walltime spent in the \exxm module; and $\braket{t_{\rm Total}}$, the total walltime associated with a given CPMD step.~\cite{note_no_invfft_time}
  To gauge the reproducibility of these timings, each CPMD simulation was run in triplicate; the observed fluctuations were always smaller than the precision reported (\ie $< 10^{-2}$~s) and were not included in the table.
  Also shown is the ratio, $\braket{t_{\exxm}}/\braket{t_{\rm GGA}}$, demonstrating that the walltime required to perform EXX-based CPMD simulations with the \exxm module are approximately $1\mathrm{-}3 \times$ that of the underlying GGA.
  All $\braket{t_{\exxm}}$ timings were further broken down into the walltime dedicated to computation events ($\braket{t_{\exxm}^{\rm comp}}$), communication overhead ($\braket{t_{\exxm}^{\rm comm}}$), and processor idling ($\braket{t_{\exxm}^{\rm idle}}$); the corresponding fractions of the $\braket{t_{\exxm}}$ walltime ($f_{\exxm}^{\rm comp}$, $f_{\exxm}^{\rm comm}$, and $f_{\exxm}^{\rm idle}$) were reported as percentages (in $\%$). 
  All timings reflect the fact that one \mpi{} process was executed per node on \textit{Mira}, and all $16$ physical cores (up to $64$ hyperthreads) per node were used for the intranode \omp{} parallelization.
  To utilize all available \mpi{} processes during the underlying GGA calculation, $N_{\rm tg} = 2^n \ge 1$ was set to the maximum possible value such that $N_{\rm tg} \times N_{\rm slab} \le N_{\rm proc}$ (see text for more details).
  }
  \begin{tabular}{ccccc|ccccc|ccc}
    \hline\hline
    \multicolumn{5}{c|}{Parameters} & \multicolumn{5}{c|}{\texttt{QE} Module Timings} & \multicolumn{3}{c}{Breakdown of $t_\exxm$} \\
    \hline
    $N_{\rm water}$ & $N_{o}$ & $N_{\rm proc}$ & $\zeta$ & $N_{\rm tg}$ & $\braket{t_{\rm GGA}}$ & $\braket{t_{\rm MLWF}}$ & $\braket{t_{\exxm}}$ & $\braket{t_{\rm Total}}$ & $\frac{\braket{t_{\exxm}}}{\braket{t_{\rm GGA}}}$ & $\braket{t_{\exxm}^{\rm comp}}$ ($f_{\exxm}^{\rm comp}$) & $\braket{t_{\exxm}^{\rm comm}}$ ($f_{\exxm}^{\rm comm}$) & $\braket{t_{\exxm}^{\rm idle}}$ ($f_{\exxm}^{\rm idle}$) \\
    \hline
    $64 $  & $256$  & $ 128$ & $1/2$ & $1     $ & $2.81$ & $0.16$ & $7.34$ &            $10.31$ & $2.6$ & $4.07$\quad($55.4$) & $0.96$\quad($13.1$) & $2.31$\quad($31.5$) \\
    $64 $  & $256$  & $ 256$ & $1$   & $1     $ & $1.97$ & $0.17$ & $3.83$ & \phantom{1}$ 5.97$ & $1.9$ & $2.05$\quad($53.4$) & $0.52$\quad($13.5$) & $1.27$\quad($33.1$) \\
    $64 $  & $256$  & $ 512$ & $2$   & $2     $ & $1.02$ & $0.16$ & $2.74$ & \phantom{1}$ 3.92$ & $2.7$ & $1.06$\quad($38.9$) & $0.39$\quad($14.3$) & $1.28$\quad($46.8$) \\
    $64 $  & $256$  & $1024$ & $4$   & $4     $ & $0.63$ & $0.16$ & $1.70$ & \phantom{1}$ 2.49$ & $2.7$ & $0.54$\quad($32.0$) & $0.37$\quad($21.6$) & $0.79$\quad($46.5$) \\
    \hline
    $128$  & $512$  & $ 256$ & $1/2$ & $1     $ & $5.19$ & $1.43$ & $8.27$ &            $14.89$ & $1.6$ & $4.60$\quad($55.6$) & $1.21$\quad($14.7$) & $2.46$\quad($29.8$) \\
    $128$  & $512$  & $ 512$ & $1$   & $2     $ & $2.64$ & $0.43$ & $4.35$ & \phantom{1}$ 7.42$ & $1.7$ & $2.35$\quad($54.1$) & $0.64$\quad($14.8$) & $1.36$\quad($31.2$) \\
    $128$  & $512$  & $1024$ & $2$   & $4     $ & $1.57$ & $0.41$ & $3.04$ & \phantom{1}$ 5.02$ & $1.9$ & $1.25$\quad($41.0$) & $0.51$\quad($16.9$) & $1.28$\quad($42.1$) \\
    $128$  & $512$  & $2048$ & $4$   & $8     $ & $0.96$ & $0.41$ & $1.96$ & \phantom{1}$ 3.33$ & $2.0$ & $0.67$\quad($34.4$) & $0.48$\quad($24.8$) & $0.80$\quad($40.9$) \\
    \hline
    $256$  & $1024$ & $ 512$ & $1/2$ & $2     $ & $6.39$ & $2.77$ & $8.34$ &            $17.50$ & $1.3$ & $4.19$\quad($50.2$) & $1.58$\quad($18.9$) & $2.57$\quad($30.8$) \\
    $256$  & $1024$ & $1024$ & $1$   & $4     $ & $3.59$ & $1.20$ & $4.80$ & \phantom{1}$ 9.59$ & $1.3$ & $2.23$\quad($46.5$) & $1.11$\quad($23.2$) & $1.46$\quad($30.4$) \\
    $256$  & $1024$ & $2048$ & $2$   & $8     $ & $2.23$ & $1.13$ & $3.33$ & \phantom{1}$ 6.69$ & $1.5$ & $1.26$\quad($38.0$) & $0.76$\quad($22.9$) & $1.30$\quad($39.1$) \\
    $256$  & $1024$ & $4096$ & $4$   & $16    $ & $1.59$ & $1.08$ & $2.41$ & \phantom{1}$ 5.08$ & $1.5$ & $0.77$\quad($31.9$) & $0.82$\quad($34.2$) & $0.82$\quad($33.9$) \\
    \hline\hline
  \end{tabular}
  \label{tab:timings}
\end{table*}

Initial structures for \ce{(H2O)64}, \ce{(H2O)128}, and \ce{(H2O)256} were systematically generated using the following procedure: (\textit{i}) randomly packing $N_{\rm water}=64,128,256$ water molecules into simple cubic unit cells with lattice parameters chosen to match the targeted density of $0.993$~g/cm$^{3}$; (\textit{ii}) equilibrating each of these randomly packed structures \via MD simulations in the $NVT$ ensemble using the TIP4P2005 force field~\cite{abascal_general_2005} at $300$~K for $1.0$~ns in \texttt{GROMACS}~\cite{abraham_gromacs:_2015}; (\textit{iii}) further equilibrating each of the TIP4P2005 structures \via CPMD simulations in the $NVT$ ensemble using the PBE0 hybrid xc functional~\cite{perdew_rationale_1996,adamo_toward_1999} at $330$~K for $\approx 250$~fs (\ie $500$~CPMD steps).
The computational timings reported herein are meant to reflect the walltime spent during EXX-based CPMD simulations in the $NVT$ ensemble, and were therefore averaged over $50$ additional CPMD steps starting from the equilibrated structures obtained using this three-step procedure.
Here we note that the TIP4P2005 MD simulations performed in step (\textit{ii}) were used to equilibrate the \textit{intermolecular} degrees of freedom in these systems, and the additional CPMD simulations performed in step (\textit{iii}) were used to ensure that the \textit{intramolecular} degrees of freedom (\ie the \ce{OH} bonds and \ce{HOH} angles) were equilibrated at the PBE0 level.
Since the temperature of these systems will rapidly increase once the rigid-molecule TIP4P2005 constraint is lifted, this additional CPMD equilibration step is important when determining a representative average for the walltime cost during EXX-based CPMD simulations in the $NVT$ ensemble.
During the CPMD simulations (\ie the $500$ equilibration steps and subsequent $50$ production steps), the temperature (of the ions) was controlled using massive Nos\'{e}-Hoover thermostats, each with a chain length of $4$~\cite{martyna_nosehoover_1992,tobias_molecular_1993}.
The nuclear and electronic degrees of freedom were integrated using the standard Verlet algorithm and a time step of $2.0$~au ($\approx 0.05$~fs); to ensure a clear adiabatic separation between the electronic and nuclear degrees of freedom during the CP dynamics, we used a fictitious electronic mass of $100$~au and the nuclear mass of deuterium for each hydrogen atom.
Interactions between the valence electrons and the ions (consisting of the nuclei and their corresponding frozen-core electrons) were treated using the Hamann-Schl{\" u}ter-Chiang-Vanderbilt (HSCV) type norm-conserving pseudopotentials~\cite{hamann_norm-conserving_1979,vanderbilt_optimally_1985} distributed with the \texttt{Qbox} package~\cite{gygi_architecture_2008}.
The valence electronic pseudo-wavefunctions were expanded in a planewave basis set which includes planewaves with a kinetic energy up to $85$~Ry.
Mass preconditioning was applied to all Fourier components of the electronic pseudo-wavefunctions with a kinetic energy above $25$~Ry~\cite{tassone_acceleration_1994}.
To enable distributed storage of all real-space quantities according to the \texttt{GRID} data distribution scheme in \texttt{QE} (see Fig.~\ref{fig:redist}), the real-space (and simple-cubic) grids utilized in these calculations were partitioned into $N_{\rm slab} = 140,176,220$ slabs along the $z$-direction for \ce{(H2O)64}, \ce{(H2O)128}, and \ce{(H2O)256}, respectively.
All computational timings were generated using an in-house development version of \texttt{QE} (based on \texttt{v5.0.2})~\cite{note_our_code_repo}.

Computational timings for each of these $12$ CPMD simulations of liquid water at the hybrid PBE0 level on \textit{Mira} are presented in Table~\ref{tab:timings}.
In this table, all timings have been averaged over $50$ CPMD steps and are reported (in s/step) for the following four \texttt{QE} modules: (\textit{i}) the walltime associated with the underlying GGA calculation ($\braket{t_{\rm GGA}}$); (\textit{ii}) the walltime associated with MLWF localization between each CPMD step \via nested SODD optimization of the Marzari-Vanderbilt functional ($\braket{t_{\rm MLWF}}$, see Eq.~\eqref{eq:cpE_relocalize}); (\textit{iii}) the walltime spent in the \exxm module ($\braket{t_{\exxm}}$, see Fig.~\ref{fig:flowchart}); and (\textit{iv}) the total walltime associated with a given CPMD step ($\braket{t_{\rm Total}}$)~\cite{note_no_invfft_time}.
To ensure a fair comparison between $\braket{t_{\rm GGA}}$ and $\braket{t_{\exxm}}$, the underlying GGA calculation utilized all \mpi{} processes available to the \exxm module.
This was accomplished using an existing two-tier parallelization scheme in \texttt{QE}, which allows for a computationally efficient execution of the \texttt{fwdFFT}/\texttt{invFFT} operation (\ie a typical bottleneck during GGA-based CPMD simulations).
The first parallelization tier takes advantage of the fact that the real-space grid has been partitioned into $N_{\rm slab}$ slabs along the $z$-direction (with each slab distributed to a particular \mpi{} process); this allows one to split the 3D \texttt{fwdFFT} and \texttt{invFFT} operations into 2D intra-slab FFT operations (which are executed in parallel without the need for additional communication) and 1D inter-slab FFT operations (which are also executed in parallel, but require communication among the pool of \mpi{} processes).
At the first tier, the underlying GGA calculation can utilize up to $N_{\rm proc} = N_{\rm slab}$ \mpi{} processes.
To enable the use of $N_{\rm proc} > N_{\rm slab}$ \mpi{} processes (when available), the second parallelization tier (\ie task-group parallelization) is employed to further distribute the independent 3D \texttt{FFT} operations associated with the $N_{o}$ orbitals.
At the second tier, $N_{\rm tg} = 2^n \ge 1$ can be set to the maximum possible value such that $N_{\rm tg} \times N_{\rm slab} \le N_{\rm proc}$, thereby enabling the underlying GGA calculation to utilize up to $N_{\rm proc} = N_{\rm tg} \times N_{\rm slab}$ \mpi{} processes.
We note in passing that the scalability of task-group parallelization depends on the communication bandwidth, and will often deteriorate when $N_{\rm tg} \gg 4$.
By using this approach, the underlying GGA calculation is able to utilize the pool of available \mpi{} processes, thereby ensuring a reasonably fair comparison between the $\braket{t_{\rm GGA}}$ and $\braket{t_{\exxm}}$ timings.

In the \texttt{QE} module timings in Table~\ref{tab:timings}, we observed that the \exxm module is still the overall bottleneck during hybrid DFT based CPMD simulations.
By exploiting both the natural sparsity of the exchange interaction and our massively parallel implementation, the walltime cost to evaluate all EXX-related quantities is comparable to that of the underlying GGA (\ie $\braket{t_\exxm} / \braket{t_{\rm GGA}}$ is now within the range of $\approx 1$--$3$).
We further stress that this ratio steadily decreases with increasing system size due to the more favorable scaling of the \exxm module (see below).
Since the \exxm module requires MLWFs (and the underlying GGA does not), we now discuss the additional cost needed to perform the nested SODD optimization of the Marzari-Vanderbilt functional ($\braket{t_{\rm MLWF}}$) between each CPMD step.
In all CPMD simulations performed herein, this MLWF refinement procedure (see Sec.~\ref{subsec:MLWFpropagation}) only required $3$--$4$ SODD steps (on average) per CPMD step.
As such, $\braket{t_{\rm MLWF}}$ only represents a minor contribution to $\braket{t_{\rm Total}}$ for systems containing $< 500$~atoms (\eg \ce{(H2O)64} and \ce{(H2O)128}).
For larger systems (\eg \ce{(H2O)256}), $\braket{t_{\rm MLWF}}$ can become quite substantial ($\approx 10\mathrm{-}20\%$ of $\braket{t_{\rm Total}}$) as the MLWF procedure requires cubic-scaling matrix operations.
As such, a more efficient MLWF localization procedure (which takes advantage of the sparsity of the MLWFs) will be required to efficiently utilize the \exxm module for system sizes that are significantly larger than \ce{(H2O)256}.

\begin{figure}[t!]
  \includegraphics[width=\linewidth,draft=false]{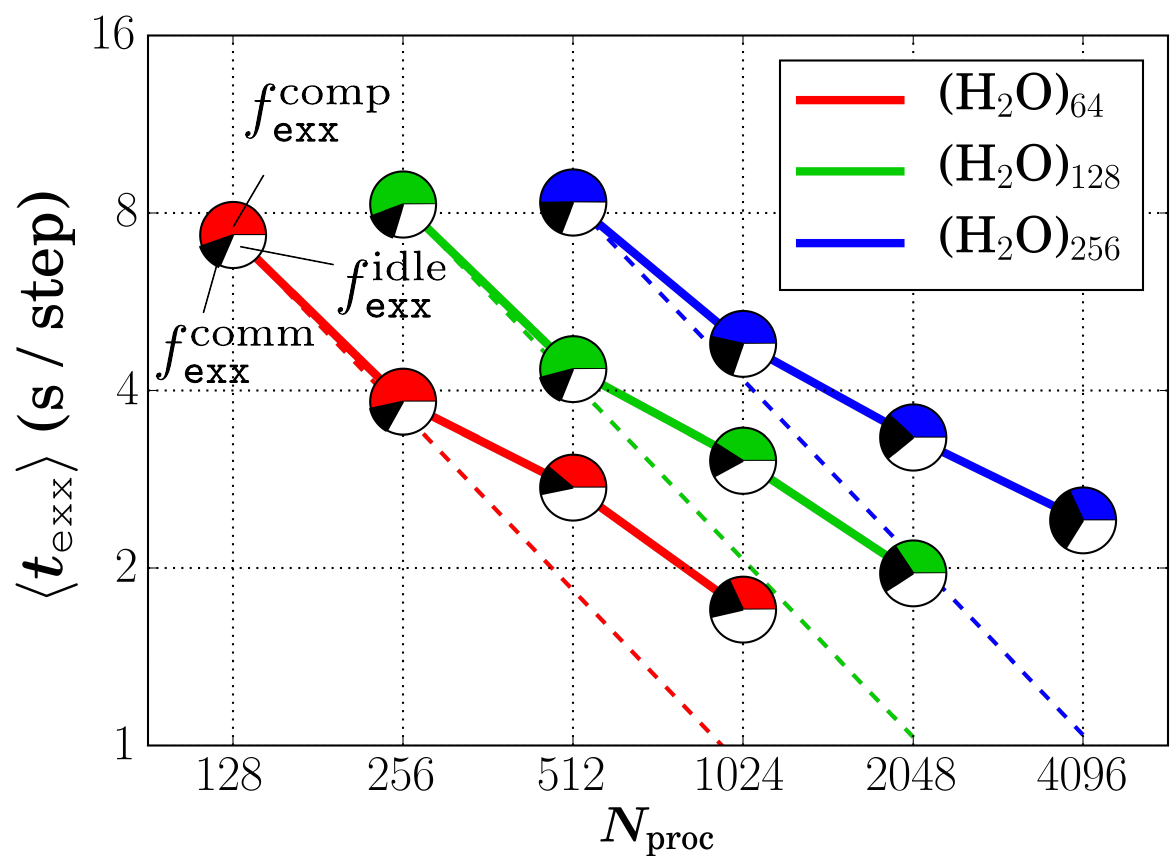}
  \caption{
  Strong-scaling analysis of the \mpi{} internode parallelization level in \exxm during CPMD simulations of liquid water at the hybrid PBE0 level on the \textit{Mira} IBM Blue Gene/Q platform.
  For a fixed system size (\ie \ce{(H2O)64} (red line), \ce{(H2O)128} (green line), and \ce{(H2O)256} (blue line)), the walltime spent in the \exxm module ($\braket{t_{\exxm}}$ in s/step, averaged over $50$ CPMD steps) is plotted versus the number of \mpi{} processes ($N_{\rm proc}$), which were varied to include $\zeta = N_{\rm proc}/N_{o} = \sfrac{1}{2},1,2,4$.
  For reference, ideal strong-scaling timings were plotted as dashed lines for each system size, and were computed with respect to the corresponding $\zeta_{\rm ref}=\sfrac{1}{2}$ timings.
  Pie plots were used to illustrate the fraction of $\braket{t_{\exxm}}$ dedicated to computation events ($f_{\exxm}^{\rm comp}$, colored), communication overhead ($f_{\exxm}^{\rm comm}$, black), and processor idling ($f_{\exxm}^{\rm idle}$, white).
  }
  \label{fig:mpi_scaling_st}
\end{figure}

Based on the timings in Table~\ref{tab:timings}, we assessed the strong-scaling behavior of the \exxm module by analyzing how $\braket{t_{\exxm}}$ changes as the number of processing elements was varied for a fixed problem size.
This was accomplished by changing $\zeta = N_{\rm proc}/N_{o} = \sfrac{1}{2},1,2,4$ when simulating \ce{(H2O)64}, \ce{(H2O)128}, and \ce{(H2O)256} (see Fig.~\ref{fig:mpi_scaling_st}).
For each system size, we computed the strong-scaling efficiency \via 
\begin{align}
  \eta^{\rm strong}_{\mpi{}}(\zeta) &\equiv 
  \frac{\zeta_{\rm ref} \cdot \braket{t_{\exxm}}_{\zeta_{\rm ref}}}{\zeta \cdot \braket{t_{\exxm}}_{\zeta}} = \frac{ \frac{1}{2} \cdot \braket{t_{\exxm}}_{\zeta = 1/2}}{\zeta \cdot \braket{t_{\exxm}}_{\zeta}} ,
  \label{eq:eff_mpi_st}
\end{align}
in which $\zeta_{\rm ref}=\sfrac{1}{2}$ was chosen as the reference (or baseline) $\zeta$ value (as this represents a realistic computational setup) and $\braket{t_{\exxm}}_{\zeta}$ is the walltime spent in the \exxm module for a given $\zeta$.
When averaged over all three systems, we find that $\eta^{\rm strong}_{\mpi{}}$ decreases to $\approx 93\%$ ($\zeta=1$), $\approx 66\%$ ($\zeta=2$), and $\approx 50\%$ ($\zeta=4$) as the number of processing elements is increased (see below for a more detailed discussion).
For even higher $\zeta$, the number of \mpi{} processes becomes comparable to the number of overlapping $\braket{ij}$ pairs; as such, $\eta^{\rm strong}_{\mpi{}}$ is expected to deteriorate even further for $\zeta \gg 4$.

\begin{figure}[t!]
  \includegraphics[width=\linewidth,draft=false]{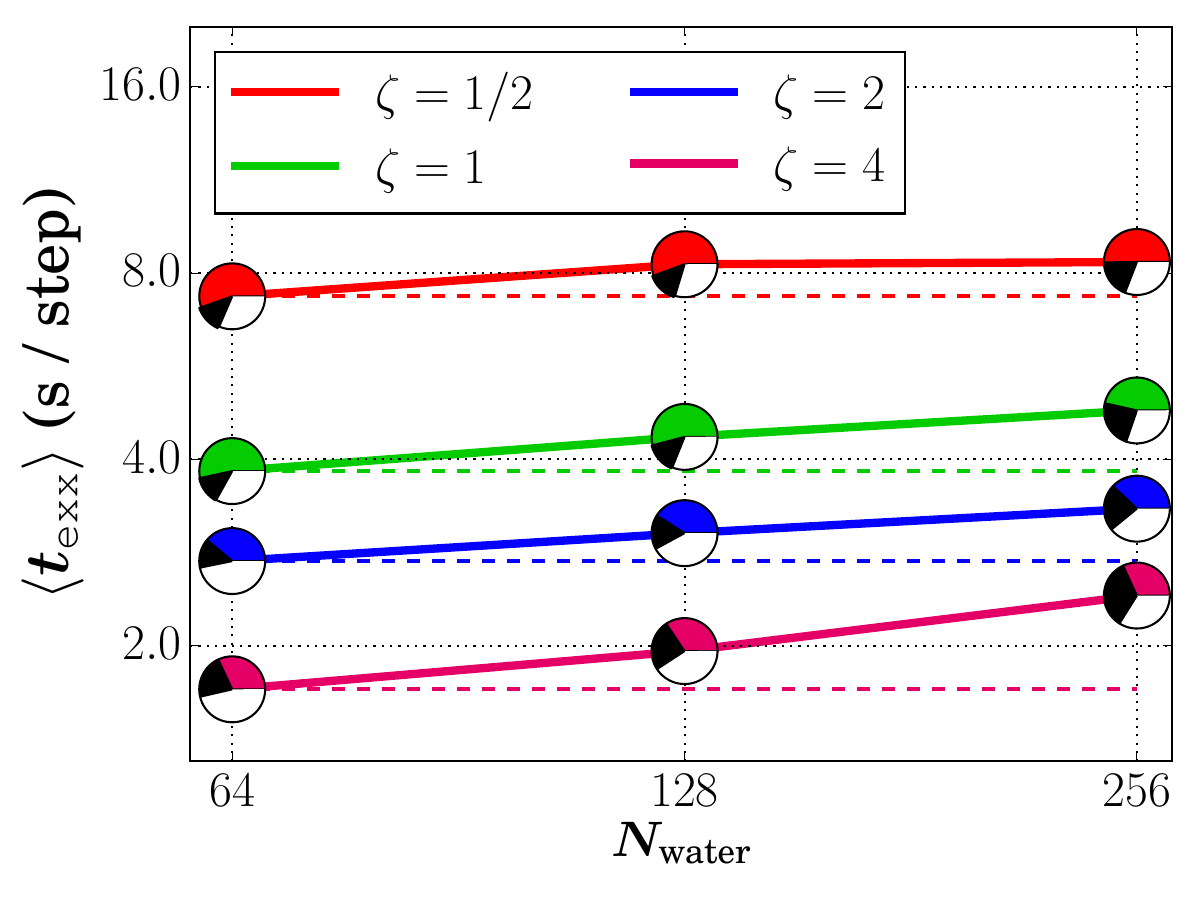}
  \caption{
  Weak-scaling analysis of the \mpi{} internode parallelization level in \exxm during CPMD simulations of liquid water at the hybrid PBE0 level on the \textit{Mira} IBM Blue Gene/Q platform.
  For a fixed ratio of system size to number of processing elements (\ie $\zeta=\sfrac{1}{2}$ (red line), $\zeta=1$ (green line), $\zeta=2$ (blue line), and $\zeta=4$ (magenta line)), the walltime spent in the \exxm module ($\braket{t_{\exxm}}$ in s/step, averaged over $50$ CPMD steps) is plotted versus the system size, which was varied to include $N_{\rm water}=64,128,256$ water molecules.
  For reference, ideal weak-scaling timings were plotted as dashed lines for each $\zeta$, and were computed with respect to the corresponding \ce{(H2O)64} timings.
  Pie plots were again used to illustrate the fraction of $\braket{t_{\exxm}}$ dedicated to computation events ($f_{\exxm}^{\rm comp}$, colored), communication overhead ($f_{\exxm}^{\rm comm}$, black), and processor idling ($f_{\exxm}^{\rm idle}$, white).
  }
  \label{fig:mpi_scaling_wk}
\end{figure}

Based on the timings in Table~\ref{tab:timings}, we also assessed the weak-scaling behavior of the \exxm module by analyzing how $\braket{t_{\exxm}}$ changes as the problem (system) size was varied for a fixed ratio of problem size to number of processing elements.
This was accomplished by considering \ce{(H2O)64}, \ce{(H2O)128}, and \ce{(H2O)256} for fixed values of $\zeta = N_{\rm proc}/N_{o} \in \{ \sfrac{1}{2},1,2,4\}$ (see Fig.~\ref{fig:mpi_scaling_wk}).
For each $\zeta$ value, we computed the weak-scaling efficiency \via 
\begin{equation}
  \eta^{\rm weak}_{\mpi{}}(N_{\rm water}) \equiv \frac{\braket{t_{\exxm}}_{N_{\rm water}^{\rm ref}}}{\braket{t_{\exxm}}_{N_{\rm water}}} = \frac{\braket{t_{\exxm}}_{N_{\rm water}=64}}{\braket{t_{\exxm}}_{N_{\rm water}}} , 
  \label{eq:eff_mpi_wk}
\end{equation}
in which $N_{\rm water}^{\rm ref}=64$ was chosen as the reference (or baseline) system size (as this represents a realistic computational setup) and $\braket{t_{\exxm}}_{N_{\rm water}}$ is the the walltime spent in the \exxm module for a given $N_{\rm water}$.
When averaged over all four $\zeta$ values, we find that $\eta^{\rm weak}_{\mpi{}}$ decreases to $\approx 89\%$ ($N_{\rm water}=128$) and $\approx 81\%$ ($N_{\rm water}=256$) as the system size is increased.
As shown in Fig.~\ref{fig:mpi_scaling_wk}, the \exxm module is quite scalable as the system size is increased, and the time-to-solution can be kept (relatively) constant for systems as large as \ce{(H2O)256} provided that a consistent (\ie fixed $\zeta$) amount of computational resources are available (see below for a more detailed discussion).

Despite the fact that the strong- and weak-scaling efficiencies of the \exxm module are not perfect, our algorithm is still able to furnish all EXX-related quantities in $\approx 2.4$~s for the largest system considered herein, \ie \ce{(H2O)256} with $\zeta = 4$.
As such, the \exxm module in \texttt{QE} enables relatively long (\eg $10$--$100$~ps) CPMD simulations for large-scale condensed-phase systems consisting of $500\mathrm{-}1000$~atoms at the hybrid DFT level of theory.
Quite interestingly, the overall cost of the \exxm module in this case (see Table~\ref{tab:timings} and Figs.~\ref{fig:mpi_scaling_st}--\ref{fig:mpi_scaling_wk}) can be decomposed into roughly equal contributions from computation ($f_{\exxm}^{\rm comp} \equiv \braket{t_{\exxm}^{\rm comp}}/\braket{t_{\exxm}} \approx \sfrac{1}{3}$), communication ($f_{\exxm}^{\rm comm} \equiv \braket{t_{\exxm}^{\rm comm}}/\braket{t_{\exxm}}\approx \sfrac{1}{3}$), and idling ($f_{\exxm}^{\rm idle} \equiv \braket{t_{\exxm}^{\rm idle}}/\braket{t_{\exxm}}\approx\sfrac{1}{3}$).
This breakdown of $\braket{t_{\exxm}}$ demonstrates that the \exxm algorithm is \textit{not} computation bound (as one might expect for the relatively large number of computation events required for hybrid DFT).
As discussed below, there still remains significant room for algorithmic improvements which would combat the relatively high cost associated with the communication overhead and processor idling, both of which are currently under development by our group and will be the topic of future work.
When combined with state-of-the-art preconditioners during the CG solution of the PE (which are also under intense development by our group), the computational cost of the current \exxm algorithm can be significantly sped up, which would further enable hybrid DFT-based AIMD simulations of large-scale condensed-phase systems across sufficiently longer timescales.

\textit{Computation Events}.
When further breaking down the computation events in the \exxm module (see Fig.~\ref{fig:flowchart}) that contribute to $\braket{t_{\exxm}^{\rm comp}}$, we find that the computational costs associated with Step IV (\texttt{Solution of Poisson's Equation}) and Step V (\texttt{Computation of Energy and Forces}) scale nearly ideally with $N_{\rm proc}$ (\eg $\eta^{\rm strong}_{\mpi{}} > 90\%$) and $N_{\rm water}$ (\eg $\eta^{\rm weak}_{\mpi{}} > 99\%$).
However, the computational effort in Step II (\texttt{Construction of Pair List}) required for determining the unique pair list (see Fig.~\ref{fig:pair_list} and Sec.~\ref{subsubsec:impl_spdomain}) was implemented in serial (in the current version of the \exxm module) and does not scale with $N_{\rm proc}$.
In addition, the computational cost associated with Step II grows quadratically with system size ($\mathcal{O}(N_o^2)$); although this step is quite cheap for smaller system sizes (\eg \ce{(H2O)64} and \ce{(H2O)128}), this cost can become more substantial for larger systems (\eg \ce{(H2O)256}). 
As a result, Step II (in its current form) leads to some of the deterioration (particularly for \ce{(H2O)256}) seen in $\eta^{\rm strong}_{\mpi{}}$ and $\eta^{\rm weak}_{\mpi{}}$ (see Figs.~\ref{fig:mpi_scaling_st}--\ref{fig:mpi_scaling_wk}).
In future versions of \exxm, we plan to mitigate this unnecessary computational cost by parallelizing Step II over \mpi{} processes and using a Verlet list (which will be updated periodically throughout a given CPMD simulation) to avoid unnecessary consideration of distant MLWF pairs; as such, these improvements will increase both the strong- and weak-scaling efficiencies of \exxm.
Although Step IV does scale nearly ideally with $N_{\rm proc}$ and $N_{\rm water}$, the computational cost associated with solving the PE for each overlapping $\braket{ij}$ pair still remains the dominant contribution to $\braket{t_{\exxm}^{\rm comp}}$.
In the near future, we plan to significantly reduce this primary source of computational effort by using more sophisticated guesses (for $\tv{ij}$) in conjunction with novel preconditioners during the CG solution of the PE.
An additional future direction to overcome this computational hurdle might also involve offloading all computation events (not necessarily limited to the solution of the PE) to graphical processing units (GPUs), which provide significantly higher computational throughput than CPUs.

\textit{Communication Overhead}. 
In addition to the computation events described above, the communication overhead in Fig.~\ref{fig:flowchart} also contributes to the degradation of $\eta^{\rm strong}_{\mpi{}}$ and $\eta^{\rm weak}_{\mpi{}}$ observed in Figs.~\ref{fig:mpi_scaling_st}--\ref{fig:mpi_scaling_wk}.
Here, this non-ideal scaling behavior mainly originates from: (\textit{i}) the sending/receiving of MLWFs ($\{\tphi{i}\}$) in Step III (\texttt{Communication of MLWFs}) and the sending/receiving of wavefunction forces ($\{\dxx{i}\}$) at the conclusion of Step V (\texttt{Computation of Energy and Forces}), as well as (\textit{ii}) the redistribution of $\{\tphi{i}\}$ in Step I (\texttt{Redistribution of MLWFs}) and the redistribution of $\{\dxx{i}\}$ in Step VI (\texttt{Redistribution of Wavefunction Forces}).
In this regard, the former is more important for relatively smaller systems (\eg \ce{(H2O)64} and \ce{(H2O)128}) employing fewer processing elements (\eg $\zeta = \sfrac{1}{2}$ and $\zeta = 1$) due to the fact that each \mpi{} process needs to send/receive significantly more MLWFs and wavefunction forces (see Secs.~\ref{subsubsec:impl_commphi} and \ref{subsubsec:impl_compute_esf}).
In the same breath, the latter dominates for larger systems (\eg \ce{(H2O)256}) employing more processing elements (\eg $\zeta = 2$ and $\zeta = 4$) due to the \texttt{ALL-TO-ALL} communication events in Steps I and VI (see Fig.~\ref{fig:redist}, and Secs.~\ref{subsubsec:impl_redist_phi} and \ref{subsubsec:impl_redist_d}).
As a result, these communication events lead to a noticeable upward tilt in the strong- and weak-scaling curves in Figs.~\ref{fig:mpi_scaling_st}--\ref{fig:mpi_scaling_wk}, which is particularly evident in the large system and $\zeta$ limit.
To attack the communication overhead associated with the sending/receiving of $\{\tphi{i}\}$ and $\{\dxx{i}\}$, we plan to implement an asynchronous (non-blocking) communication protocol that will overlap with the computation events in Steps IV--V.
By doing so, the serial communication--computation--communication process in Steps III--V (\ie communication of $\{\tphi{i}\}$, followed by the solution of the PE for all overlapping pairs and computation of the $\braket{ij}$ contribution to $\exx$ and $\{\dxx{i}\}$, followed by communication of $\{\dxx{ij}\}$) can be overlapped to effectively mask the communication overhead (see the right panel of Fig.~\ref{fig:pair_list}).
To attack the communication overhead associated with the redistribution of $\{\tphi{i}\}$ and $\{\dxx{i}\}$, we plan to exploit the locality of the MLWFs by only performing the redistribution based on the compact supports of each MLWF.
By doing so, this algorithmic improvement has the potential to completely eliminate the unnecessary \texttt{ALL-TO-ALL} communication events in Steps I and VI, which would significantly reduce the communication overhead in the \exxm module.
As an added bonus, this approach would also allow for a more accurate evaluation of $\{\dxx{ij}\}$ on $\Omega_j$ \via Eq.~\eqref{eq:Dxx_mlwf_omega}, thereby eliminating any residual error in the wavefunction forces (see Fig.~\ref{fig:conv_dxx}).
In addition to the above strategies, we also plan to port the \exxm module to parallel GPU architectures (\eg with NVLink technology), which will allow us to exploit faster peer-to-peer connections and further reduce the communication overhead.

\textit{Processor Idling}.
The last and most critical issue that limits the strong- and weak-scaling efficiency in the \exxm algorithm is processor idling due to workload imbalance.
This imbalance mainly originates from: (\textit{i}) the imperfect distribution of overlapping $\braket{ij}$ pairs across the pool of \mpi{} processes (see Fig.~\ref{fig:pair_list} and Sec.~\ref{subsubsec:impl_spdomain}), and (\textit{ii}) the variability in the number of CG steps required during the solution of the PE for each overlapping $\braket{ij}$ pair.
In the current \exxm module, these issues are primarily due to the static load-balancing algorithm described in Sec.~\ref{subsubsec:impl_spdomain}, which assumes that the computational workload (\ie the number of CG steps) associated with the solution of the PE is equivalent for each $\braket{ij}$ pair, and limits this computation to the $P_{i}$ or $P_{j}$ \mpi{} processes only (and not $P_{k}$ for example). 
To attack the processor idling, we plan to remove these limitations by employing a task-based load-balancing algorithm which will account for the workload imbalance using a dynamic scheduler, and has the flexibility to assign (and even reassign) a given $\braket{ij}$ task to any available \mpi{} process.
In addition to the load-balancing algorithm, we also plan to implement more intelligent initial guesses for $\tv{ij}$ (to aid in the convergence of the CG solution to the PE) as well as the aforementioned asynchronous communication protocol (which will overlap with the computational events in Step IV), and expect that both of these algorithmic improvements will also mitigate the processor idling in the \exxm module.
The current \exxm module also faces challenges associated with processor idling when there is an inherent workload imbalance (in the number of overlapping pairs per MLWF) due to the physical nature of the problem, \ie the (inherent and transient) heterogeneity present in systems containing interfaces as well as disordered systems (like liquids).
To balance the workload in such heterogeneous systems, one could employ a different parallelization level for each MLWF (\ie $\zeta = \zeta(i)$), which would allow the \exxm module to dynamically adopt the number of processing elements dedicated to a given MLWF based on its number of overlapping $\braket{ij}$ pairs.

\subsubsection{Intranode Parallelization \via \omp{} \label{subsubsec:f-g_para}}

\begin{figure}[t!]
  \includegraphics[width=\linewidth,draft=false]{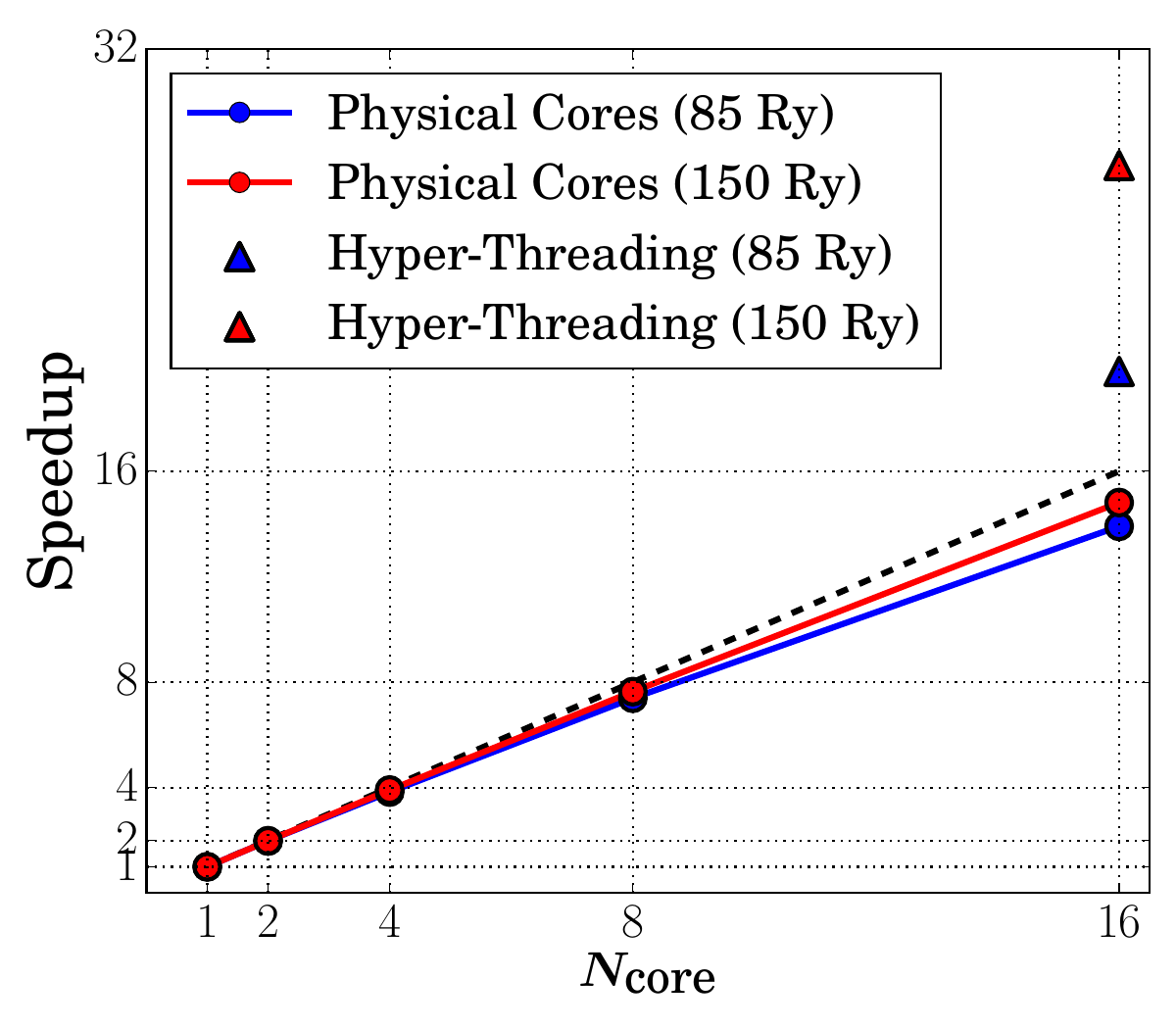}
  \caption{
  Strong-scaling analysis of the \omp{} intranode parallelization level in \exxm during CPMD simulations of \ce{(H2O)64} at the hybrid PBE0 level on the \textit{Mira} IBM Blue Gene/Q platform.
  For a fixed system and basis set size (\ie \ce{(H2O)64} with a planewave kinetic energy cutoff of $85$~Ry (blue line) and $150$~Ry (red line)), the speedup in the walltime spent in Step IV: Solution of Poisson's Equation of Fig.~\ref{fig:flowchart} ($\braket{t_{\exxm}^{\rm Step IV}}$, averaged over $50$ CPMD steps) is plotted versus the number of physical cores, which were varied to include $N_{\rm thread} = N_{\rm core} = 1,2,4,8,16$.
  Beyond the maximum number of physical cores ($N_{\rm core} = 16$) per node on \textit{Mira}, the \omp{} intranode parallelization level can further utilize hyperthreading technology to access up to $N_{\rm thread} = 64$ (hyper)threads (depicted by the blue and red triangles).
  For reference, the ideal strong-scaling performance (using up to $N_{\rm core} = 16$) was plotted as a dashed line, and normalized to unity for $N_{\rm core} = 1$.
  }
  \label{fig:omp_strong_scaling}
\end{figure}
Within each \mpi{} process, the \exxm module uses \omp{} threading to further parallelize the following operations for each overlapping $\braket{ij}$ pair: (\textit{i}) the CG solution of the PE for the near-field $\tv{ij}$ (Step IV in Fig.~\ref{fig:flowchart}), (\textit{ii}) the multipole expansion for the far-field $\tv{ij}$ (Step IV), and (\textit{iii}) other (less computationally intensive) operations (\eg proto-subdomain construction in Step II, $\{\tphi{i}\}$ loading/off-loading in Step III, and $\exx$ integration in Step V).
To critically assess the strong-scaling performance of the intranode \omp{} parallelization (in analogy to the internode \mpi{} parallelization in Sec.~\ref{subsubsec:c-g_para}), we analyzed how $\braket{t_{\rm Step \, IV}}$ (the typical computational bottleneck in the \exxm module) changes as the number of \omp{} threads ($N_{\rm thread}$) was varied during a CPMD simulation of \ce{(H2O)64} with $\zeta = 1$.
To maintain a consistent internode communication pattern, we used one \mpi{} process per node on the \textit{Mira} IBM Blue Gene/Q architecture; since each node contains $16$ physical cores, five different levels of \omp{} parallelization were assessed by varying $N_{\rm thread} \in \{ 1,2,4,8,16 \}$ threads across $N_{\rm core} = N_{\rm thread}$ physical cores (see Fig.~\ref{fig:omp_strong_scaling}).
For each $N_{\rm thread}$ value, we computed the strong-scaling efficiency \via 
\begin{equation}
  \eta^{\rm strong}_{\omp{}}(N_{\rm thread}) \equiv \frac{\braket{t_{\exxm}^{\rm Step \, IV}}_{N_{\rm thread}=1}}{N_{\rm thread} \cdot \braket{t_{\exxm}^{\rm Step \, IV}}_{N_{\rm thread}}} ,
  \label{eq:eff_omp_st}
\end{equation}
in which $N_{\rm thread} = N_{\rm core} = 1$ was chosen as the reference (or baseline) \omp{} setting and $\braket{t_{\exxm}^{\rm Step \, IV}}_{N_{\rm thread}}$ is the walltime spent in Step IV of the \exxm module for a given $N_{\rm thread}$.
Here, we find that the computational costs associated with Step IV scale very well with $N_{\rm thread}$ (\eg $\eta^{\rm strong}_{\omp{}} = 84\%$ when using all $16$ physical cores) when using a typical constant-volume ($NVT$) planewave basis setting (\ie $85$~Ry kinetic energy cutoff).
With a heavier workload (\ie a typical constant-pressure ($NpT$) planewave basis setting with a $150$~Ry kinetic energy cutoff), we find that the strong-scaling efficiency of the \exxm module is significantly better and maintains a nearly ideal efficiency of $\eta^{\rm strong}_{\omp{}} = 92\%$ when using all $16$ physical cores.
We note in passing that hyperthreading each physical core on \textit{Mira} into four logical cores yields an additional boost (\ie $30\mathrm{-}40\%$ speedup) in the computational performance of \exxm.

\begin{table*}[ht!]
  \centering
  \caption{
  Computational timings profile for CPMD simulations of liquid water at the hybrid PBE0 level on the \textit{Mira} IBM Blue Gene/Q, \textit{Cori} Haswell, and \textit{Cori} KNL platforms using the \exxm module in \texttt{QE}. All timings (in s/step) correspond to an average over $50$ CPMD steps for \ce{(H2O)128} with $\zeta=1$ and $N_{\rm tg} = 2$ (see Table~\ref{tab:timings} and the text for more details). All timings reflect the fact that one \mpi{} process was executed per node on the given architecture, and all available physical cores per node were used for the intranode \omp{} parallelization; unless otherwise specified, hyperthreading was fully activated on each physical core.
  }
  \begin{tabular}{cc|ccccc|ccc}
    \hline\hline
    \multicolumn{2}{c|}{Architecture}    & \multicolumn{5}{c|}{\texttt{QE} Module Timings} & \multicolumn{3}{c}{Breakdown of $t_\exxm$} \\
    \hline
    Machine & CPU & $\braket{t_{\rm GGA}}$ & $\braket{t_{\rm MLWF}}$ & $\braket{t_{\exxm}}$ & $\braket{t_{\rm Total}}$ & $\frac{\braket{t_{\exxm}}}{\braket{t_{\rm GGA}}}$ & $\braket{t_{\exxm}^{\rm comp}}$ ($f_{\exxm}^{\rm comp}$) & $\braket{t_{\exxm}^{\rm comm}}$ ($f_{\exxm}^{\rm comm}$) & $\braket{t_{\exxm}^{\rm idle}}$ ($f_{\exxm}^{\rm idle}$) \\
    \hline
    \textit{Mira} & IBM BlueGene/Q           & $2.64$ & $0.43$ & $4.35$ & $\phantom{1}7.42$ & $1.7$ & $2.35$\quad($54.1$) & $0.64$\quad($14.8$) & $1.36$\quad($31.2$) \\
    \textit{Cori} & Haswell              & $1.07$ & $0.72$ & $1.66$ & $\phantom{1}3.45$ & $1.6$ & $0.84$\quad($50.8$) & $0.37$\quad($22.2$) & $0.45$\quad($27.0$) \\
    \textit{Cori} & KNL                  & $4.45$ & $1.76$ & $6.53$ & $12.74$           & $1.5$ & $3.51$\quad($53.6$) & $1.50$\quad($23.0$) & $1.53$\quad($23.4$) \\
    \hline
    \textit{Cori} & KNL (no hyperthreading) & $5.38$ & $1.15$ & $3.57$ & $10.10$           & $0.7$ & $1.84$\quad($51.7$) & $1.03$\quad($28.8$) & $0.70$\quad($19.5$) \\
    \hline\hline
  \end{tabular}
  \label{tab:arch_bench}
\end{table*}
\subsubsection{Performance on Other Supercomputer Architectures \label{subsubsec:cori-performance}}
To provide an additional assessment of the performance of the \exxm module, we performed an analogous computational analysis on the \textit{Cori} Haswell and KNL supercomputer architectures housed at the National Energy Research Scientific Computing Center (NERSC).
For simplicity, we considered the \ce{(H2O)128} test case described above, and limited our analysis to the most common $\zeta = 1$ case (in which $N_{\rm proc} = N_{o} = 512$).
For the \mpi{} parallelization level, we employed one \mpi{} process per node on the \textit{Cori} Haswell and KNL architectures.
For the \omp{} parallelization level, we used all physical cores on each node (\ie $24$ and $68$ for the Haswell and KNL architectures, respectively).
When specified, we fully activated hyperthreading on each physical core, which corresponds to a maximum total of $48$ and $272$ \omp{} threads for each Haswell and KNL node, respectively.
As shown in Table~\ref{tab:arch_bench}, the \exxm module behaves quite consistently across all three architectures considered (\ie \textit{Mira} IBM BlueGene/Q, \textit{Cori} Haswell, and \textit{Cori} KNL).
Here, we first not that $\braket{t_{\exxm}}/\braket{t_{\rm GGA}}$ is fairly constant and fluctuates between $1.5\mathrm{-}1.7$; as such, the \exxm module enables hybrid DFT calculations with a walltime cost that is comparable to semi-local DFT on all three architectures.
We further note that the fractional breakdown of $\braket{t_{\exxm}}$ into computation, communication, and processor idling is also similar; as such, our comprehensive three-pronged strategy (\textit{vide supra}) to attack each of these contributions is likely to lead to a robust \exxm module with significantly improved performance across a wide array of HPC architectures.
When comparing $\braket{t_{\exxm}}$ across these architectures, we find that \textit{Cori} Haswell (with $512 \times 24 = 12,288$ physical cores) has a faster turnaround (by $\approx 2.6\times$) than \textit{Mira} IBM BlueGene/Q (with $512 \times 16 = 8,192$ physical cores), while \textit{Cori} KNL (with $512 \times 68 = 34,816$ physical cores) is noticeably slower (by $\approx 0.67\times$).
We note in passing that hyperthreading introduces noticeable performance improvements on the IBM BlueGene/Q and Haswell architectures, but leads to a significant decrease in the performance of \exxm on the KNL architecture.
For instance, deactivating the hyperthreading option on KNL leads to an $\approx 80\%$ speedup in $\braket{t_{\exxm}}$ and an $\approx 20$\% slowdown in $\braket{t_{\rm GGA}}$; in doing so, $\braket{t_{\exxm}}/\braket{t_{\rm GGA}} = 0.7$ and the EXX calculation is now faster than the corresponding GGA calculation.

\section{Conclusions and Future Outlook \label{sec:conclusion}}

In this work, we presented a detailed discussion of the theoretical framework, algorithmic implementation, and computational performance of a linear scaling approach that exploits sparsity in the real-space evaluation of the EXX interaction in finite-gap condensed-phase systems by utilizing a localized (MLWF) representation of the occupied orbitals.
Our theoretical discussion focused on the integration of this approach into CPMD, and highlighted the central role played by $\tv{ij}$---the MLWF-product potential obtained \via the CG solution of Poisson's equation for the corresponding MLWF-product density $\trho{ij}$---in the evaluation of the EXX energy and wavefunction forces.
We then provided a comprehensive description of the \exxm algorithm, which has been implemented in the \texttt{CP} module of the open-source \texttt{QE} package, and employs a hybrid \mpi{}/\omp{} parallelization scheme to efficiently utilize the HPC resources available on current- and next-generation supercomputer architectures.
This was followed by a critical assessment of the accuracy and parallel performance (\eg strong and weak scaling) of this approach when performing large-scale AIMD simulations of liquid water in the canonical ($NVT$) ensemble.
With access to HPC resources, we demonstrated that \exxm enables us to compute the EXX contribution to the energy and wavefunction forces for \ce{(H2O)256}, a condensed-phase system containing $\approx 750$~atoms, in just under $2.4$~s.
With a walltime cost that is comparable to semi-local DFT, the \exxm module takes us one step closer to routinely performing AIMD simulations of complex and large-scale condensed-phase systems for sufficiently long timescales at the hybrid DFT level of theory.

In its current form, the \exxm module can also be used for high-throughput applications such as the generation of high quality AIMD data for training, developing, and testing next-generation machine-learning and neural-network based force fields~\cite{han2017deep,zhang2018deepmd,zhang_end--end_2018,ko_isotope_2019}.
Despite the favorable scalability and computational performance of \exxm, however, we found that this algorithm is not computation-bound for larger systems (such as \ce{(H2O)256)}) with an overall walltime that can be roughly split into three equivalent contributions: computation events, communication overhead, and processor idling (due to workload imbalance).
As such, there still exits significant room for improving the performance of the \exxm module, and we are currently in the process of implementing a comprehensive three-pronged strategy that will attack each of these contributions and significantly reduce the overall walltime cost.
Inspired by the work of Gygi and coworkers~\cite{gygi_compact_2009,gygi_efficient_2013,dawson_performance_2015}, we are also implementing an MLWF-specific domain strategy (as outlined in Secs.~\ref{sec:RealSpaceEXX} and \ref{sec:transferability}) that will allow the $\beta$-version of \exxm to perform accurate and efficient hybrid DFT simulations of condensed-phase systems ranging from large-gap homogeneous systems like liquid water (with a narrow distribution of MLWF spreads) to small-gap heterogeneous systems like solvated semiconducting nanoparticles (with a wide distribution of MLWF spreads).

In addition to improving the performance and generality of \exxm, our group is also in the process of extending this approach to perform BOMD simulations, as well as generalizing this MLWF-based framework to enable linear-scaling and highly accurate evaluations of screened and range-separated exchange~\cite{heyd_hybrid_2003,gerber_hybrid_2005,vydrov_assessment_2006,janesko_screened_2009,baer_tuned_2010,kronik_excitation_2012,karolewski_using_2013}.
Other improvements can also be straightforwardly incorporated into \exxm such as alternative localization schemes that are better suited to treat heterogeneous systems~\cite{gygi_efficient_2013,dawson_performance_2015} and metals~\cite{damle_variational_2018,cornean_localised_2019}, or furnish localized orbitals in a non-iterative fashion to avoid convergence issues~\cite{damle_compressed_2015,damle_computing_2017,damle_scdm-k:_2017,damle_disentanglement_2018}.
From a wavefunction theory point of view, \exxm is also a viable approach for the mean-field HF approximation, and is a logical starting point for enabling condensed-phase AIMD with local electron correlation methods.

Interested users can find \exxm implemented in the most recent version of \texttt{QE}~\cite{giannozzi_advanced_2017}, with additional information (including a detailed description of the \exxm keywords) available in the \texttt{QE} user's manual online~\cite{footnote_qe_doc_url}.
In the next paper in this series, we will generalize our MLWF-based EXX approach to treat arbitrary Bravais lattice based simulation cells, as well as derive (and implement) the EXX contributions to the cell forces (\ie the stress tensor).
These extensions to \exxm are needed for performing constant-pressure AIMD simulations in the $NpH$ and $NpT$ ensembles, and will enable us to model large condensed-phase systems under realistic thermodynamic conditions (\ie at finite $T$ and $p$) at the hybrid DFT level of theory.

\begin{acknowledgements}
  H-YK, JJ, and RD acknowledge partial support from Cornell University through start-up funding and the Center for Alkaline Based Energy Solutions (CABES), an Energy Frontier Research Center funded by the U.S. Department of Energy, Office of Science, Office of Basic Energy Sciences, under Award No. DE-SC0019445.
  H-YK and RC gratefully acknowledge support from the U.S. Department of Energy under Grant No. DE-SC0019394.
  XW acknowledges support from  National Science Foundation through Award No. DMR-1552287.
  This research used resources of the National Energy Research Scientific Computing (NERSC) Center, which is supported by the Office of Science of the U.S. Department of Energy under Contract No. DE-AC02-05CH11231.
  This research used resources of the Argonne Leadership Computing Facility at Argonne National Laboratory, which is supported by the Office of Science of the U.S. Department of Energy under Contract No. DE-AC02-06CH11357.
  Additional resources were provided by the Terascale Infrastructure for Groundbreaking Research in Science and Engineering (TIGRESS) High Performance Computing Center and Visualization Laboratory at Princeton University.
\end{acknowledgements}


\begin{mcitethebibliography}{158}
\bibitem[not()]{note_no_invfft_time}
Despite the fact that the real-space MLWFs had already been generated (with no
  additional cost) during the formation of the real-space electron density
  (needed to evaluate the semi-local xc energy), we found an additional (and
  completely unnecessary) \texttt{invFFT} call that specifically re-generates
  the real-space MLWFs in the current version of \texttt{QE}. As such, we have
  not included this additional cost in $\braket{t_{\rm Total}}$, and plan to
  remove this unnecessary routine from future versions of \texttt{QE}.\relax
\mciteBstWouldAddEndPunctfalse
\mciteSetBstMidEndSepPunct{\mcitedefaultmidpunct}
{}{\mcitedefaultseppunct}\relax
\EndOfBibitem
\end{mcitethebibliography}

\providecommand{\latin}[1]{#1}
\makeatletter
\providecommand{\doi}
  {\begingroup\let\do\@makeother\dospecials
  \catcode`\{=1 \catcode`\}=2\doi@aux}
\providecommand{\doi@aux}[1]{\endgroup\texttt{#1}}
\makeatother
\providecommand*\mcitethebibliography{\thebibliography}
\csname @ifundefined\endcsname{endmcitethebibliography}
  {\let\endmcitethebibliography\endthebibliography}{}

\end{document}